\documentclass{article}

\usepackage{arxiv}
\usepackage[utf8]{inputenc}
\usepackage[american]{babel}
\usepackage[style=apa,backend=biber]{biblatex}
\DeclareLanguageMapping{american}{american-apa}
\usepackage{amsmath, amssymb, amsfonts, amsthm, amsbsy}
\usepackage{graphicx}
\usepackage{caption}
\usepackage{subcaption}
\usepackage{float}
\usepackage{rotating}
\usepackage{tabularx}
\usepackage{longtable}
\usepackage{multirow}
\usepackage{numprint}
\usepackage[T1]{fontenc}
\usepackage{hyperref}
\usepackage{url}
\usepackage{booktabs}
\usepackage{amsfonts}
\usepackage{nicefrac}
\usepackage{microtype}
\usepackage{cleveref}
\usepackage{csquotes}

\usepackage{xcolor}
\definecolor{darkred}{RGB}{165,30,55}
\definecolor{darkblue}{RGB}{0, 0, 139}
\definecolor{darkgreen}{RGB}{1, 100, 32}

\usepackage{tikz}
\tikzstyle{ov}=[shape=rectangle,draw=black!80,minimum height=0.6cm,
                minimum width=0.6cm,thick]
\tikzstyle{lv}=[shape=circle,draw=black!80,thick,minimum width=0.7cm]
\tikzstyle{lr}=[shape=circle,draw=black!00,thick,minimum width=0.7cm]

\newtheorem{theorem}{Theorem}

\newtheoremstyle{assumptionstyle}
  {3pt}{3pt}{\normalfont}{}{\bfseries}{:}{.5em}
  {\thmname{#1} \thmnumber{#2} -\thmnote{ #3}}
\theoremstyle{assumptionstyle}
\newtheorem{causalassumption}{Assumption}

\newtheorem{structuralassumption}{Assumption}

\newtheorem{statassumption}{Assumption}

\title{RAPSEM: Identifying Latent Mediators Without Sequential Ignorability via a Rank-Preserving Structural Equation Model}

\author{
 Sofia Morelli \\
  Methods Center\\
  Eberhard Karls Universität Tübingen\\
  72074 Tübingen \\
  \texttt{sofia.morelli@uni-tuebingen.de} \\
   \And
 Roberto Faleh \\
  Methods Center\\
  Eberhard Karls Universität Tübingen\\
  72074 Tübingen \\
  \And
 Holger Brandt \\
  Methods Center\\
  Eberhard Karls Universität Tübingen\\
  72074 Tübingen \\
}

\begin{document}
\maketitle
\begin{abstract}
Standard structural equation models (SEMs) are often used to identify latent mediators. However, valid inference typically relies on the strong, frequently violated Sequential Ignorability assumption. We introduce the Rank-Preserving Structural Equation Model (RAPSEM), which increases robustness through $G$-estimation while maintaining the measurement model's integrity through a two-stage method of moments (2SMM) for factor score corrections. RAPSEM replaces the no unmeasured mediator-outcome confounding with the weaker no unobserved effect modification assumption. By leveraging treatment randomization, RAPSEM achieves identification in a manner equivalent to instrumental variable estimation through structurally emerging instruments. Specifically, identification relies on treatment-covariate interactions that influence the mediator but have no direct effect on the outcome, allowing researchers to utilize natural heterogeneity in treatment response as a testable source of identification. We provide a robustness assessment for the core identifying assumption and establish the consistency and asymptotic normality of the resulting estimator. Simulation studies demonstrate that RAPSEM remains unbiased under unobserved confounding, whereas standard SEM yields biased results. RAPSEM achieves reasonable power for sample sizes above $500$, depending on the strength of the structural instruments. The method is implemented in the accompanying \texttt{rapsem} R package, and its practical utility is illustrated through an empirical example from educational research.
\end{abstract}


Mediation analysis is a central topic in applied psychological research, particularly when designing and evaluating intervention studies \parencite{Windgassen2016}. 
A mediator variable represents part of the causal pathway connecting an intervention to an outcome \parencite{Holland1988}. That is, the intervention influences the mediator, which in turn affects the outcome. Positioned temporally and conceptually between intervention and outcome, mediators explain how and why treatments work, which offers several advantages. 

First, mediators provide insight into the mechanisms through which interventions exert their effects. For example, cognitive behavioral therapy (CBT) may reduce social anxiety symptoms by first reducing maladaptive beliefs, suggesting that cognitive change is the driving force behind therapeutic improvement \parencite{Boden2012, DeCastella2015}. Second, understanding which variables serve as mediators can guide the refinement of interventions: if a specific mechanism is identified, treatment protocols can be more precisely tailored to target it. Third, mediators can act as early indicators of treatment success because they precede the outcome in the causal sequence. When such variables are easier or less costly to measure, this allows for more timely and resource-efficient evaluations.

A difficulty in identifying mediators is potential unobserved confounders. While conducting a randomized controlled trial (RCT) protects the treatment-outcome and treatment-mediator pathways, it does not shield the mediator-outcome relationship from confounding. Experimentally isolating this effect would require the direct manipulation and randomization of the mediator itself. However, such an approach is typically challenging or even impossible for psychological constructs. Consequently, most applied studies rely on the assumption of Sequential Ignorability, which presumes the absence of unmeasured mediator-outcome confounding \parencite{Imai2010}. \textcite{Vo2020} found that $96\%$ of mediation studies employ traditional methods based on this assumption \parencite[for examples see][]{Brandt2024}.

Sequential ignorability requires that all relevant covariates influencing both the mediator and the outcome are included in the model, a standard that is rarely achievable in practice. Researchers might adjust for variables such as baseline symptom severity, age, gender, or duration of symptoms. Yet other plausible confounders, such as interpersonal sensitivity, peer rejection history, or recent social stressors, are more difficult to assess. Biological markers, such as genetic predispositions and cortisol reactivity, may be prohibitively expensive to obtain, and contextual factors, like population density or seasonal social activity levels, are easily overlooked. Indeed, \textcite{Vo2020} found that only $50\%$ of mediation studies adjusted for mediator-outcome confounders, and those that did often relied solely on baseline values of the mediator and outcome.

The failure to account for confounding has serious statistical consequences. Omitting even a single relevant confounder can severely bias mediation estimates, typically resulting in an overestimation of the mediated effect \parencite{Fritz2016}. This compromises the validity of the findings and may lead to overly optimistic conclusions about treatment mechanisms.

These limitations have motivated researchers in fields such as biomedicine to develop methods that are robust to omitted variable bias, with $G$-estimation being a particularly promising direction that remains underused due to its complex implementation \parencite{VansteelandtJoffe2014}.
\textcite{Tenhave2007} formulated causal mediation analysis within the potential outcomes framework using $G$-estimation under the rank-preserving (RP) assumption from which the approach in this paper takes its name. Others have built on this approach, developed similar methods in parallel, and pointed out connections to the instrumental variable (IV) framework \parencite{Dunn2007, Albert2008, Lynch2008, Emsley2010, Small2012, Zheng2015a}.

\textcite{Zheng2015a} introduced a relaxed version of the rank-preserving assumption that does not require the strict individual-level constancy of the original RPM, while retaining the core identification strategy, which we adopt here. The corresponding identifying assumption is weaker than sequential ignorability \parencite{Small2012, Zheng2015a}: it requires only that unmeasured confounders do not modify the mediator effect or the direct effect of treatment. That is, unmeasured variables are permitted as long as they do not interact with treatment or mediator on the outcome. Identification is achieved by requiring that at least one baseline covariate interact with treatment to predict the mediator but not the outcome. Simulation results further illustrate that this $G$-estimation approach yields approximately unbiased estimates of direct and mediated effects under unmeasured mediator–outcome confounding, whereas standard regression and structural equation approaches show systematic bias under the same conditions and even maintained robustness to unobserved effect modification \parencite{Zheng2015a, Brandt2020}.

Compared to alternative IV-based approaches to mediation, the $G$-estimation framework occupies a distinct position. A rich literature proposes the use of external instruments to identify mediation effects \parencite[e.g.,][]{Frohelich2017}. In this context, the exclusion restriction requires that the instrument affects the outcome only through the mediator, with no unobserved path connecting the instrument directly to the outcome. This assumption is not empirically testable and must be justified through domain knowledge.\footnote{Over-identification tests can provide partial indirect evidence when multiple instruments are available \parencite{Angrist1996}.} Finding credible external instruments remains a fundamental challenge in psychological research, where the absence of natural experiments and the interconnectedness of psychological constructs make it difficult to identify variables that plausibly satisfy the exclusion restriction \parencite{Bullock2010}. The $G$-estimation approach avoids the need for external instruments by leveraging the standard structure of an RCT, specifically the treatment randomization and the baseline covariates typically collected in such trials.

Despite their advantages, RPMs have not yet been extended to latent variable models. In psychological research, variables of interest are often measured indirectly using self-report questionnaires. Standard regression cannot properly handle such measures, especially when multiple indicators are highly correlated or represent a multidimensional construct \parencite{Bollen1989}. Structural equation modeling (SEM) overcomes these limitations by explicitly modeling latent variables through measurement models. By estimating relationships at the latent level, SEM accounts for measurement error and the shared variance among indicators, thereby increasing statistical power to detect effects.

To overcome this significant limitation for fields like psychology, where measurement error is common, we introduce the Rank-Preserving Structural Equation Model (RAPSEM). This framework integrates the methods developed by \textcite{Tenhave2007, Zheng2015a} with factor score regression and interaction corrections \parencite{Wall2000, Wall2003}, incorporating latent constructs while maintaining the relaxed identifying assumption. RAPSEM thus enables robust mediation analysis even in the presence of measurement error and omitted confounders between the mediator and outcome.

Thereby, we make the following contributions. We formally extend $G$-estimation for mediation analysis to the latent variable case, moving beyond observed-variable models to account for measurement error in mediators, covariates, and outcomes. We explicate the identification strategy underlying RAPSEM, detailing the required assumptions, its origins in Structural Nested Mean Models (SNMMs), and its connections to the IV framework. We describe the estimation procedure and its implementation in an \texttt{R} package and formally establish consistency and asymptotic normality of the estimator. We show how to access the robustness to effect modification by unobserved confounders within the RAPSEM framework, following \textcite{Zheng2015a}. Model performance is then evaluated in two simulation studies: the first compares the robustness of RAPSEM to standard SEM under violations of the underlying causal assumptions, and the second examines the statistical power of RAPSEM across varying scenarios. An empirical example from educational research further illustrates the practical differences between approaches.

\section{Setting and Causal Effect Definitions}

We model the mediation mechanism illustrated in \Cref{fig:path}, assuming the data are collected in an RCT in which the treatment $R$ is randomly assigned.\footnote{While an RCT setting is preferred, the framework can be extended to observational data under appropriate covariate adjustment, for example via balancing methods, provided that the Ignorability Assumption (see \ref{assump:randomization}) holds.}
The path diagram depicts the mediator, outcome, and covariates using the latent variables $\eta_M$, $\eta_Y$, and $\eta_X$, each measured by three observed indicators ($m_1$--$m_3$, $y_1$--$y_3$, and $x_1$--$x_3$).

\begin{figure}
    \begin{center}
        \resizebox{0.5\textwidth}{!}{
            \begin{tikzpicture}[>=stealth,semithick]
\node[ov] (r1) at (-1,0)  {$R$};
\node[lv, draw=darkblue] (u1) at (5,3)  {$U$};

\node[lv] (m) at (2,2)  {$\eta_M$};
\node[ov] (m2) [above of=m]  {$m_2$};
\node[ov] (m1) [left of=m2]  {$m_1$};
\node[ov] (m3) [right of=m2]  {$m_3$};

\node[lv] (y) at (5,0)  {$\eta_Y$};
\node[ov] (y2) [right of=y]  {$y_2$};
\node[ov] (y1) [above of=y2]  {$y_1$};
\node[ov] (y3) [below of=y2]  {$y_3$};

\node[lv] (x) at (1.5,-1.5)  {$\eta_{X}$};
\node[ov] (x2) [below of=x]  {$x_2$};
\node[ov] (x3) [left of=x2]  {$x_3$};
\node[ov] (x1) [right of=x2]  {$x_1$};

\node[lr] (epsm1) at (0.5,2)  {$\zeta_{m}$};
\node[lr] (epsy1) at (5,-1.5)  {$\zeta_{y}$};
\node[lr] (epsx) at (0,-1.5)  {$\zeta_{x}$};

\path[->] 
(r1) edge node[above,scale=.8] {$\gamma_{r}$} coordinate(bmr1) (m);

\path[->, color=darkgreen]
(r1) edge node[below,scale=.8] {$\theta_{r}$} coordinate(byr1) (y)
(m) edge node[below,scale=.8] {$\theta_{m}$} coordinate(bym1) (y);

\path[->] 
(m) edge (m1)
(m) edge (m2)
(m) edge (m3);

\path[->] 
(y) edge (y1)
(y) edge (y2)
(y) edge (y3);

\path[->] 
(x) edge (x1)
(x) edge (x2)
(x) edge (x3);

\path[->, color=darkblue] 
(u1) edge (m)
(u1) edge node[right,scale=0.8] {$\neq0$} (y);

\path[->, color=darkred] 
(u1) edge node[right,scale=0.8] {$=0$} (bym1)
(u1) edge (byr1);

\path[->] 
(epsm1) edge (m)
(epsy1) edge (y)
(epsx) edge node[below,scale=0.6] {} (x);

\path[->] 
(x) edge coordinate(bmx) (m)
(x) edge (y);

\end{tikzpicture}
        }
    \end{center}
    \caption{Path diagram of the RAPSEM model estimating the controlled direct effect ($\theta_r$) of treatment $R$ on outcome $\eta_Y$ and the mediation effect ($\theta_m$) of mediator $\eta_M$ on outcome $\eta_Y$, adjusting for covariates $\eta_X$. Each latent factor ($\eta_M, \eta_Y$ and $\eta_X$) is measured by three observed indicators ($m_1$–$m_3$, $y_1$–$y_3$, and $x_1$–$x_3$, respectively). The model accounts for an unobserved confounder $U$ between mediator and outcome, with the No Unobserved Effect Modification assumption highlighted in red.}
    \label{fig:path}
\end{figure}

The goal is to estimate the controlled direct effect (CDE) of treatment $R$ on the outcome $\eta_Y$, and the controlled mediation effect (CME) of the mediator $\eta_M$ on the outcome $\eta_Y$ \parencite{Robins1992, Pearl2001, Vanderweele2011}, adjusting for measured covariates $\eta_X$. We define these effects within the potential outcomes framework \parencite{Rubin1974}, using the latent potential outcome $\eta_{Y_i}^{r,\eta_m}$ for individual $i$ receiving treatment $R = r$ when the mediator takes value $\eta_{M_i} = \eta_m$.\footnote{\textcite{Zheng2015a} show how the controlled effects relate to natural direct and indirect effects.}

The controlled direct effect captures the contrast between potential outcomes under different treatment levels while holding the mediator $\eta_m$ fixed and conditioning on covariates $\boldsymbol{\eta}_{\mathbf{x}_i}$. Under an additive linear model with a binary treatment $r\in\{0,1\}$ and without any higher-order structural terms including treatment–mediator interactions, it reduces to the difference between the treated and control group captured by $\theta_{r}$:
\begin{equation}
\label{eq:cde}
    \operatorname{CDE} = \operatorname{E}\left[\eta_{Y}^{1,\eta_m} - \eta_{Y}^{0,\eta_m} \mid \boldsymbol{\eta}_{\mathbf{X}} = \boldsymbol{\eta}_{\mathbf{x}}\right] = \theta_{r}.
\end{equation}

The controlled mediation effect captures the contrast between potential outcomes when the mediator is set to $\eta_{m'}$ rather than $\eta_m$, while holding the treatment $r$ fixed and conditioning on covariates $\boldsymbol{\eta}_{\mathbf{x}_i}$. For a discrete mediator, it corresponds to the difference between two mediator categories; for a continuous mediator, it represents the effect of a finite change in the mediator from $\eta_m$ to $\eta_{m'}$. In the additive linear case, it is given by:
\begin{equation}
\label{eq:cme}
    \operatorname{CME} = \operatorname{E}\left[\eta_{Y}^{r,\eta_{m'}} - \eta_{Y}^{r,\eta_m} \mid \boldsymbol{\eta}_{\mathbf{X}} = \boldsymbol{\eta}_{\mathbf{x}}\right] = \theta_{m} (\eta_{m'} - \eta_m).
\end{equation}
\section{Model Formulation}

RAPSEM comprises a measurement component and a structural component, both detailed below. While mathematically equivalent to classical SEM \parencite{Bollen1989}, the framework diverges significantly in its interpretive logic and the scope of its theoretical claims. This distinction is rooted in the causal nature of RAPSEM, which derives its structural part from the Potential Outcomes Framework and utilizes the causal inference technique of $G$-estimation. A notation table is provided in \Cref{sec:notation}.

\subsection{Measurement Model}

We consider a measurement model involving three types of latent variables: a mediator $\eta_M$ measured by a vector of observed indicators $\mathbf{m}$\footnote{While multiple latent mediators may be incorporated, such an extension would necessitate a modification of the underlying causal estimands.}, a set of latent covariates $\boldsymbol{\eta}_X$ measured by indicators $\mathbf{x}$, and an outcome variable $\eta_Y$ measured by indicators $\mathbf{y}$. Let $K$ be the total number of latent variables and $P$ the total number of indicators. For subject $i$, the $P \times 1$ vector of all observed indicators is given by\footnote{Throughout, $^{\top}$ applied to a row of blocks denotes vertical stacking, i.e.\ $\begin{pmatrix} \mathbf{A} & \mathbf{B} \end{pmatrix}^{\top} := \begin{pmatrix} \mathbf{A} \\ \mathbf{B} \end{pmatrix}$, the blocks themselves not being transposed; applied to a single matrix or vector, $^{\top}$ denotes the ordinary transpose.}
\begin{equation} \label{eq:observed_vars_xi}
    \mathbf{z}_i =  
    \begin{pmatrix} 
        \mathbf{m}_i &  \mathbf{x}_i & \mathbf{y}_i 
    \end{pmatrix}^{\top},
\end{equation}
and the corresponding $K \times 1$ vector of latent variables is
\begin{equation} \label{eq:latent_vars_eta}
    \boldsymbol{\eta}_i = 
    \begin{pmatrix} 
        \eta_{M_i} & \boldsymbol{\eta}_{\mathbf{X}_i} & \eta_{Y_i}
    \end{pmatrix}^{\top}.
\end{equation}

The measurement model links the observed indicators to their respective latent constructs. In general, our framework accommodates a broad class of latent variable models of the form
\begin{equation}  \label{eq:measuremenq_model_general}
    \begin{aligned}
        \mathbf{z}_i \mid \boldsymbol{\eta}_i &\sim F\big(\operatorname{E}[\mathbf{z}_i \mid \boldsymbol{\eta}_i], \operatorname{Var}(\mathbf{z}_i \mid \boldsymbol{\eta}_i)\big), \\
        \text{with} \quad f\big(\operatorname{E}[\mathbf{z}_i \mid \boldsymbol{\eta}_i]\big) &= \boldsymbol{\tau} + \boldsymbol{\Lambda} \boldsymbol{\eta}_i.
    \end{aligned}
\end{equation}
Here, $F$ denotes the conditional distribution of $\mathbf{z}_i$ given the latent variables $\boldsymbol{\eta}_i$, characterized by its conditional mean $\operatorname{E}[\mathbf{z}_i \mid \boldsymbol{\eta}_i]$ and, where applicable, its conditional covariance matrix $\operatorname{Var}(\mathbf{z}_i \mid \boldsymbol{\eta}_i)$. The link function $f$ maps the conditional expectation of the indicators to the latent scale, representing the observed means as a linear function of the latent variables with intercepts $\boldsymbol{\tau}$ and a factor loading matrix $\boldsymbol{\Lambda}$.

The specification of $f$ and $F$ depends on the nature of the observed indicators. For continuous indicators, $f$ is the identity link and $F$ is typically multivariate normal. For dichotomous indicators, $f$ is a logit or probit link, which extends to multinomial or ordered models for categorical or ordinal indicators and can incorporate item-specific parameters, as in item response theory (IRT) models. In these cases, $F$ represents the joint distribution of the resulting discrete outcomes, typically Bernoulli or multinomial.

In what follows, we focus on the linear Gaussian specification, which is well-suited for Likert-type item responses that can be treated as approximately continuous:
\begin{equation} \label{eq:measurement_model}
    \mathbf{z}_i = \boldsymbol{\tau} + \boldsymbol{\Lambda} \boldsymbol{\eta}_i + \boldsymbol{\epsilon}_i,
\end{equation}
where $\boldsymbol{\epsilon}_i$ represents measurement error with $\operatorname{Var}(\mathbf{z}_i \mid \boldsymbol{\eta}_i) = \operatorname{Var}(\boldsymbol{\epsilon}_i)$.
To ensure model identification, we adopt standard constraints by fixing the scale of each latent variable. Specifically, for each latent factor, one corresponding indicator is selected as a reference, with its loading fixed to 1 and its intercept to 0. Under this identification scheme, we partition the intercept vector $\boldsymbol{\tau}$ and loading matrix $\boldsymbol{\Lambda}$ as
\begin{equation} \label{eq:cfa_parameters}
    \boldsymbol{\tau} = 
    \begin{pmatrix} 
        \boldsymbol{\tau}_{\mathrm{free}} & \mathbf{0_{K \times 1}}
    \end{pmatrix}^{\top}, \quad
    \boldsymbol{\Lambda} = 
    \begin{pmatrix} 
        \boldsymbol{\Lambda}_{\mathrm{free}} & \mathbf{I}_K 
    \end{pmatrix}^{\top},
\end{equation}
where $\boldsymbol{\tau}_{\mathrm{free}}$ is a $(P-K) \times 1$ vector of free intercepts, $\boldsymbol{\Lambda}_{\mathrm{free}}$ is a $(P-K) \times K$ matrix of free factor loadings, $\mathbf{0}_{K \times 1}$ is a $K \times 1$ vector of zeros, and $\mathbf{I}_K$ is the $K \times K$ identity matrix.

\subsection{Structural Model}

In line with \parencite{Zheng2015a}, the general latent variable form of the potential outcome for person $i$ under treatment $r$ and mediator value $\eta_m$ can be expressed as a linear combination of the structural component, a nuisance term, and a remaining error term:
\begin{equation} \label{eq:potential_outcome}
    \eta_{Y_i}^{r,\eta_m} = \underbrace{g(\boldsymbol{\eta}_{\mathbf{X}_i})}_{\text{nuisance term}} + \underbrace{\sum_{s=1}^{S}\theta_{s} \cdot h_{s}(r, \eta_{M}, \boldsymbol{\eta}_{\mathbf{X}_i})}_{\text{structural component}} + \underbrace{\zeta_{Y_i}^{r,\eta_m}}_{\text{error term}}.
\end{equation}

As detailed in \textcite{Zheng2015a}, no inherent restrictions need to be imposed on the functional form of $h(\cdot)$, provided that the resulting parameters can be mapped to well-defined causal contrasts under the prevailing identification assumptions. Furthermore, the model may, in principle, include interactions between the treatment and the mediator, as well as interactions between treatment or mediator and covariates, thus explicitly capturing effect heterogeneity. We focus, however, on the additive linear structure represented in \Cref{fig:path} and consistent with the original model of \textcite{Tenhave2007}. This choice preserves the interpretability of the $\operatorname{CDE}$ and $\operatorname{CME}$ by ensuring that $\theta_r$ and $\theta_m$ function as invariant structural parameters.\footnote{As noted by \textcite{Robins2003}, the identification of natural effects typically requires a no-interaction assumption to decompose the total effect. While controlled effects remain identifiable in the presence of interaction, this assumption is still necessary to represent them as single, invariant parameters. Including an interaction term $\theta_{rm}(r \cdot \eta_M)$ would transform these effects into functions of the levels of other variables.}

The nuisance term $g(\boldsymbol{\eta}_{\mathbf{X}_i}) = \operatorname{E}\!\left[\eta_{Y_i}^{0,0}
\mid \boldsymbol{\eta}_{\mathbf{X}_i}\right]$ represents the covariate-conditional mean of the
baseline outcome under no treatment and no mediation. In principle, it may be modeled as an arbitrary function of the covariates, denoted by $g(\boldsymbol{\eta}_{\mathbf{X}_i})$, including nonparametric forms \parencites[e.g., using splines in][]{Brandt2020}[or a deep neural network in][]{Faleh2026}. Here, however, we approximate $g(\cdot)$ as a linear combination of covariates, potentially including polynomial terms and interactions. Such transformations can be accommodated using the factor score corrections proposed by \textcite{Wall2000}, whereas more general nonlinear specifications are not supported by existing latent factor score correction methods \parencite{Hayes2020}. Consequently, we model the nuisance term as $g(\boldsymbol{\eta}_{\mathbf{X}_i}) = q_{yx}(\boldsymbol{\eta}_{\mathbf{X}_i})\,\boldsymbol{\beta}_{x}$, where $q_{yx}(\cdot)$ denotes the admissible transformations.

The error term $\zeta_{Y_i}^{r,\eta_m}$ may explicitly depend on covariates $\boldsymbol{\eta}_{\mathbf{X}_i}$ as well as on unobserved confounders $U$, being only subject to $\operatorname{E}\!\left[\zeta_{Y_i}^{r,\eta_m} \mid \boldsymbol{\eta}_{\mathbf{X}_i}\right] = 0$ for every $(r,\eta_m)$. We return to this point when discussing the identification of the causal estimands.

Assuming the conditions in the next section hold, the additive linear potential outcome model for a continuous outcome can be represented by the following structural model:
\[
    \eta_{Y_i} = \theta_{r} \cdot R_i + \theta_{m} \cdot \eta_{M_i} + q_{yx}(\boldsymbol{\eta}_{\mathbf{X}_i})\,\boldsymbol{\beta}_{x} + \zeta_{Y_i},
\]
which corresponds to the model specification supported by RAPSEM. This model specifies the latent outcome for each individual under their realized treatment and mediator values. To facilitate the formulation of the estimation procedure, we adopt matrix notation
\begin{equation} \label{eq:observed_outcome}
    \boldsymbol{\eta}_{Y} = \boldsymbol{\Xi}_y \boldsymbol{\alpha} + \boldsymbol{\zeta}_y
\end{equation}
with parameter vector
\[
    \boldsymbol{\alpha} =
    \begin{pmatrix}
        \theta_{r} &
        \theta_{m} &
        \boldsymbol{\beta}_{x}
    \end{pmatrix}^{\top}
\]
and design matrix $\boldsymbol{\Xi}_y$ composed of row vectors containing the predictors of each subject $i$
\[
    \boldsymbol{\xi}_{Y_i} =
    \begin{pmatrix}
        R_i & \eta_{M_i} & q_{yx}(\boldsymbol{\eta}_{\mathbf{X}_i})
    \end{pmatrix}.
\]

In addition to the outcome model, we specify a separate structural equation for the latent mediator $\eta_{M_i}$. In this specification, we include interaction terms between the treatment and covariates to account for treatment effect heterogeneity, which is relevant for the power of RAPSEM as discussed in the assumptions and identification sections and further illustrated in the simulation study. The mediator model can then directly be written in matrix notation as
\begin{equation} \label{eq:mediator_model}
    \boldsymbol{\eta}_{M} = \boldsymbol{\Xi}_m\boldsymbol{\gamma} + \boldsymbol{\zeta}_{m},
\end{equation}
with parameter vector
\[
    \boldsymbol{\gamma} =
    \begin{pmatrix}
        \gamma_{r} &
        \boldsymbol{\gamma}_{x} &
        \boldsymbol{\gamma}_{xr}
    \end{pmatrix}^{\top}
\]
and the rows of the design matrix given by
\[
    \boldsymbol{\xi}_{M_i} =
    \begin{pmatrix}
        R_i & q_{mx}(\boldsymbol{\eta}_{\mathbf{X}_i}) & R_i \cdot q_{mrx}(\boldsymbol{\eta}_{\mathbf{X}_i})
    \end{pmatrix},
\]
where $q_{mx}(\cdot)$ and $q_{mrx}(\cdot)$ denote transformations of the latent covariates in the main and interaction effects, respectively.

For the simulation study, we introduce a further simplification by considering two covariates that enter both the outcome and mediator models without transformation. Specifically, we set
\[
    q_{yx}(\boldsymbol{\eta}_{\mathbf{X}_i}) = q_{mx}(\boldsymbol{\eta}_{\mathbf{X}_i}) = q_{mrx}(\boldsymbol{\eta}_{\mathbf{X}_i}) =
    \begin{pmatrix}
        \eta_{X_{1i}} & \eta_{X_{2i}} 
    \end{pmatrix},
\]
which yields the full structural model\footnote{Note that factor score corrections are still required for the weight calculation (see the estimation section).}
\begin{equation} \label{eq:concrete_model}
    \begin{aligned}
        \eta_{Y_i} &= \theta_r R_i + \theta_m \eta_{M_i} + \beta_{x1} \eta_{X_{1i}} + \beta_{x2} \eta_{X_{2i}} + \zeta_{Y_i}, \\
        \eta_{M_i} &= \gamma_r R_i + \gamma_{x1} \eta_{X_{1i}} + \gamma_{x2} \eta_{X_{2i}}
        + \gamma_{x1r} R_i \eta_{X_{1i}} + \gamma_{x2r} R_i \eta_{X_{2i}} + \zeta_{M_i}.
    \end{aligned}
\end{equation}
\section{Assumptions}

The assumptions listed in \Cref{tab:assumptions} are required to identify contrasts in \Crefrange{eq:cde}{eq:cme} from the observed data and to ensure the consistency and asymptotic normality of the effect estimates. We discuss the most relevant assumptions for identification, \ref{assump:randomization}, \ref{assump:nem} and \ref{assump:treat_cov_inter}, in the following and define the remaining ones in \Cref{sec:appendix_assumption}.

\begin{table}[ht]
    \centering
    \caption{Overview of the assumptions required for identification, consistency, and asymptotic normality, grouped by category.}
    \label{tab:assumptions}
    \begin{tabular}{ll}
    \toprule
    \textbf{Category} & \textbf{Assumption Names} \\
    \midrule
    \multirow{5}{*}{Causal Assumptions} 
        & \ref{assump:randomization} Ignorability \\
        & \ref{assump:nem} Unobserved Heterogeneity Mean Independence \\
        & \ref{assump:consistency} Consistency \\
        & \ref{assump:positivity} Positivity \\
        & \ref{assump:structural_measurement} Causal Efficacy\\
    \midrule
    \multirow{1}{*}{Structural Assumption} 
        & \ref{assump:treat_cov_inter} Covariate-Treatment Interaction\\      
    \midrule
     \multirow{6}{*}{Regularity Assumptions}  
        & \ref{assump:iid_sampling} IID Sampling \\
        & \ref{assump:model_spec} Correct Model Specification \\
        & \ref{assump:regularity_measurement_model} Measurement Model Regularity \\
        & \ref{assump:measurement_error} Measurement Error \\
        & \ref{assump:finite_moments} Factor Score Finite Moments \\
        & \ref{assump:structural_error} Structural Equation Error \\
    \bottomrule
    \end{tabular}
\end{table}

\subsection*{Causal Assumptions}

We focus here on the causal assumptions regarding unobserved confounding, while the remaining standard assumptions of the potential outcomes framework are detailed in \Cref{sec:appendix_assumption}.

\begin{causalassumption}[Ignorability] \label{assump:randomization}
    Conditional on measured baseline covariates $\eta_x$, the treatment assignment is independent of all potential outcomes and potential mediator values:
    \[
        R \perp\!\!\!\perp (\eta_M^{r}, \eta_Y^{r,\eta_m}) \mid \boldsymbol{\eta}_{\mathbf{X}} \quad \forall\, r, \eta_m .
    \]
    This assumption states that, after adjusting for observed covariates, there are no unmeasured confounders between the treatment and the mediator and no unmeasured confounders between the treatment and the outcome. This assumption holds for randomized interventions.
\end{causalassumption}

\begin{causalassumption}[Unobserved Heterogeneity Mean Independence] \label{assump:nem}
    As in \textcite{Zheng2015a}, we assume the conditional expectation of the potential outcome error is invariant to the potential treatment and mediator levels. Specifically, there exists a function $F$ such that
    \begin{equation}
    E\left[
    \zeta_{Y_i}^{r,\eta_m}(U_i,\boldsymbol{\eta}_{X_i})
    \mid R_i,\eta_{M_i},\boldsymbol{\eta}_{X_i}
    \right]
    =
    F(R_i,\eta_{M_i},\boldsymbol{\eta}_{X_i}),
    \quad \forall r,\eta_m,
    \end{equation}
    where $F$ does not depend on the potential the potential treatment
and mediator levels $(r,\eta_m)$. This implies that the realized treatment and mediator value assignments carry no additional information about unobserved outcome heterogeneity beyond the measured covariates.

    Another way to interpret this assumption is through the lens of effect modification by unmeasured confounders. If an unmeasured confounder $U$ modifies the treatment-outcome and/or the mediator-outcome effect, the effects would differ across strata defined by $U$. Since this would cause the mean of $\zeta^{rm}$ to shift with the observed treatment and mediator, the above assumption would be violated. This rules out interaction effects between the unobserved confounder and the mediator in determining the outcome, unless explicitly modeled \parencite[see][for such extensions]{Zheng2015a}, as indicated by the red paths in \Cref{fig:path}. A comparison with the original Rank Preserving and the No Essential Heterogeneity \parencite{Heckman2006} assumption can be found in \Cref{sec:appendix_rp_nem}.
\end{causalassumption}

Assumption~\ref{assump:nem} is central to our framework, as it replaces the commonly invoked Sequential Ignorability assumption. It can be considered weaker for the following reason: In the context of an RCT, Sequential Ignorability reduces to requiring independence between the potential outcomes of the mediator and outcome, conditional on covariates and treatment, which formally demands the error terms of the outcome and mediator equations to be uncorrelated. This rules out any unmeasured mediator-outcome confounding and any effect modification by such confounders. In contrast, \ref{assump:nem} permits the error terms of the outcome and mediator equations to be correlated, allowing for unmeasured mediator-outcome confounding. It only rules out that this correlation manifests as a systematic shift in the mean of the outcome error across treatment-mediator conditions, conditional on observed covariates.

As noted in the introduction, no unmeasured confounding represents a strong, often unrealistic requirement. We argue that no effect modification by unobserved confounders may be more plausible in the same settings. For instance, consider genetic predisposition as an unobserved confounder in the studies mentioned in the introduction, where CBT reduces the outcome social anxiety symptoms by first decreasing the mediator maladaptive beliefs. While genetics likely influences both the baseline frequency of maladaptive belief and the physiological threshold for anxiety, it is less likely to act as an effect modifier of the mediator-outcome relationship. That is, while genetics may determine a patient's initial symptom levels (the intercept), it is unlikely to fundamentally alter the psychological mechanism (the slope) through which a reduction in maladaptive beliefs translates into symptom relief. Once conditioned on observed covariates like age and baseline severity, this mechanism is assumed to be a stable process that operates independently of the genetic factors that established the initial symptom levels.

\subsection*{Structural Assumption}

Next, we define the critical structural requirement necessary to relax the Sequential Ignorability assumption. Other standard regularity conditions are detailed in \Cref{sec:appendix_assumption}.

\begin{structuralassumption}[Treatment-Covariate Interaction]
\label{assump:treat_cov_inter}
    There exists at least one observed covariate that interacts with the treatment to significantly affect the mediator (e.g., $r \times \eta_{x_k} \to \eta_m$), providing sufficient variation to ensure that the weight matrix introduced in the next section has full column rank.
    This treatment--covariate interaction affects the outcome solely through its effect on the mediator and has no direct influence on the outcome.
\end{structuralassumption}

\section{Identification}
\label{sec:identification}

In the following, we establish the identification of $\theta_m$ and $\theta_r$ as causal effects. First, we review the derivation of orthogonality conditions via $G$-estimation, as developed by \textcite{Tenhave2007} and \textcite{Zheng2015a}. Then, we examine the formal connections between these conditions and the use of instrumental variables (IVs). Throughout this section, we treat the latent variables as if they were observed and establish identification of $\boldsymbol{\theta}=(\theta_r,\theta_m)$ from the joint distribution of $(R,\eta_M,\eta_Y,\boldsymbol{\eta}_{\mathbf X})$. By the measurement part of Assumption~\ref{assump:model_spec}, the moments of this distribution required for the orthogonality conditions are identified from the observed indicators. Assumption~\ref{assump:structural_measurement} ensures that replacing the latent variables by their measurement-model-based estimates does not alter the interpretation of $\boldsymbol{\theta}$ as causal effects.

\subsection{Orthogonality Conditions}

Mediation analysis inherently follows a longitudinal structure, involving three temporally ordered measurement occasions following the assessment of baseline covariates: treatment, mediator, and outcome. From this perspective, it is natural to employ generalized causal inference methods, commonly referred to as $G$-methods, which were developed to address time-varying exposures and confounding in longitudinal settings \parencite{Robins2008}. While the broader class of $G$-methods includes the $G$-formula (or $G$-computation) and inverse probability weighting, the approach of \textcite{Tenhave2007} is based on $G$-estimation. 

$G$-estimation proceeds by relating the observed outcome $\eta_{Y_i}$ to the potential outcome $\eta_{Y_i}^{r,\eta_m}$ in \Cref{eq:potential_outcome} via a Structural Nested Mean Model (SNMM). This relation is established by sequentially "blipping down" the mediator and treatment effects to remove their respective influences from the observed data:
\begin{equation*}
    \eta_{Y_i}^{0,0}(\boldsymbol{\theta}) = \eta_{Y_i} - \theta_m \eta_{M_i} - \theta_r R_i.\footnote{In general $\eta_{Y_i}^{0,0}(\boldsymbol{\theta}) = \eta_{Y_i} - \sum_{s=1}^{S}\theta_{s} \cdot h_{s}(r, \eta_{M}, \boldsymbol{\eta}_{\mathbf{X}_i}).$}
\end{equation*}
Under Assumptions \ref{assump:consistency}, \ref{assump:positivity}, and the outcome part of \ref{assump:model_spec}, the blipped-down outcome evaluated at the true parameter values
$\boldsymbol{\theta}^* = (\theta_r^*, \theta_m^*)$
satisfies
\begin{equation*}
    \eta_{Y_i}^{0,0}(\boldsymbol{\theta}^*)
    = g(\boldsymbol{\eta}_{\mathbf{X}_i}) + \zeta_{Y_i}.
\end{equation*}
Subtracting the nuisance term yields
\begin{equation*}
    \zeta_{Y_i} = 
   \eta_{Y_i}^{0,0}(\boldsymbol{\theta}^*)
    - g(\boldsymbol{\eta}_{\mathbf{X}_i}).
\end{equation*}
Assumptions \ref{assump:randomization} and \ref{assump:nem} imply that $R_i$ carries no additional information about $\zeta_{Y_i}$ beyond what $\boldsymbol{\eta}_{\mathbf{X}_i}$ already provides:
\begin{equation*}
    \operatorname{E}\big[\zeta_{Y_i} \mid R_i, \boldsymbol{\eta}_{\mathbf{X}_i}\big]
    = \operatorname{E}\big[\zeta_{Y_i} \mid \boldsymbol{\eta}_{\mathbf{X}_i}\big]
    = 0.
\end{equation*}
While this resembles a standard ignorability statement, it serves here as the basis for the orthogonality condition
\begin{equation*}
    \operatorname{Cov}\big(\zeta_{Y_i},\,
    \mathbf{A}(R_i, \boldsymbol{\eta}_{\mathbf{X}_i}) \mid \boldsymbol{\eta}_{\mathbf{X}_i}\big) = 0,
\end{equation*}
defined for any weighting function $\mathbf{A}$. Since the adjusted residual is mean-centered, this condition translates to the population estimation equation
\begin{equation}
    \operatorname{E} \left[\mathbf{A}(R_i, \boldsymbol{\eta}_{\mathbf{X}_i})^{\top}
    \left(\eta_{Y_i}^{0,0}(\boldsymbol{\theta}^*)
    - g(\boldsymbol{\eta}_{\mathbf{X}_i})\right)\right] = 0
\end{equation}
provided all entries of $\mathbf{A}$ satisfy the centering constraint $\operatorname{E}\big[a_s(R_i, \boldsymbol{\eta}_{\mathbf{X}_i}) \mid \boldsymbol{\eta}_{\mathbf{X}_i}\big] = 0$. The corresponding sample estimating equation is
\begin{equation}\label{eq:moment_condition}
    \sum_{i=1}^{N}
    \mathbf{A}(R_i,\boldsymbol{\eta}_{\mathbf{X}_i})^\top
    \left\{
    \eta_{Y_i}^{0,0}(\boldsymbol{\theta})
    -\hat g(\boldsymbol{\eta}_{\mathbf{X}_i})
    \right\}
    =0.
\end{equation}
where $\hat g$ is a working model for the nuisance function $g$.

\textcite{Zheng2015a} showed that the most efficient weights for the outcome model in \Cref{eq:potential_outcome} are given by
\begin{equation} \label{eq:optimal_weights}
    \mathbf{a}_s = \left(\operatorname{E}\left[h_s(R_i, \eta_{M_i}, \boldsymbol{\eta}_{\mathbf{X}_i}) \mid \boldsymbol{\eta}_{\mathbf{X}_i}, R_i\right] - \operatorname{E}\left[h_s(R_i, \eta_{M_i}, \boldsymbol{\eta}_{\mathbf{X}_i}) \mid \boldsymbol{\eta}_{\mathbf{X}_i}\right]\right)\cdot \Omega_{\boldsymbol{\eta}_{\mathbf{X}_i}}^{-1},
\end{equation}
if the covariance of the error term conditioned on the covariates $\Omega_{\boldsymbol{\eta}_{\mathbf{X}_i}}$ does not depend on $r$ and $\eta_m$. These weights capture the variation in $h_s(R_i, \eta_{M_i}, \boldsymbol{\eta}_{\mathbf{X}_i})$ that is attributable to the treatment, conditioned on the covariates.

\subsection{Instrumental Variable Interpretation}

This $G$-estimation approach aligns closely with the theory of instrumental variables \parencite{Angrist1996}, a connection established by \textcite{Dunn2007}. In the classical IV framework, an external variable serves as a valid instrument if it satisfies two core conditions: it must be correlated with the endogenous predictor (relevance) and uncorrelated with the outcome error term (exclusion restriction).

Under the $G$-estimation framework, the randomized treatment $R_i$ functions as an instrument for the mediator $\eta_{M_i}$. Relevance is satisfied via the interaction term required in the first sentence of Assumption~\ref{assump:treat_cov_inter}, while the exclusion restriction is maintained through the conditional independence between the treatment and the potential outcome (Assumption~\ref{assump:randomization}) and the structural constraint in the second sentence of Assumption~\ref{assump:treat_cov_inter}. Consequently, the instrument is constructed from the model structure by leveraging covariates that moderate the mediator but not the outcome effect. Notably, the strength of these moderating relationships can be empirically tested, e.g., using the first-stage $F$-statistic from the regression of the mediator on treatment, covariates, and their interaction.

The orthogonality conditions in \Cref{eq:moment_condition} directly mirror IV moment conditions, with the weighting functions in \Cref{eq:optimal_weights} acting as residualized instruments. As demonstrated by \textcite{Lee2021, Lee2024}, using such centered instruments provides the advantage of maintaining consistency even for non-continuous outcomes.

\section{Estimation}
\label{sec:estimation}

RAPSEM adopts a limited-information estimation strategy in which latent variable effects are recovered through a two-step procedure. In the first stage, factor scores are estimated based solely on the measurement model, treating it independently from the structural component. In the second stage, these estimated scores are used as observed proxies for the latent constructs in the structural equation. In this last step, we replace ordinary least squares with latent $G$-estimation. Details on implementation in the R package \texttt{rapsem} \parencite{rapsem}, available at \url{https://github.com/PsychometricsMZ/rapsem}., available on GitHub, are provided in \Cref{sec:appendix_implementation}. Refer to \Cref{sec:notation} for a table of all used symbols.

\subsection{Notation}

Let $\boldsymbol{\eta}_i$ denote the true underlying factor scores in \Cref{eq:latent_vars_eta}, which are used in \Cref{eq:observed_outcome}. We later partition these scores into the outcome factor $\eta_{Y_i}$ and the predictor factors $\boldsymbol{\eta}_{\text{pred},i}$, which include the mediator factor $\eta_{{M}_i}$ and the covariate factors $\boldsymbol{\eta}_{\mathbf{X}_i}$. The theoretical estimator of the factor scores based on the true measurement parameters in \Cref{eq:measurement_model} is denoted by $\tilde{\boldsymbol{\eta}}_i$, while $\hat{\boldsymbol{\eta}}_i$ refers to the practical estimation of $\tilde{\boldsymbol{\eta}}_i$ obtained by plugging in the estimated measurement parameters.

\subsection{First Stage}

In the first stage, we adopt the method proposed by \textcite{Wall2000} to estimate the latent factor scores $\boldsymbol{\eta}_i$. We collect the free CFA parameters of the measurement model specified in \Crefrange{eq:measurement_model}{eq:cfa_parameters} in
\begin{equation}\label{eq:upsilon1}
    \boldsymbol{\Upsilon}_1
    = \begin{pmatrix}
    \boldsymbol{\tau}_{\mathrm{free}} &
    \operatorname{vec} \boldsymbol{\Lambda}_{\mathrm{free}} &
    \operatorname{vech} \operatorname{Var}(\boldsymbol{\epsilon}_i)
    \end{pmatrix}
\end{equation}
where $\operatorname{vec}$ denotes the column-wise vectorization of a matrix and $\operatorname{vech}$ the vectorization of the non-redundant elements of a symmetric matrix.
Given $\boldsymbol{\Upsilon}_1$, the estimator for the latent factor scores of subject $i$ is defined as
\begin{equation} \label{eq:factor_score_estimator}
    \tilde{\boldsymbol{\eta}}_i = 
    \begin{pmatrix}
        -\mathbf{H} & \mathbf{I}_K + \mathbf{H} \boldsymbol{\Lambda}_{\mathrm{free}} 
    \end{pmatrix}
    \left[
        \mathbf{z}_i - 
        \begin{pmatrix} 
            \boldsymbol{\tau}_{\mathrm{free}} & \mathbf{0}_{K \times 1} 
        \end{pmatrix}^{\top}
    \right],
\end{equation}
where $\mathbf{z}_i$ denotes the vector of observed indicators, and the matrix $\mathbf{H}$ is given by
\[
    \mathbf{H} = 
    \begin{pmatrix} 
        \mathbf{0}_{K \times (P - K)} & \mathbf{I}_K 
    \end{pmatrix} 
    \operatorname{Var}(\boldsymbol{\epsilon}_i)
    \begin{pmatrix} 
        \mathbf{I}_{(P - K)} & -\boldsymbol{\Lambda}_{\mathrm{free}}^{\top} 
    \end{pmatrix}^{\top}
    \left[ 
        \begin{pmatrix} 
            \mathbf{I}_{(P - K)} & -\boldsymbol{\Lambda}_{\mathrm{free}} 
        \end{pmatrix} 
        \operatorname{Var}(\boldsymbol{\epsilon}_i)
        \begin{pmatrix} 
            \mathbf{I}_{(P - K)} & -\boldsymbol{\Lambda}_{\mathrm{free}}^{\top} 
        \end{pmatrix}^{\top}
    \right]^{-1}.
\]

To account for the propagation of measurement error, we treat the factor score estimator $\tilde{\boldsymbol{\eta}}_i$ as a noisy proxy of the true latent variables $\boldsymbol{\eta}_i$, assuming the relationship
\begin{equation} \label{eq:fact_score_proxy}
    \tilde{\boldsymbol{\eta}}_i = \boldsymbol{\eta}_i + \mathbf{e}_i,
\end{equation}
where the estimation error $\mathbf{e}_i$ is linearly related to the measurement error $\boldsymbol{\epsilon}_i$ via
\begin{equation} \label{eq:fact_score_error}
    \mathbf{e}_i = 
    \begin{pmatrix}
       -\mathbf{H} &  \mathbf{I}_K + \mathbf{H} \boldsymbol{\Lambda}_{\mathrm{free}} 
    \end{pmatrix}
    \boldsymbol{\epsilon}_i.
\end{equation}
For use in the errors-in-variables estimation in the subsequent stage, we also need the second moment of $\mathbf{e}_i$, i.e., its covariance matrix
\begin{equation} \label{eq:sigma_ee}
    \boldsymbol{\Sigma}_{ee} = 
    \begin{pmatrix}
        -\mathbf{H} & \mathbf{I}_K + \mathbf{H} \boldsymbol{\Lambda}_{\mathrm{free}}
     \end{pmatrix}
     \operatorname{Var}(\boldsymbol{\epsilon}_i)
     \begin{pmatrix}
        \mathbf{0}_{(P-K) \times K} & \mathbf{I}_K
     \end{pmatrix}^{\top}.
\end{equation}

The parameters in $\boldsymbol{\Upsilon}_1$ are estimated using standard confirmatory factor analysis (CFA), which minimizes the discrepancy between the model-implied and observed covariance matrices \parencite{Bollen2002}. Substituting the CFA estimates $\hat{\boldsymbol{\Upsilon}}_1$ into~\Cref{eq:factor_score_estimator,eq:sigma_ee} yields the estimated factor scores $\hat{\boldsymbol{\eta}}_i$ and the associated error covariance structure $\hat{\boldsymbol{\Sigma}}_{ee}$, which serve as inputs for the second stage.

\subsection{Second stage}

In the second stage, the structural equation in \Cref{eq:potential_outcome} is solved with the $G$-estimation approach introduced in the identification section with an additional two-stage method-of-moments (2SMM) correction for latent variable interactions, as proposed by \textcite{Wall2000}. In our implementation, we adopt their modified 2SMM estimator and incorporate an additional ridge-inspired variance term to improve numerical stability \parencite{Hoerl2000}. 

\subsubsection[Solving the G-Estimation Equation]{Solving the $G$-Estimation Equation}

In the identification section, we derived the orthogonality condition in \Cref{eq:moment_condition}. Here, we show how to solve it for the observed outcome model in \Cref{eq:observed_outcome}. We begin by specifying the optimal weights introduced in \Cref{eq:optimal_weights}.

For a binary treatment that is randomized independently of the covariates, the treatment weight simplifies to the centered treatment indicator:\footnote{In the general formulation, the optimal weight additionally involves the covariance matrix $\Omega_{\boldsymbol{\eta}_{\mathbf{X}_i}}$, which accounts for heteroskedasticity of the outcome error across covariate values. For simplicity, we set it to the identity matrix, yielding a consistent, though not necessarily fully efficient, estimator.}
\begin{equation}
    a_{r,i} = R_i - \operatorname{E}[R].
\end{equation}

The mediator weight in this setting is given by the treatment-induced shift in the conditional expectation of the mediator:
\begin{equation} \label{eq:med_weight_general}
    a_{m,i} = \left(\operatorname{E}\left[\eta_{M_i} \mid \boldsymbol{\eta}_{\mathbf{X}_i}, R_i=1\right]
    - \operatorname{E}\left[\eta_{M_i} \mid \boldsymbol{\eta}_{\mathbf{X}_i}, R_i=0\right]\right)
    \cdot a_{r,i}.
\end{equation}

To construct this weight, we require the difference in the expected value of the mediator conditional on the covariates under treatment and control. This quantity can be obtained by fitting the mediator model in \Cref{eq:mediator_model} and computing the predicted mediator values under both treatment conditions. Considering for example the specification from \Cref{eq:concrete_model}, we have
\[
    \begin{aligned}
        \operatorname{E}\left[\eta_{M_i} \mid \boldsymbol{\eta}_{\mathbf{X}_i}, R_i=1\right] &= \gamma_r + \gamma_{x1}\eta_{X_{1i}} + \gamma_{x2}\eta_{X_{2i}} + \gamma_{x1r} \eta_{X_{1i}} + \gamma_{x2r} \eta_{X_{2i}}, \\
        \operatorname{E}\left[\eta_{M_i} \mid \boldsymbol{\eta}_{\mathbf{X}_i}, R_i=0\right] &= \gamma_{x1}\eta_{X_{1i}} + \gamma_{x2}\eta_{X_{2i}},
    \end{aligned}
\]
so that the difference simplifies to
\begin{equation} \label{eq:med_weight_explicit}
    \operatorname{E}\left[\eta_{M_i} \mid \boldsymbol{\eta}_{\mathbf{X}_i}, R_i=1\right]
    - \operatorname{E}\left[\eta_{M_i} \mid \boldsymbol{\eta}_{\mathbf{X}_i}, R_i=0\right]
    = \gamma_r + \gamma_{x1r} \eta_{X_{1i}} + \gamma_{x2r} \eta_{X_{2i}}.
\end{equation}

In addition, we introduce pseudo-weights for the covariate terms, given by the transformed baseline covariates $q_{yx}(\boldsymbol{\eta}_{\mathbf{X}_i})$ themselves. Since the nuisance term is modeled as a linear combination of these transformed covariates, we can define an extended weight matrix $\mathbf{W}$ that incorporates both the structural weights and the pseudo-weights.\footnote{For a non-linear nuisance term, defining a common weight function is not possible; instead, an iterative estimation procedure is required, alternating between the estimation of the nuisance term and the weights for the structural component.} Each row of $\mathbf{W}$ is defined as
\[
    \mathbf{w}_{i} =
    \begin{pmatrix}
        a_{r,i} & a_{m,i} & q_{yx}(\boldsymbol{\eta}_{\mathbf{X}_i})
    \end{pmatrix}.
\]
The estimating equation then becomes
\begin{equation} \label{eq:g_estimation}
    \mathbf{W}^\top\boldsymbol{\zeta}_Y = 0,
\end{equation}
which can be solved in a single step by replacing expectations with their sample analogues and solving for the structural parameters. This yields the closed-form, sample-based estimator
\begin{equation} \label{eq:observed_gparm}
    \bar{\boldsymbol{\alpha}} = (\mathbf{W}^\top \boldsymbol{\Xi}_y)^{-1} \mathbf{W}^\top \boldsymbol{\eta}_{Y},
\end{equation}
which corresponds to a linear IV estimator, with $\mathbf{W}$ serving as the matrix of residual (structural) instruments.

\subsubsection{Factor Score Correction}

The structural form of the outcome factor score $\eta_{Y_i}$ may depend on polynomial terms of the covariate and mediator scores $\boldsymbol{\eta}_{\text{pred},i}$. Any interactions or higher-order terms involving $\boldsymbol{\eta}_{i}$ must be corrected using the moments of the measurement error $\mathbf{e}_i$. We define these moments of $\mathbf{e}_i$ needed for the correction as $\boldsymbol{\Upsilon}_2$. Assuming normally distributed measurement errors $\boldsymbol{\epsilon}_i$, they reduce to linear combinations of elements in the estimated covariance matrix $\hat{\boldsymbol{\Sigma}}_{ee}$.

Instead of using the naive sample-based estimator $\bar{\boldsymbol{\alpha}}$ in \Cref{eq:observed_gparm}, which ignores measurement error in the factor scores, we estimate the true population parameter
\begin{equation} \label{eq:param_gestimation}
    \boldsymbol{\alpha} = \operatorname{E}[\mathbf{w}_i^\top \boldsymbol{\xi}_{Y_i}]^{-1}\operatorname{E}[\mathbf{w}_i^\top \eta_{Y_i}]
\end{equation}
with the two stage method-of-moments estimator from \textcite{Wall2000}
\begin{equation} \label{eq:structural_para_estimator}
    \hat{\boldsymbol{\alpha}} = \hat{\mathbf{M}}^{-1} \hat{\mathbf{m}},
\end{equation}
where
\begin{equation} \label{eq:oberserved_moments}
\begin{aligned}
    \hat{\mathbf{M}} &= \frac{1}{N} \sum_{i=1}^{N} \mathbf{M}(\hat{\boldsymbol{\eta}}_{\text{pred},i}, \hat{\boldsymbol{\Upsilon}}_{2}), \\
    \hat{\mathbf{m}} &= \frac{1}{N} \sum_{i=1}^{N} \mathbf{m}(\hat{\boldsymbol{\eta}}_i, \hat{\boldsymbol{\Upsilon}}_{2})
\end{aligned}
\end{equation}
are computed from the estimated factor scores $\hat{\boldsymbol{\eta}}_i$ and the estimated measurement error moments $\hat{\boldsymbol{\Upsilon}}_2$. The moment functions $\mathbf{M}(\cdot)$ and $\mathbf{m}(\cdot)$ are defined such that their conditional expectations given the true factor scores $\boldsymbol{\eta}_i$ and the treatment $R_i$ satisfy
\begin{equation}\label{eq:m_conditional_expectation}
  \begin{aligned}
    \operatorname{E}\!\left[\mathbf{M}(\tilde{\boldsymbol{\eta}}_{\mathrm{pred},i},
      \boldsymbol{\Upsilon}_2)\mid\boldsymbol{\eta}_i,R_i\right]
      &= \mathbf{w}_i^{\top}\boldsymbol{\xi}_{Y_i},\\
    \operatorname{E}\!\left[\mathbf{m}(\tilde{\boldsymbol{\eta}}_i,
      \boldsymbol{\Upsilon}_2)\mid\boldsymbol{\eta}_i,R_i\right]
      &= \mathbf{w}_i^{\top}\eta_{Y_i},
  \end{aligned}
\end{equation}
where $\tilde{\boldsymbol{\eta}}_i$ denotes the theoretical factor score estimator based on the true measurement parameters, so that, by iterated expectations, their means equal the population moments in \Cref{eq:param_gestimation}.

Let $J$ denote the highest order in which any component of $\boldsymbol{\eta}_{\text{pred},i}$ appears in $\mathbf{W}$ or $\boldsymbol{\Xi}_{y}$. Then $\mathbf{M}$ and $\mathbf{m}$ are constructed via the expansion
\begin{equation} \label{eq:moment_construction}
    \begin{pmatrix}
        \mathbf{M}(\tilde{\boldsymbol{\eta}}_{\text{pred},i}, \boldsymbol{\Upsilon}_2) & \mathbf{m}(\tilde{\boldsymbol{\eta}}_i, \boldsymbol{\Upsilon}_2)
    \end{pmatrix}
    = \sum_{j=0}^{J} (-1)^j D_j(\tilde{\boldsymbol{\eta}}_i, \boldsymbol{\Upsilon}_2),
\end{equation}
where the correction terms are recursively defined as
\begin{equation} \label{eq:correction_term_construction}
    \begin{aligned}
        D_0(\tilde{\boldsymbol{\eta}}_i, \boldsymbol{\Upsilon}_2) &= \tilde{\mathbf{w}}_i^\top
        \begin{pmatrix}
            \tilde{\boldsymbol{\xi}}_{Y_i} & \tilde{\eta}_{Y_i}
        \end{pmatrix}, \\
        D_j(\tilde{\boldsymbol{\eta}}_i, \boldsymbol{\Upsilon}_2) &= \operatorname{E} \left[ D_{j-1}(\tilde{\boldsymbol{\eta}}_i, \boldsymbol{\Upsilon}_2) \mid \boldsymbol{\eta}_i, R_i \right] - D_{j-1}(\boldsymbol{\eta}_i, \boldsymbol{\Upsilon}_2) \quad \text{for } j \in \{1,\dots,J\}.
    \end{aligned}
\end{equation}
Each $D_j$ term captures contributions of order up to $2(J-j)$ in $\boldsymbol{\eta}_{\text{pred},i}$ and up to order $j$ in the measurement error moments. The term $D_0$ corresponds to the uncorrected estimates, while $D_j$ for $j > 0$ serves as a correction based on higher-order moments of $\mathbf{e}_i$. The specification for the model in \Cref{eq:concrete_model} with $J = 1$ is detailed in \Cref{sec:appendix_fact_corr}.
Plugging in the estimates $\hat{\boldsymbol{\eta}}_i$ and $\hat{\boldsymbol{\Sigma}}_{ee}$ from the first stage, we obtain $\hat{\boldsymbol{\alpha}}$ in \Cref{eq:structural_para_estimator}.

\subsubsection{Modifications for Numerical Stability}

To improve efficiency and ensure that $\hat{\mathbf{M}}$ keep the corrected moment matrix well conditioned, particularly in small-sample settings, we adopt the modification of the 2SMM estimator in \Cref{eq:oberserved_moments} proposed by \textcite{Wall2000}. We separate the uncorrected estimates
\begin{equation*}
    \begin{pmatrix}
        \hat{\mathbf{M}}_1 & \hat{\mathbf{m}}_1
    \end{pmatrix}
    = \frac{1}{N} \sum_{i=1}^{N} D_0(\hat{\boldsymbol{\eta}}_i, \hat{\boldsymbol{\Upsilon}}_2),
\end{equation*}
from the correction terms
\begin{equation*}
    \begin{pmatrix}
        \hat{\mathbf{M}}_2 & \hat{\mathbf{m}}_2
    \end{pmatrix}
    = \frac{1}{N} \sum_{i=1}^{N} \sum_{j=1}^{J} (-1)^j D_j(\hat{\boldsymbol{\eta}}_i, \hat{\boldsymbol{\Upsilon}}_2).
\end{equation*}
Then, we define
\begin{equation*}
    \mathbf{R}_1 =
    \begin{pmatrix}
        \hat{\mathbf{M}}_1 & \hat{\mathbf{m}}_1 \\
        \hat{\mathbf{m}}_1^{\top} & \tfrac{1}{N} \sum_{i=1}^{N}\hat{\eta}_{Y_i}^2
    \end{pmatrix}
    \text{ and }
    \mathbf{R}_2 =
    \begin{pmatrix}
        -\hat{\mathbf{M}}_2 & -\hat{\mathbf{m}}_2 \\
        -\hat{\mathbf{m}}_2^{\top} & \hat{\boldsymbol{\Sigma}}_{ee,k_yk_y}
    \end{pmatrix},
\end{equation*}
where $\hat{\boldsymbol{\Sigma}}_{ee,k_yk_y}$ denotes the estimated error variance of the outcome factor score, which is the last of the $k$ latent variables in \Cref{eq:latent_vars_eta}.
Let $\hat{\lambda}$ denote the largest eigenvalue of
\begin{equation*}
    \mathbf{R}_1^{-1}\mathbf{R}_2.
\end{equation*}
Then, the modified estimator is given by
\begin{equation*}
    \begin{aligned}
    (\check{\mathbf{M}}, \check{\mathbf{m}})=
    \begin{cases}
    \left(\hat{\mathbf{M}}_1, \hat{\mathbf{m}}_1\right) + \left(1-\tfrac{c}{N}\right)\left(\hat{\mathbf{M}}_2, \hat{\mathbf{m}}_2\right), & \text{if } \tfrac{1}{\hat{\lambda}} \geq 1+\tfrac{1}{N}, \\[6pt]
    \left(\hat{\mathbf{M}}_1, \hat{\mathbf{m}}_1\right) + \left(\tfrac{1}{\hat{\lambda}}-\tfrac{1}{N}-\tfrac{c}{N}\right)\left(\hat{\mathbf{M}}_2, \hat{\mathbf{m}}_2\right), & \text{otherwise},
    \end{cases}
    \end{aligned}
\end{equation*}
where $c$ $\in [0, \dim(\boldsymbol{\alpha})+5]$ is an empirically chosen tuning parameter \parencite{Wall2000}. 

While this adjustment bounds the measurement-error correction in finite samples, the resulting matrix may still be poorly conditioned, particularly in small samples or when the moments are highly collinear (e.g., if the effect modification required by Assumptions~\ref{assump:treat_cov_inter} is weak). To improve numerical stability when inverting the matrix, we therefore introduce an additional regularization term by adding a small variance $v$ to the diagonal of $\check{\mathbf{M}}$, analogous to a ridge penalty in linear regression. The final estimator then becomes
\begin{equation} \label{eq:final_param_est_theta}
    \check{\boldsymbol{\alpha}} = (\check{\mathbf{M}} + v \mathbf{I})^{-1} \check{\mathbf{m}}.
\end{equation}

\section{Asymptotic Properties}

We first consider the measurement model parameters, collected in $\hat{\boldsymbol{\Upsilon}} = (\hat{\boldsymbol{\Upsilon}}_1, \hat{\boldsymbol{\Upsilon}}_2)$.

\begin{theorem}\label{thm:upsilon}
    Under Assumptions~\ref{assump:iid_sampling}, \ref{assump:model_spec},
    \ref{assump:regularity_measurement_model}, and \ref{assump:measurement_error} the Z-estimator $\hat{\boldsymbol{\Upsilon}}$ 
    is $\sqrt{N}$-consistent and satisfies
    \[
        \hat{\boldsymbol{\Upsilon}} - \boldsymbol{\Upsilon}
        = \frac{1}{N} \sum_{i=1}^N \boldsymbol{\Delta}_i + o_p\!\left(N^{-1/2}\right),
    \]
    where $\{\boldsymbol{\Delta}_i\}$ are i.i.d.\ with $\operatorname{E}
    [\boldsymbol{\Delta}_i] = 0$ and finite variance.
\end{theorem}

Next, we regard the behavior of the factor score estimator $\hat{\boldsymbol{\eta}}_i$ relative to $\tilde{\boldsymbol{\eta}}_i$ defined in \Cref{eq:factor_score_estimator}.

\begin{theorem}\label{thm:eta}
    Under \Cref{thm:upsilon},
    \[
        \hat{\boldsymbol{\eta}}_i - \tilde{\boldsymbol{\eta}}_i = \hat{\mathbf{B}}_i \, (\hat{\boldsymbol{\Upsilon}}_1 - \boldsymbol{\Upsilon}_1),
    \]
    where
    \[
        \mathbf{B}_i = 
        \begin{pmatrix} 
            \mathbf{H} & \mathbf{H}(\mathbf{q}_i^\top \otimes \mathbf{I}_{p-k}) & -(\mathbf{p}_i^\top \otimes \mathbf{I}_k)
        \end{pmatrix}
    \]
    with
    \[
         \mathbf{q}_i = 
        \begin{pmatrix}
            0 & \mathbf{I}_k 
        \end{pmatrix} \mathbf{z}_i
        \quad \text{and} \quad
        \mathbf{p}_i =
        \begin{pmatrix} 
            \mathbf{I}_{p-k} & -\boldsymbol{\Lambda}_{\mathrm{free}} 
        \end{pmatrix} 
        \left[ 
            \mathbf{z}_i - 
            \begin{pmatrix}
                \boldsymbol{\tau}_{\mathrm{free}} & \mathbf{0}_{k \times 1}
            \end{pmatrix}^{\top}
        \right].
    \]
    Consequently, $\hat{\boldsymbol{\eta}}_i$ is $\sqrt{N}$-consistent.
\end{theorem}

Finally, we can formulate the properties of the structural parameter estimator $\hat{\boldsymbol{\alpha}}$ defined in \Cref{eq:structural_para_estimator}.

\begin{theorem}\label{thm:theta}
Under Assumption~\ref{assump:treat_cov_inter} and \ref{assump:iid_sampling}--\ref{assump:structural_error}, $\hat{\boldsymbol{\alpha}}$ is consistent and asymptotically normally distributed with asymptotic variance
\[
    \mathbf{G}^{-1} \mathbf{S} \mathbf{G}^{-\top},
\]
where
\[
    \mathbf{G} = \operatorname{E}\!\left[ \mathbf{w}_i \boldsymbol{\xi}_{y_i}^\top \right],
\]
\[
    \mathbf{S} = \operatorname{Var}\!\left[ \mathbf{d}_i \right], 
    \qquad 
    \mathbf{d}_i = \mathbf{l}(\tilde{\boldsymbol{\eta}}_i, \boldsymbol{\Upsilon}_{2}, \boldsymbol{\alpha}) 
    + \overline{\mathbf{C}}\, \boldsymbol{\Delta}_i,
\]
\[
    \mathbf{l}(\boldsymbol{\eta}_i, \boldsymbol{\Upsilon}_{2}, \boldsymbol{\alpha}) 
    = \mathbf{m}(\boldsymbol{\eta}_i, \boldsymbol{\Upsilon}_{2}) 
    - \mathbf{M}(\boldsymbol{\eta}_{\mathrm{pred},i}, \boldsymbol{\Upsilon}_{2})\, \boldsymbol{\alpha},
\]
\[
    \overline{\mathbf{C}} = \operatorname{E}\!\left[
        \begin{pmatrix}
            \left. \frac{\partial \mathbf{l}}{\partial \boldsymbol{\eta}_i^{\top}} 
            \right|_{\tilde{\boldsymbol{\eta}}_i, \boldsymbol{\Upsilon}_{2}, \boldsymbol{\alpha}} \mathbf{B}_i &
            \left. \frac{\partial \mathbf{l}}{\partial \boldsymbol{\Upsilon}_{2}^{\top}} 
            \right|_{\tilde{\boldsymbol{\eta}}_i, \boldsymbol{\Upsilon}_{2}, \boldsymbol{\alpha}}
        \end{pmatrix}
    \right].
\]
\end{theorem}

All proofs are provided in \Cref{sec:append_proofs}.
\Cref{thm:theta} also holds for the modified 2SMM estimator $\check{\mathbf{M}}^{-1} \check{\mathbf{m}}$ \parencite{Wall2000}. 
Our adapted version in \Cref{eq:final_param_est_theta}, however, introduces a deliberate bias towards zero, whereby the reduction in variance increases overall efficiency in line with the standard rationale of regularization. Specifically, the bias derivation based on Singular Value Decomposition (SVD) shows that the ridge penalty $v$ shrinks the coordinates of the true coefficient along the principal components of the data \parencite{Hastie2020}. For a standardized principal component with unit variance, this bias simplifies to:
\[
    \operatorname{Bias}(\check{\alpha}_j) = \frac{v}{1 + v} \alpha^*_j.
\]

\section{Robustness Assessment}
\label{sec:sensitvity}

Causal assumptions cannot be fully tested as they concern unobserved quantities. However, we can assess the vulnerability to violations of Assumption~\ref{assump:nem}, following the ideas in \textcite{Zheng2015a}.\footnote{For sensitivity analyses regarding Sequential Ignorability, see, for example, \textcite{Imai2010, VanderWeele2010}.}

Let $U$ be an unobserved confounder that interacts with the latent mediator $\eta_M$, and let this unobserved interaction, $\eta_M U$, affect the outcome with effect size $\delta$. If this interaction is present but omitted from the estimation, the corresponding asymptotic bias of the estimator $\check{\boldsymbol{\alpha}}$ defined in \Cref{eq:final_param_est_theta} is given by
\begin{equation}\label{eq:bias_formla}
    \operatorname{Bias}_{U}(\check{\boldsymbol{\alpha}})=\delta\, (\mathbf{M}')^{-1} \operatorname{E}\big[\mathbf{w}_i^{\top} \eta_{M_i} U_i\big]
\end{equation}
with $\mathbf{M}'=\operatorname{E}[\mathbf{w}_i^{\top}\boldsymbol{\xi}_{Y_i}]+v \mathbf{I}$. The second term $\operatorname{E}[\mathbf{w}_i^{\top} \eta_{M_i} U_i]$ captures the dependence between the weights and the interaction term. Its $\ell$-th element can be decomposed as
\[
    \operatorname{E}[w_{i\ell}\, \eta_{M_i} U_i] = \operatorname{Cov}(w_{i\ell}, \eta_{M_i} U_i) + \operatorname{E}[w_{i\ell}] \operatorname{E}[\eta_{M_i} U_i].
\]
In our setting, this simplifies to the covariance $\operatorname{Cov}[w_{i\ell}, \eta_{M_i} U_i]$. This simplification arises because the elements of $\mathbf{W}$ are centered by construction: the structural weights $a_{r,i}$ and $a_{m,i}$ must fulfill the centering condition to serve as residualized instruments, and we additionally center the covariate terms $q_{yx}(\boldsymbol{\eta}_{\mathbf{X}_i})$. Consequently, $\operatorname{E}[w_{i\ell}]=0$ for all $\ell$, which causes the second term of the decomposition to vanish.

Bias emerges if this covariance is non-zero. The plausibility of such a higher-order nonlinear dependence is generally impossible to evaluate empirically. Quantifying this bias would require specifying the exact functional form of the dependency between the weights, the mediator, and the unobserved confounder. This requirement precludes the use of standard sensitivity analysis techniques, such as those by \textcite{Imai2010}, which rely on bounded parameters (e.g., the correlation between residuals $\rho\in [-1, 1]$) to provide a finite, interpretable space for simulation. In contrast, our bias term depends on a higher-order interaction involving $\mathbf{u}$ that lacks such natural bounds, as the scale of $\mathbf{u}$ and the functional form of its interaction with $\mathbf{w}_j$ and $\boldsymbol{\eta}_M$ are entirely unconstrained. Consequently, unlike traditional sensitivity analysis, which explores a known range of possible violations, the magnitude of this third-order dependence remains practically unquantifiable without making assumptions about the joint distribution of $U$ that are rarely justifiable.\footnote{\textcite{Zheng2015a} bound the bias by the maximum or total violation across interaction terms. However, the coefficients of these bounds again depend on the joint distribution of the unobserved confounder and are computable only in simple models.}

Instead, we focus on the inverse matrix $(\check{\mathbf{M}}')^{-1}$ in \Cref{eq:bias_formla}, which is a calculable and informative quantity. This matrix serves as a scaling factor that determines the extent to which an unobserved interaction is amplified and distributed across the parameter vector. Specifically, the diagonal elements reflect the direct vulnerability of each coefficient to the omitted moderation, whereas the off-diagonal elements indicate the degree of bias transmission between covariates.

The influence of $\check{\mathbf{M}}'$ is fundamentally tied to the stability of its inversion: as the matrix approaches singularity, the entries of the inverse grow significantly, boosting the impact of unobserved effect modification. This allows us to draw several links to our structural specifications. First, Assumption~\ref{assump:treat_cov_inter} ensures that $\check{\mathbf{M}}'$ has full rank and is thus invertible, a condition that fails if the mediator weight becomes collinear with the treatment. weight. For example, if $\gamma_{x1r} = \gamma_{x2r} = 0$ in \Cref{eq:med_weight_explicit}, the mediator weight reduces to $\gamma_r a_{r,i}$, creating collinearity with the treatment weight, leaving the $G$-estimation equation without a unique solution. Consequently, a stronger influence of the covariate-treatment interactions on the mediator not only ensures identification but also serves as a structural defense against bias by improving the conditioning of the system. Second, the numerical stability modifications mitigate the risk of extreme bias in settings of weak identification by bounding the elements of the inverse matrix, although the ridge term cannot restore identification.

To quantify the sensitivity due to structural instability, \textcite{Zheng2015a} suggest considering the condition number, which for RAPSEM is given by:
\begin{equation}
    \kappa (\check{\mathbf{M}}') = \|\check{\mathbf{M}}'\| \cdot \|\check{\mathbf{M}}'^{-1}\| = \frac{\sigma_{\text{max}}(\check{\mathbf{M}}')}{\sigma_{\text{min}}(\check{\mathbf{M}}')}
\end{equation}
where $\sigma_{\text{max}}$ and $\sigma_{\text{min}}$ represent the maximum and minimum singular values, respectively. For improved interpretability, we propose the reciprocal condition number, $1/\kappa \in [0,1]$, as a robustness index: A value near 1 indicates a well-conditioned system where the potential influence of unobserved effect modification is naturally attenuated. Conversely, a value approaching zero signals that the identification is fragile, warning that the causal estimates are highly sensitive to even minor violations of Assumption~\ref{assump:nem}.

\section{Simulation Studies}

We evaluate the proposed method through two simulation studies. The first compares the response of standard SEM and RAPSEM to unobserved confounding, while the second investigates RAPSEM’s statistical power. In the latter, we focus on the structural requirements of Assumption~\ref{assump:treat_cov_inter}, specifically exploring the relationship between the covariate-treatment interaction strength and the robustness coefficient. The entire code is available on GitHub.

\subsection{Data Generation}

We generated data based on \Cref{eq:concrete_model} with an additional unobserved confounding variable $u_i$
\begin{equation}
    \begin{aligned}
        \eta_{y_i} &= 0.224\cdot r_i + \theta_{m}\cdot \eta_{m_i} + 0.240 \cdot \eta_{x1_i} +0.240 \cdot \eta_{x2_i} + \delta_{u}\cdot u_i + \delta_{um} \cdot u_i \eta_{m_i} + \zeta_{y_i}, \\
        \eta_{m_i} &= 0.316 \cdot r_i + 0.277 \cdot \eta_{x1_i} + 0.277 \cdot \eta_{x2_i} + \gamma_{x1r}\cdot r_i\eta_{x1_i}+\gamma_{x2r}\cdot r_i\eta_{x2_i}+ \delta_{u} \cdot u_i + \zeta_{m_i},
    \end{aligned}
\end{equation}
where the parameters $\theta_r$, $\boldsymbol{\theta}_x$, $\gamma_r$, $\boldsymbol{\gamma}_x$ were fixed to standardized effect sizes corresponding to $R^2$ of $5\%$, $5\%$, $15\%$ and $20\%$, respectively, across all simulations. When varying the other effect sizes, we employed Cohen’s d values of $0.2, 0.5, 0.65$, and $0.8$. These represent small, medium, and large effects \parencite{Cohen1977} and correspond to $R^2$ values of $1\%, 6\%, 10\%,$ and $14\%$, respectively. For each parameter combination, we generated $100$ datasets, applying the same seeds across conditions to ensure comparability. 

The treatment $R$ was sampled from a discrete uniform distribution on $\{-1, 1\}$.\footnote{Note that the CDE thus corresponds to $\frac{1}{2}\theta_r$.} Covariates $\eta_{x_1}$ and $\eta_{x_2}$ were drawn from a multivariate standard normal distribution with correlation $\rho = 0.3$, while the confounder $U$ followed an independent standard normal distribution. Residuals $\zeta_m$ and $\zeta_y$ were sampled from $N(0, \psi)$, where the residual variance $\psi = 1 - R^2$ was determined by population parameters to maintain a theoretical unit variance for $\eta_m$ and $\eta_y$.

Each latent variable was measured by three indicators with parameters randomized per dataset to enhance generalizability. Item-specific reliabilities $\text{rel}_j$ were sampled from $U(\text{rel} - 0.1, \text{rel} + 0.1)$, where $\text{rel}$ denotes the reliability level defined by the simulation condition. Factor loadings were set to $\lambda_j = \sqrt{\text{rel}_j}$ with corresponding residual variances of $1 - $ \\ $\text{rel}_j$, and item intercepts $\nu_j$ were drawn from $U(-1, 1)$.

\subsection{Simulation Study 1: Robustness against Unobserved Confoundedness and Effect Modification}

We compared three estimation methods to evaluate their performance in the presence of unobserved confounding and effect modification:
\begin{enumerate}
    \item \textbf{RAPSEM}: Our proposed framework utilizing latent $G$-estimation implemented as R package \texttt{rapsem}
    \item \textbf{Uncorrected}: A two-stage latent regression approach using the predictor matrix $\boldsymbol{\Xi}_y$ in place of the weight matrix $\mathbf{W}$ in \Cref{eq:observed_gparm}. This represents factor score regression with the same correction for measurement error but without $G$-estimation corrections.
    \item \textbf{Lavaan}: A standard \texttt{lavaan} implementation utilizing a two-group model to capture treatment-covariate interactions, with factor loadings and intercepts constrained to equality across groups.
\end{enumerate}

\paragraph{Data conditions} 

The confounding effect sizes $\delta_{u}$ and $\delta_{um}$ were varied across different sample sizes $N$ according to the conditions in \Cref{tab:study1conditions}. When varying the interaction effect $\delta_{um}$, we held the main effect $\delta_{u}$ at its maximum ($0.374$) to respect the hierarchical principle that interactions manifest alongside their corresponding main effects. To evaluate the type I error rate, the mediator-outcome effect $\theta_{m}$ was fixed at zero. Reliability $\text{rel}$ was set to $0.9$ and interaction parameters $\boldsymbol{\gamma}_{\mathbf{x}xr}$ were fixed at $0.232$, representing a substantial treatment-covariate interaction ($R^2 = 14\%$) to ensure high statistical power of RAPSEM.

\begin{table}[ht]
    \caption{Varied parameters in Study 1: unobserved confounding effect size ($\delta_{u}$), unobserved-confounder-mediator interaction effect size ($\delta_{um}$), and sample size ($N$).}
    \centering
    \begin{tabular}{lcccc}
    \toprule
        \rule{0pt}{3ex} \textbf{Variable} & \multicolumn{4}{c}{\textbf{Levels}} \\ [0.5ex] \hline
        \rule{0pt}{3ex}
        $\delta_{u}$ & $0.100$ & $0.245$ & $0.316$ & $0.374$ \\
        $\delta_{um}$ & $0.100$ & $0.245$ & $0.316$ & $0.374$ \\
        $N$ & $250$ & $500$ & $750$ & $1000$ \\ [1ex]
    \bottomrule
    \end{tabular}
    \label{tab:study1conditions}
\end{table}

\paragraph{Results}

\begin{figure}
     \centering
      \begin{subfigure}[b]{\textwidth}
      \centering
     \includegraphics[width=0.65\textwidth]{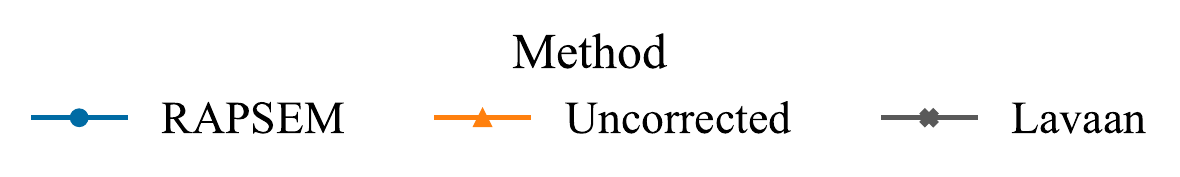}
     \end{subfigure}
     \begin{subfigure}[b]{0.31\textwidth}
         \centering
         \includegraphics[height=5.05cm]{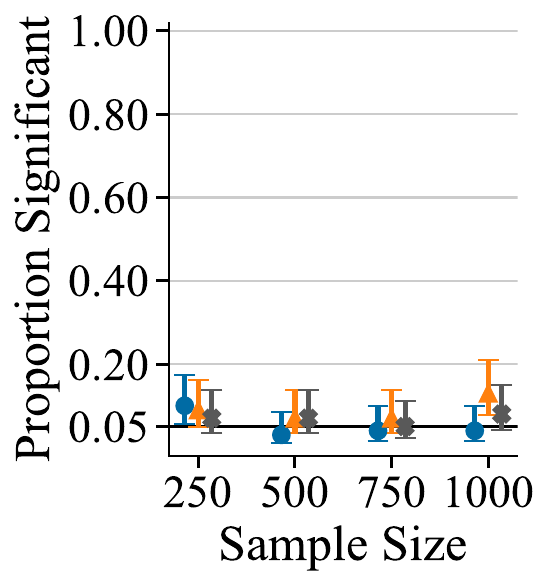}
         \caption{\centering $\delta_{u}=0.100$}
         \label{fig:conf_0.1}
     \end{subfigure}
     \hspace{-1em}
     \begin{subfigure}[b]{0.23\textwidth}
         \centering
         \includegraphics[height=5cm]{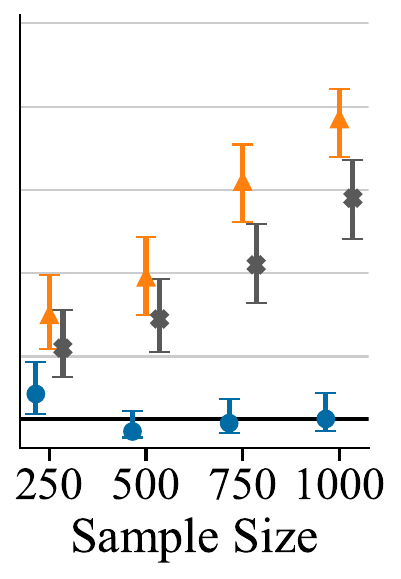}
         \caption{\centering $\delta_{u}=0.245$}
         \label{fig:conf_0.245}
     \end{subfigure}
     \hspace{-1em}
     \begin{subfigure}[b]{0.23\textwidth}
         \centering
         \includegraphics[height=5cm]{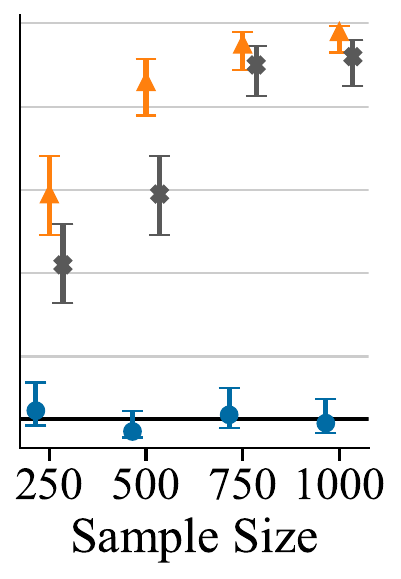}
         \caption{\centering $\delta_{u}=0.316$}
         \label{fig:conf_0.316}
     \end{subfigure}
     \hspace{-1em}
     \begin{subfigure}[b]{0.23\textwidth}
         \centering
         \includegraphics[height=5cm]{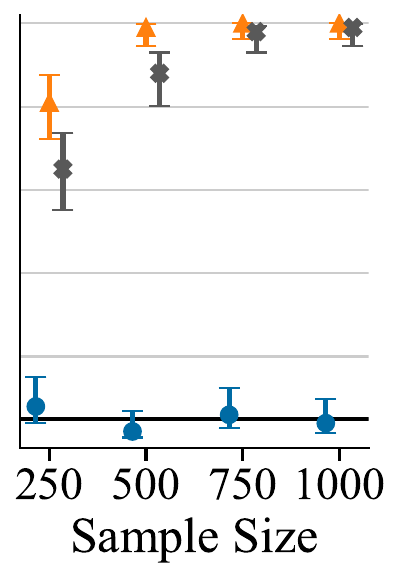}
         \caption{\centering $\delta_{u}=0.374$}
         \label{fig:conf_0.374}
     \end{subfigure}
     \caption{False positive rates of $\check{\theta}_m$ with Wilson score $95\%$ confidence intervals under varying confounding levels and sample size.}
    \label{fig:false_positive_rate}
\end{figure}

False positive rates (\Cref{fig:false_positive_rate}) were severely inflated for standard SEM approaches (Uncorrected and Lavaan) in the presence of unobserved confounding. In moderate to high confounding conditions ($\delta_u \geq 0.316$), the Type I rates exceeded $80\%$ and approached $100\%$ as the sample size increased. In contrast, RAPSEM consistently maintained the false positive error rate around the nominal $5\%$ level across all settings.

\begin{figure}
     \centering
      \begin{subfigure}[b]{\textwidth}
      \centering
     \includegraphics[width=0.65\textwidth]{figs/false_positive_rate/false_positive_rate_legend.pdf}
     \end{subfigure}
     \begin{subfigure}[b]{0.31\textwidth}
         \centering
         \includegraphics[height=5.05cm]{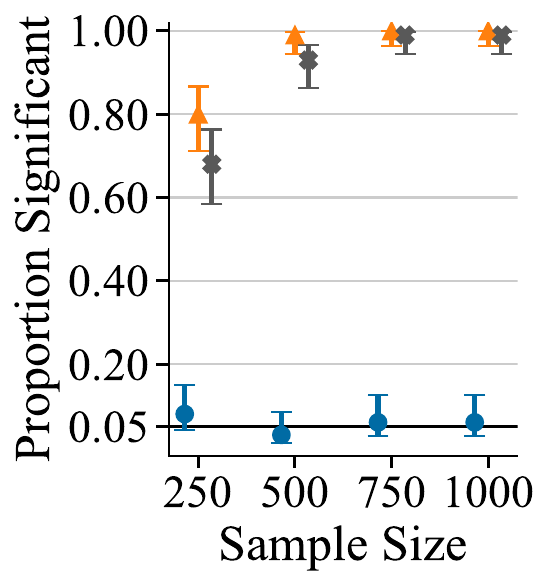}
         \caption{\centering $\delta_{um}=0.100$}
         \label{fig:conf_med_0.1}
     \end{subfigure}
     \hspace{-1em}
     \begin{subfigure}[b]{0.23\textwidth}
         \centering
         \includegraphics[height=5cm]{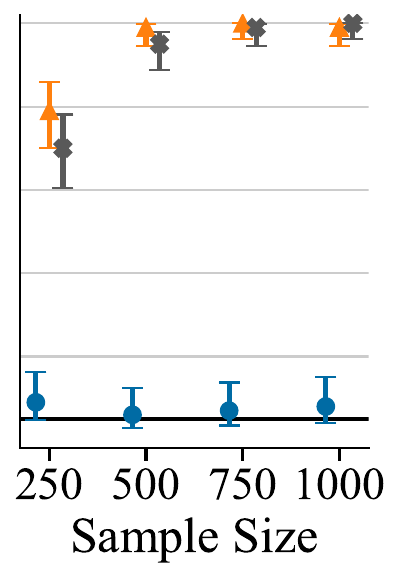}
         \caption{\centering $\delta_{um}=0.245$}
         \label{fig:conf_med_0.245}
     \end{subfigure}
     \hspace{-1em}
     \begin{subfigure}[b]{0.23\textwidth}
         \centering
         \includegraphics[height=5cm]{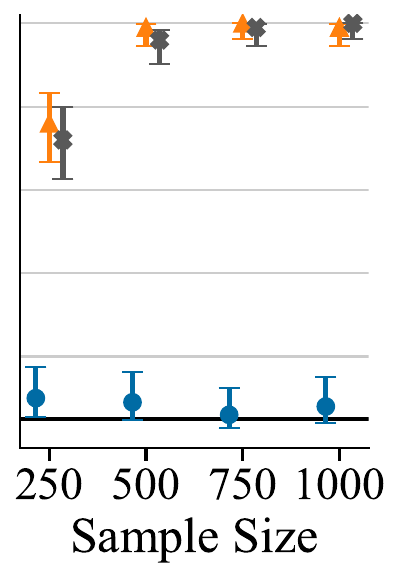}
         \caption{\centering $\delta_{um}=0.316$}
         \label{fig:conf_med_0.316}
     \end{subfigure}
     \hspace{-1em}
     \begin{subfigure}[b]{0.23\textwidth}
         \centering
         \includegraphics[height=5cm]{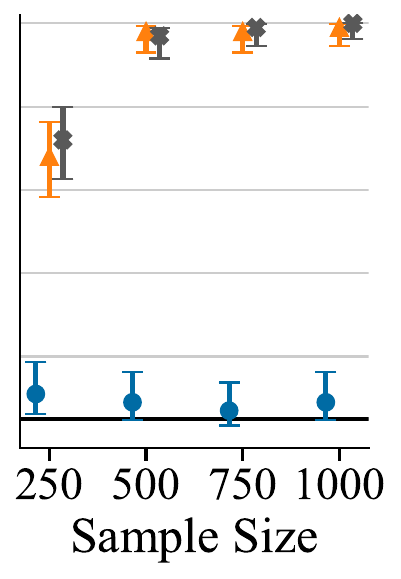}
         \caption{\centering $\delta_{um}=0.374$}
         \label{fig:conf_med_0.374}
     \end{subfigure}
     \caption{False positive rates of $\check{\theta}_m$ with Wilson score $95\%$ confidence intervals under varying confounding levels and sample size.}
    \label{fig:false_positive_rate_inter}
\end{figure}

Effect modification ($\delta_{um}$) had minimal influence on the type one error rates (\Cref{fig:false_positive_rate_inter}). For standard SEM, these rates were, interestingly, slightly reduced; for RAPSEM, they only marginally increased, such that the false positive rate stayed close to $5\%$. This robustness to the presence of unobserved effect modification provides further evidence of RAPSEM's validity, consistent with findings in \textcite{Brandt2020}.

RAPSEM estimates of $\theta_m$ remained centered around the true value of zero, whereas standard SEM estimates shifted upward with increasing $\delta_u$, indicating substantial positive bias (\Cref{fig:bias_histograms} in \Cref{sec:append_plots}). As expected, this bias is not reflected in typical SEM model fit indices, which remained within acceptable ranges despite the lack of sequential ignorability (\Cref{fig:model_fit} in \Cref{sec:append_plots}). RAPSEM’s sampling distributions also reveal increased variance in small samples, reflecting an inherent trade-off between robustness and small-sample efficiency, with the gap narrowing at $N=1000$.

\begin{figure}
     \centering
      \begin{subfigure}[b]{\textwidth}
      \centering
     \includegraphics[width=0.65\textwidth]{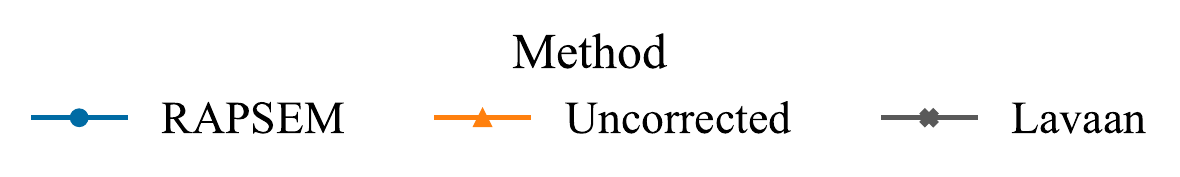}
     \end{subfigure}
     \begin{subfigure}[b]{0.31\textwidth}
         \centering
         \includegraphics[height=5cm]{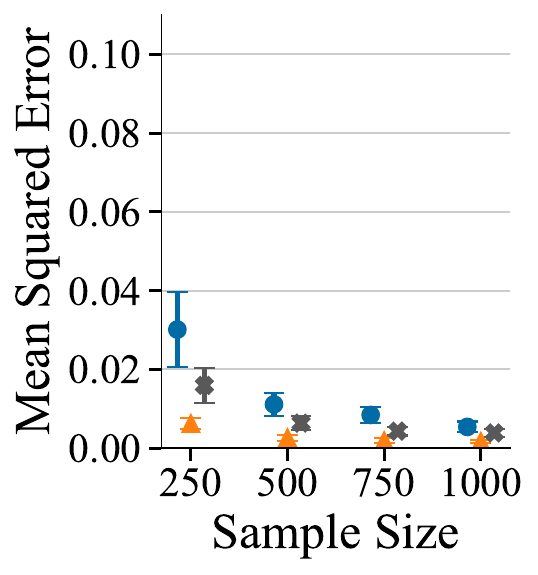}
         \caption{\centering $\delta_{u}=0.100$}
         \label{fig:mse_conf_0.1}
     \end{subfigure}
     \hspace{-1em}
     \begin{subfigure}[b]{0.236\textwidth}
         \centering
         \includegraphics[height=5cm]{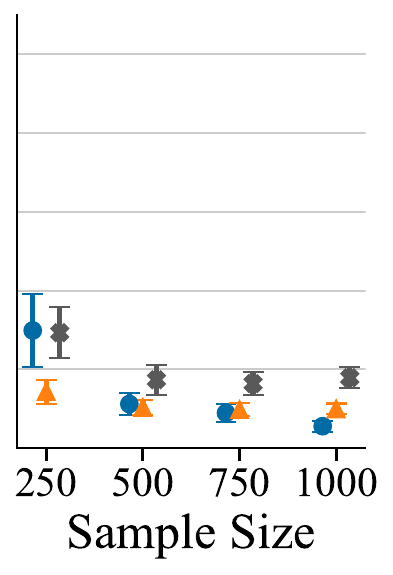}
         \caption{\centering $\delta_{u}=0.245$}
         \label{fig:mse_conf_0.245}
     \end{subfigure}
     \hspace{-1em}
     \begin{subfigure}[b]{0.236\textwidth}
         \centering
         \includegraphics[height=5cm]{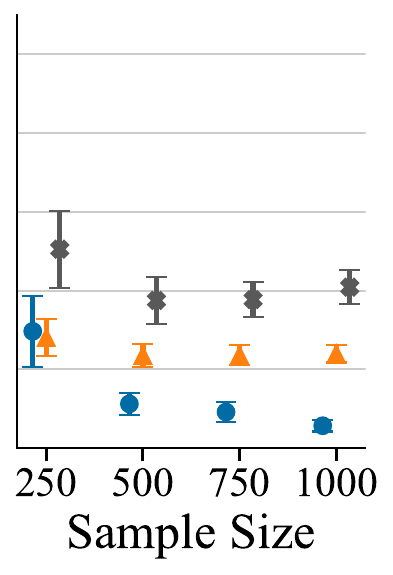}
         \caption{\centering $\delta_{u}=0.316$}
         \label{fig:mse_conf_0.316}
     \end{subfigure}
     \hspace{-1em}
     \begin{subfigure}[b]{0.236\textwidth}
         \centering
         \includegraphics[height=5cm]{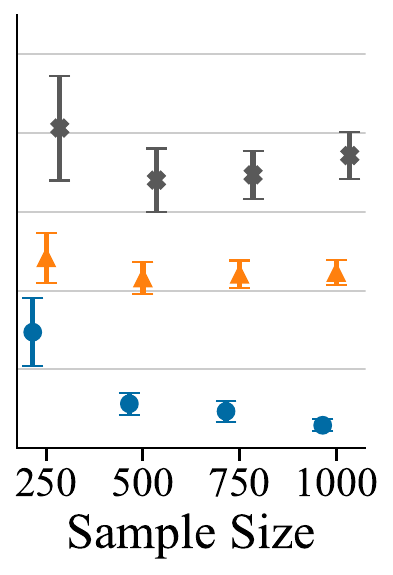}
         \caption{\centering $\delta_{u}=0.374$}
         \label{fig:mse_conf_0.374}
     \end{subfigure}
     \caption{Mean squared error of $\check{\theta}_m$ with Normal approximation $95\%$ confidence intervals under varying levels of confounding and sample size.}
    \label{fig:mean_squared_error}
\end{figure}

These two properties jointly determine MSE behavior (\Cref{fig:mean_squared_error}). In the absence of confounding, RAPSEM's higher variance dominates, yielding a larger MSE than standard SEM at small sample sizes and reaching parity only at $N=1000$. In all confounding scenarios, however, the bias term causes standard SEM's MSE to grow significantly with sample size, eventually exceeding that of RAPSEM.

\subsection{Simulation Study 2: Statistical Power}

Building on the findings of Study 1, we further investigated the efficiency--robustness trade-off by identifying the conditions under which RAPSEM achieves adequate statistical power. Additionally, we evaluated the framework's robustness to violations of \ref{assump:nem} using the condition number $\kappa$, as introduced in the sensitivity analysis section.

\paragraph{Data Conditions}

We identified four primary factors that determine the power and sensitivity of RAPSEM: the mediator effect size $\check{\theta}_m$, the strength of the covariate--treatment interaction $\boldsymbol{\gamma}_{xr}$, the reliability of the indicators $\text{rel}$, and the sample size $N$. These parameters were systematically varied according to the levels specified in \Cref{tab:study2conditions}, without unobserved confounding ($\delta_{u} = 0$). In \Cref{tab:f_stats_simulation}, we present the theoretical $F$-statistics for the joint test of the interaction terms calculated using the nested model comparison formula:
\begin{equation}
    F = \frac{\Delta R^2_{int} / m}{(1 - R^2_{total}) / (N - k - 1)}
\end{equation}
where $\Delta R^2_{int}$ is the variance uniquely explained by the $m=2$ interactions, $R^2_{total}=35\%+\Delta R^2_{int}$ is the total variance explained by the full model (including a $15\%$ treatment effect and $20\%$ covariate main effects), $N$ is the sample size, and $k=5$ represents the total number of predictors.
The robustness $1/\kappa$ was calculated in additional runs with $\text{rel}=0.9$ and $\theta_{m}=0.0$ to mirror the first study, as the mediation effect size does not influence this coefficient. $\delta_{u}$ and $\delta_{um}$ were set to zero.

\begin{table}[ht]
    \caption{Varied parameters in Study 2: levels of mediator effect size $\check{\theta}_m$, covariate--treatment interaction effect size $\boldsymbol{\gamma}_{xr}$, indicator reliability $\text{rel}$, and sample size $N$.}
    \centering
    \begin{tabular}{lcccc}
    \toprule
        \rule{0pt}{3ex} \textbf{Variable} & \multicolumn{4}{c}{\textbf{Levels}} \\ [0.5ex] \hline
        \rule{0pt}{3ex}
        $\theta_{m}$ & $0.100$ & $0.243$ & $0.316$ & $0.371$ \\ 
        $\boldsymbol{\gamma}_{xr}$ & $0.062$ & $0.152$ & $0.196$ & $0.232$ \\ 
        $\text{rel}$ & $0.5$ & $0.7$ & $0.9$ & \\
        $N$ & $250$ & $500$ & $750$ & $1000$ \\ [1ex]
    \bottomrule
    \end{tabular}
    \label{tab:study2conditions}
\end{table}

\begin{table}[ht!]
    \centering
    \caption{Theoretical $F$-statistics for the joint test of covariate--treatment interactions across conditions in Study 2.}
    \label{tab:f_stats_simulation}
    \begin{tabular}{lcccc}
    \toprule
    \textbf{\textbf{Sample Size} ($N$)} & \multicolumn{4}{c}{\textbf{\textbf{Interaction Effect Size} ($\boldsymbol{\gamma}_{xr}$)}} \\
    \cline{2-5}
     & \textbf{$0.062$} & \textbf{$0.152$} & \textbf{$0.196$} & \textbf{$0.232$} \\
    \midrule
        250  & 1.9  & 12.4  & 22.2  & 33.5 \\
        500  & 3.9  & 25.1  & 44.9  & 67.8 \\
        750  & 5.8  & 37.8  & 67.6  & 102.1 \\
        1000  & 7.8  & 50.5  & 90.4  & 136.4 \\
    \bottomrule
    \end{tabular}
\end{table}

\begin{figure}
     \centering
      \begin{subfigure}[b]{\textwidth}
      \centering
     \includegraphics[width=0.7\textwidth]{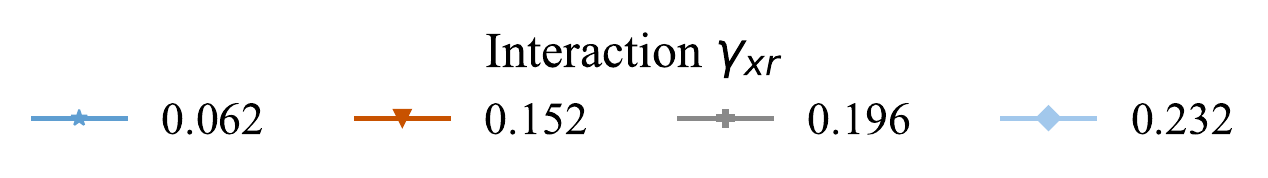}
     \end{subfigure}
     \begin{subfigure}[b]{0.31\textwidth}
         \centering
         \includegraphics[height=5.05cm]{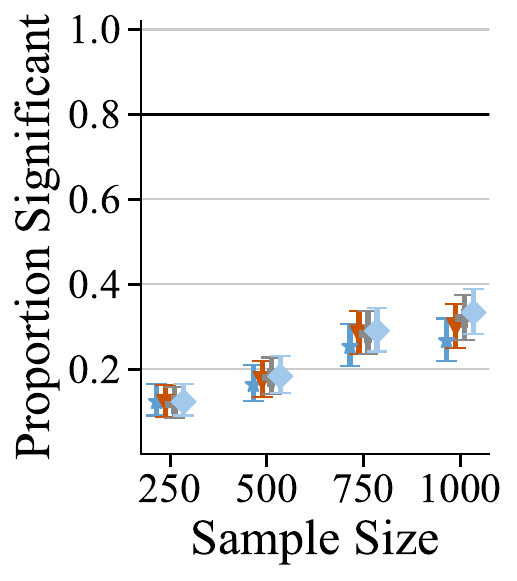}
         \caption{\centering $\theta_{m} = 0.100, $ \\ $\text{rel}= 0.5$}
         \label{fig:power_cme_0.1_rel0.5}
     \end{subfigure}
     \hspace{-1em}
    \begin{subfigure}[b]{0.23\textwidth}
         \centering
         \includegraphics[height=5cm]{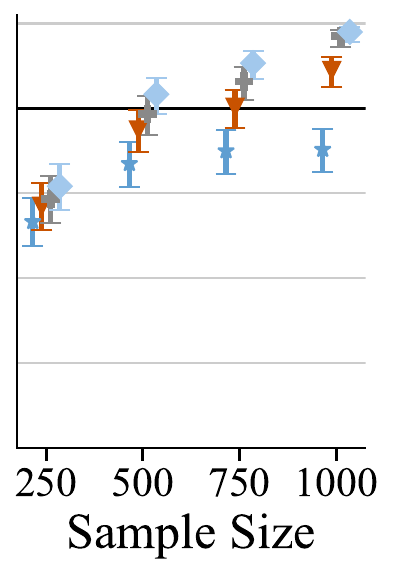}
         \caption{\centering $\theta_{m} = 0.316, $ \\ $\text{rel}= 0.5$}
         \label{fig:power_cme_0.316_rel0.5}
     \end{subfigure}
     \begin{subfigure}[b]{0.23\textwidth}
         \centering
         \includegraphics[height=5cm]{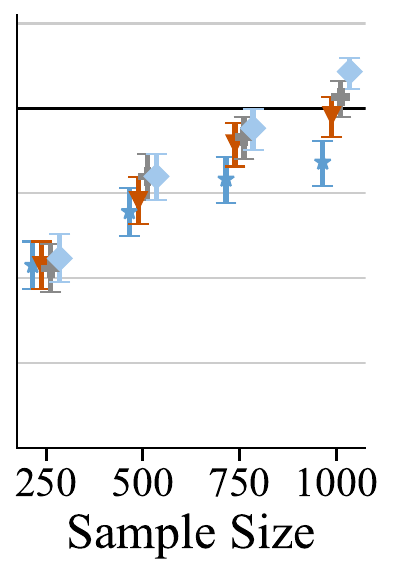}
         \caption{\centering $\theta_{m} = 0.243, $ \\ $\text{rel}= 0.5$}
         \label{fig:power_cme_0.243_rel0.5}
     \end{subfigure}
     \hspace{-1em}
     \begin{subfigure}[b]{0.23\textwidth}
         \centering
         \includegraphics[height=5cm]{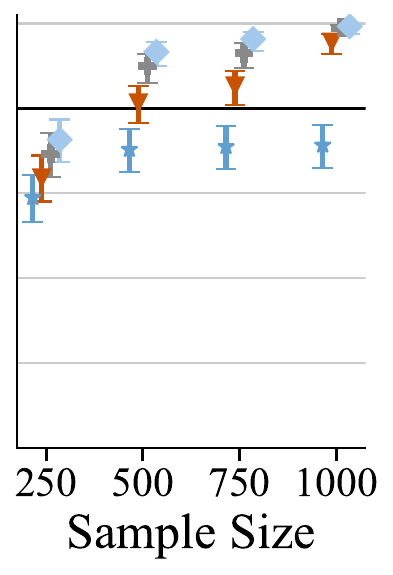}
         \caption{\centering $\theta_{m} = 0.374, $ \\ $\text{rel}= 0.5$}
         \label{fig:power_cme_0.374_rel0.5}
     \end{subfigure}
     \par
     \begin{subfigure}[b]{0.31\textwidth}
         \centering
         \includegraphics[height=5.05cm]{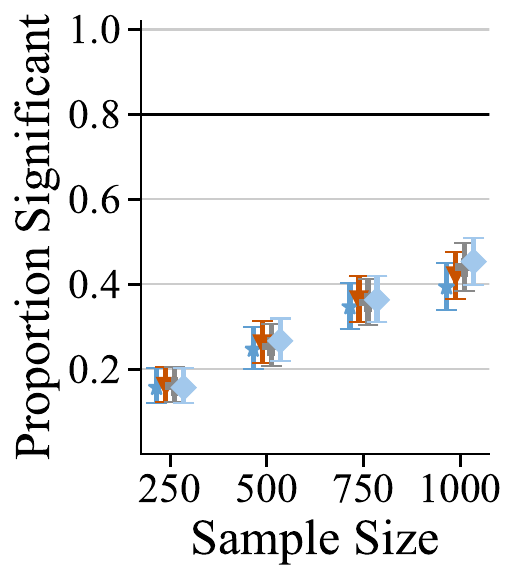}
         \caption{\centering $\theta_{m} = 0.100, $ \\ $\text{rel}= 0.7$}
         \label{fig:power_cme_0.1_rel0.7}
     \end{subfigure}
     \hspace{-1em}
    \begin{subfigure}[b]{0.23\textwidth}
         \centering
         \includegraphics[height=5cm]{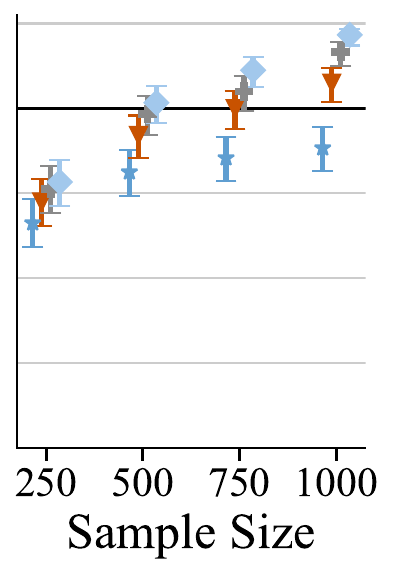}
         \caption{\centering $\theta_{m} = 0.243, $ \\ $\text{rel}= 0.7$}
         \label{fig:power_cme_0.243_rel0.7}
     \end{subfigure}
     \hspace{-1em}
     \begin{subfigure}[b]{0.23\textwidth}
         \centering
         \includegraphics[height=5cm]{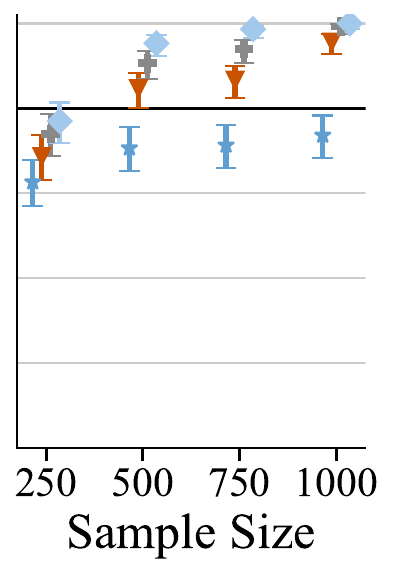}
         \caption{\centering $\theta_{m} = 0.316, $ \\ $\text{rel}= 0.7$}
         \label{fig:power_cme_0.316_rel0.7}
     \end{subfigure}
     \hspace{-1em}
     \begin{subfigure}[b]{0.23\textwidth}
         \centering
         \includegraphics[height=5cm]{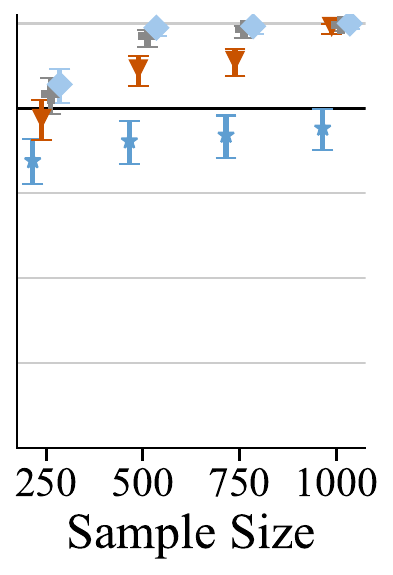}
         \caption{\centering $\theta_{m} = 0.374, $ \\ $\text{rel}= 0.7$}
         \label{fig:power_cme_0.374_rel0.7}
     \end{subfigure}
     \par
     \begin{subfigure}[b]{0.31\textwidth}
         \centering
         \includegraphics[height=5.05cm]{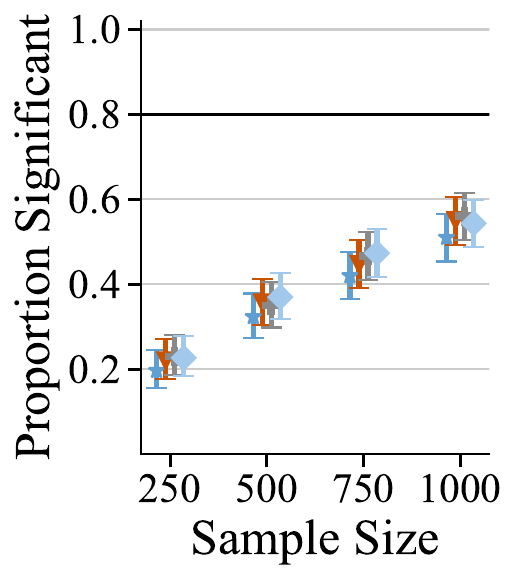}
         \caption{\centering $\theta_{m} = 0.100, $ \\ $\text{rel}= 0.9$}
         \label{fig:power_cme_0.1_rel0.9}
     \end{subfigure}
     \hspace{-1em}
     \begin{subfigure}[b]{0.23\textwidth}
         \centering
         \includegraphics[height=5cm]{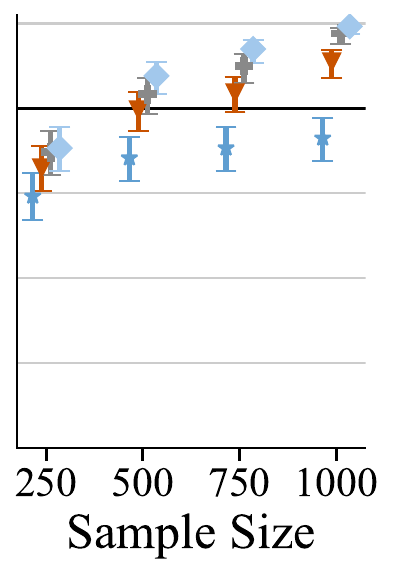}
         \caption{\centering $\theta_{m} =0.243, $ \\ $\text{rel}= 0.9$}
         \label{fig:power_cme_0.243_rel0.9}
     \end{subfigure}
     \hspace{-1em}
     \begin{subfigure}[b]{0.23\textwidth}
         \centering
         \includegraphics[height=5cm]{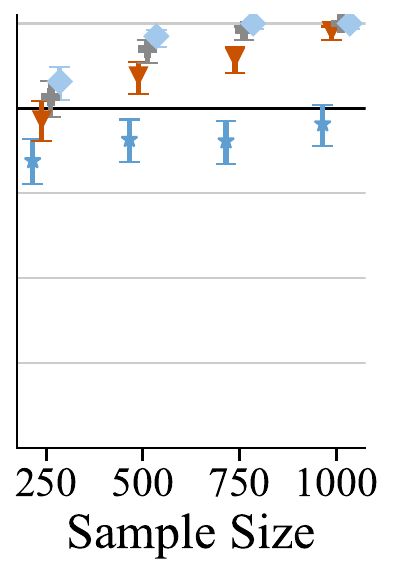}
         \caption{\centering $\theta_{m} =0.361, $ \\ $\text{rel}= 0.9$}
         \label{fig:power_cme_0.316_rel0.9}
     \end{subfigure}
     \hspace{-1em}
     \begin{subfigure}[b]{0.23\textwidth}
         \centering
         \includegraphics[height=5cm]{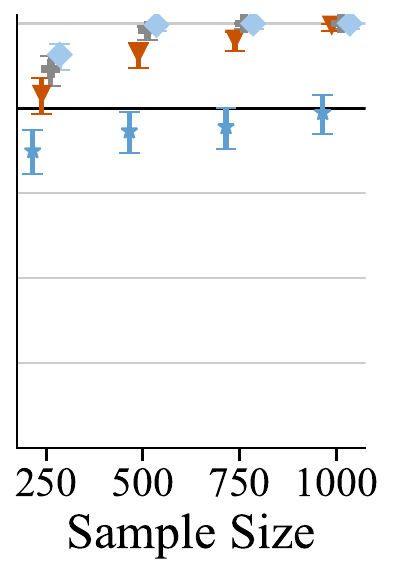}
         \caption{\centering $\theta_{m} =0.374, $ \\ $\text{rel}= 0.9$}
         \label{fig:power_cme_0.374_rel0.9}
     \end{subfigure}
     \caption{Power of $\theta_{m}$ with Wilson score $95\%$ confidence intervals for varying effect sizes, reliabilities, and covariate-treatment interaction effect sizes.}
     \label{fig:power}
\end{figure}

\begin{figure}
 \centering
    $\vcenter{\hbox{\includegraphics[height=5cm]{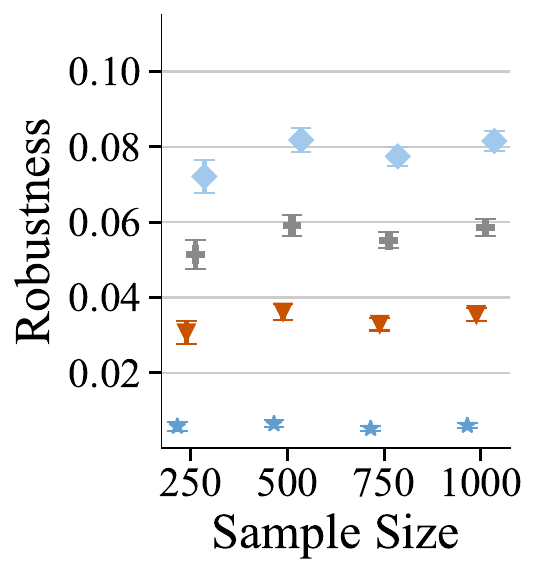}}}$%
    \hspace{0.75em}%
    $\vcenter{\hbox{\includegraphics[height=2.5cm]{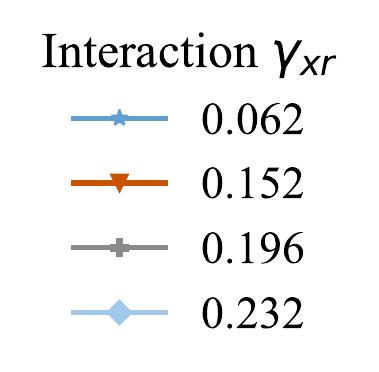}}}$
     \caption{Robustness coefficient $1/\kappa$ with Delta-method approximated $95\%$ confidence intervals for varying covariate-treatment interaction effect sizes.}
    \label{fig:sensitivity}
\end{figure}

\paragraph{Results}

The power to detect a non-zero mediation effect (\Cref{fig:power}) increased systematically with sample size ($N$), measurement reliability ($\text{rel}$), and the strength of the covariate–treatment interaction ($\gamma_{xr}$). Overall, the findings suggest that large-scale studies are essential. A sample size of $N=250$ was insufficient for detecting even large effects, while $N=500$ serves as a critical threshold where success becomes highly contingent on the remaining parameters.

Regarding mediation effect sizes, power remained insufficient for small effects ($\check{\theta}_m=0.100$) across all scenarios, failing to reach the $80\%$ threshold even at $N=1000$. While power naturally improves as $\check{\theta}_m$ increases, reaching an acceptable level for medium-to-large effects depends on the quality of the measurement and the interaction strength.

Low reliability ($\text{rel}=0.5$) significantly hinders the detection of moderate effects, often keeping power well below acceptable thresholds. While the differences between reliability levels of $0.7$ and $0.9$ are less pronounced, higher reliability consistently yields marginal gains in power.

Ultimately, the covariate–treatment interaction emerges as the most critical determinant of model performance apart from sample size. The strongest interaction levels consistently yield the highest power. Conversely, a small interaction effect ($\gamma_{xr}=0.062$) fails to reach $80\%$ power in any tested scenario. 

This interaction is not merely a matter of statistical power but a fundamental requirement for model identification. We could not include a zero interaction scenario in our simulations because, without this effect, the weight matrix becomes collinear and the inverse fails to converge, as detailed in the sensitivity analysis section. 

This structural dependency is empirically demonstrated by the robustness coefficient $1/\kappa$ (\Cref{fig:sensitivity}). While $1/\kappa$ remains approximately constant across sample sizes, it increases sharply with the interaction effect size, rising from approximately $0.006$ for a small interaction to $0.08$ for a large interaction. By a standard rule of thumb \parencite{Cheney2007}, these values correspond to a loss of about two digits and one digit of precision in the calculation of $\check{\boldsymbol{\alpha}}$, respectively, with the latter indicating greater structural robustness. This behavior confirms that stable and reliable RAPSEM estimation depends fundamentally on the interaction effect size $\gamma_{xr}$.

\section{Empirical Example}

We further illustrate the differences between our proposed method and the two alternative approaches investigated in the simulation study through an empirical example from educational psychology. We use publicly available data by \textcite{Russ2026}, who investigated the effects of various self-study strategies (generative learning activities) on cognitive learning outcomes in an RCT.

We emphasize the illustrative nature of this analysis. The modeling approach was adapted to facilitate a methodological comparison rather than to provide new evidence for or against the specific psychological relationships studied. Consequently, our modeling choices deviate substantially from those of the original authors. While their specifications were based on domain expertise and prior literature, we adapted the model specifically to demonstrate how diverging results can emerge in real-world applications when using different statistical frameworks.

Data were collected in a classroom setting involving $30$ seventh- and eighth-grade classes from $11$ secondary schools in Southwest Germany, with a resulting sample size of $N=590$ students. Contrary to the original study, which utilized manifest $z$-scores in a multilevel model, we employ SEM with a measurement model but without a hierarchical structure as implemented in RAPSEM. This involves omitting the nested structure of the data, specifically the clustering within schools and classes, a choice that leads to results differing from the original findings. This departure is useful for our illustration because, while correlated random intercepts in a multilevel model can account for some unobserved confounding, our approach is designed to address such confounding through the structural model itself. By removing the hierarchical layers, we can more clearly highlight the impact of unobserved confounding and demonstrate how the different methods studied in this paper respond to it. Consistent with the original study, missing data were handled via multiple imputation by chained equations (MICE) under the assumption of data missing at random (MAR), and all results are pooled across $20$ imputed datasets.

While the original experiment included four treatment conditions, we restricted our analysis to two: the control condition, in which students restudied the lecture content, and the treatment condition, in which they explained the content to a fictitious, non-present peer via voice messages. This results in a binary but unbalanced treatment variable, with only $150$ students in the control group. This imbalance is noteworthy because it reduces the overall statistical power to detect a mediation effect compared to a balanced design of equal total size.

The outcome, conceptual knowledge, was measured using $15$ binary items. To facilitate factor score estimation, we aggregated these items into three parcels, with item assignment randomized for each imputation. We selected task interest as the mediator, which was identified as a key predictor of conceptual knowledge in the original study, modeling it as a continuous latent variable measured by two items on a 4-point Likert scale.

While the original study collected numerous control variables, we selected three specific covariates for this illustration: students’ interest in physics, conscientiousness in physics, and perceived teacher support (each measured with four items all on a 4-point Likert scale). These specific variables were selected because they exhibited significant individual and joint interaction effects with the treatment on the mediator (see \Cref{tab:med_moderation}).
We additionally included perceived cognitive activation (six items combined to three parcels) as a confounder, given its significant main effects on both the mediator and the outcome. Notably, while the original study’s pre-test cognitive assessment would be a primary confounder to include, we deliberately omitted it to accentuate the influence of the other factors in our methodological comparison.

\begin{table}[htbp]
  \centering
  \caption{Testing individual and joint treatment-covariate interaction effects on the mediator: reported are the interaction coefficients: interaction coefficient $\gamma_{xr}$, partial $F$-statistic, and $p$-value for the comparison between models with and without interaction terms.}
  \label{tab:med_moderation}
  \begin{tabular}{lrrr}
    \toprule
    \textbf{Covariate} & \textbf{Effect Size} & \textbf{$F$-Statistic} & \textbf{$p$-value} \\
    \midrule
    interest in physics & -0.231 & 6.6 & 0.011 \\
    conscientiousness in physics & -0.220 & 5.4 & 0.020 \\
    perceived teacher support & -0.290 & 6.2 & 0.013 \\
    All Combined & --- & 5.1 & 0.002 \\
    \bottomrule
  \end{tabular}
\end{table}

We compared two model specifications:
\begin{itemize}
    \item \textbf{Full Model M1}: Included all the aforementioned variables.
    \item \textbf{Reduced Model M2}: Omitted the confounder perceived cognitive activation, artificially treating it as an unobserved confounder. This reflects a scenario with unobserved confounding where a critical dimension of the learning environment might be overlooked during study design.
\end{itemize}

\subsection{Results}

The empirical results (\Cref{tab:empirical_example}) highlight the distinct trade-offs between RAPSEM and standard SEM (Uncorrected and Lavaan).

\begin{table}[ht!]
    \centering
    \caption{Comparison of mediation effects ($\check{\theta}_m$) across estimation methods. Values represent point estimates and 95\% confidence intervals ($\check{\theta} \pm 1.96\,\check{\sigma}_{\text{pooled}}$) aggregated across imputations.}
    \label{tab:empirical_example}
    \begin{tabular}{lccc}
    \toprule
    \textbf{Model} & \multicolumn{3}{c}{\textbf{Estimation Method}} \\
    \cline{2-4}
     & \textbf{G-Estimator} & \textbf{Uncorrected} & \textbf{Lavaan} \\
    \midrule
        \textbf{M1 (Full)} & $-0.06$ $[-0.75,\,0.62]$ & $0.06$ $[-0.10,\,0.22]$ & $0.14$ $[0.00,\,0.28]$ \\
        \textbf{M2 (Reduced)} & $-0.08$ $[-0.71,\,0.55]$ & $0.09$ $[-0.01,\,0.19]$ & $0.17$ $[0.05,\,0.28]$ \\
    \bottomrule
    \end{tabular}
\end{table}

The most critical finding concerns comparative robustness. For the standard approaches, omitting confounders produces a shift in the $\check{\theta}_m$ point estimate, indicating susceptibility to omitted variable bias: the estimates of the standard factor score regression and the \texttt{lavaan} group model increase from $0.06$ to $0.09$ and from $0.14$ to $0.17$, respectively, i.e., by about half of their standard errors. RAPSEM estimates, by contrast, remain stable relative to their uncertainty, changing from $-0.06$ to $-0.08$, i.e., by less than $0.1$ standard errors.

The substantially wider confidence intervals for RAPSEM reflect greater variance relative to the standard models, consistent with our simulation findings. Given that the effective sample size is $N_{\text{eff}} = (4 \cdot N_{\text{treat}} \cdot N_{\text{control}}) / N = 447$ and McDonald's $\omega$ ranges from $0.63$ to $0.86$ (\Cref{tab:reliability} in \Cref{sec:append_rel}), measurement quality is acceptable. The excess variance is therefore primarily attributable to weak instruments. The detection of multiple interacting covariates in a dataset not originally designed to capture effect modification suggests that treatment-covariate interactions may be a realistic expectation rather than an edge case. Nevertheless, the combined $F$-statistic of $5.13$ falls below the threshold of $F > 10$ suggested by \textcite{Stock2002}. Indeed, comparing with Simulation Study 2, this value is only marginally above the theoretical one for $N = 500$ with the smallest interaction effect size ($R^2 = 1\%$) in \Cref{tab:f_stats_simulation}. This places the application in the low-power regime illustrated by the dark-blue markers in \Cref{fig:power}, which fail to reach the conventional power threshold of $0.8$. Accordingly, also the small estimated robustness coefficient ($1/\kappa = 0.007$) closely tracks our theoretical expectations (see \Cref{fig:sensitivity}). 

\begin{table}[ht!]
    \centering
    \caption{$\check{\theta}_m$ estimates under artificially increased instrument strength: The parameter $b$ represents the artificial boost applied to the interaction between treatment and task interest within the mediator equation, $F_{\text{interact}}$ denotes the $F$-statistic for all interaction terms, and $1/\kappa$ is the robustness parameter..}
    \label{tab:lambda}
    \begin{tabular}{lcccc}
    \toprule
    $b$ & $F_{\text{interact}}$ &
    $\hat{\theta}_m\ [95\,\%\ \text{CI}]$ & $\hat{\kappa}$ &
    $1/\hat{\kappa}$ \\
    \midrule
    $-0.2$ & $11.3$ & $-0.03$ $[-0.38,\,0.31]$ & $52.9$ & $0.019$ \\
    $-0.4$ & $21.0$ & $-0.02$ $[-0.28,\,0.23]$ & $34.7$ & $0.029$ \\
    $-0.6$ & $34.2$ & $-0.02$ $[-0.22,\,0.18]$ & $28.2$ & $0.035$ \\
    \bottomrule
    \end{tabular}
\end{table}

To assess how results might change if the requirements for instrument strength were fully met, we artificially increased the influence of the interaction between treatment and task interest by adding the term $b \cdot (\text{treat} \times \text{task interest})$ to the original mediator. As the magnitude of this interaction increased (\Cref{tab:lambda}), the confidence intervals for RAPSEM shrank substantially.\footnote{While the absolute effect size also increases in this scenario, this is a direct consequence of modifying the mediator's distribution rather than a lack of estimator stability.}, from a width of $0.69$ at $F_{\text{interact}} = 11.3$ to $0.40$ at $F_{\text{interact}} = 34.2$, while the point estimates remained close to zero. While $1/\kappa$ increases alongside instrument strength, it remains numerically small. Even with large interactions, it does not approach the theoretical upper bound of 1.

This exercise confirms that the wide intervals observed in our empirical example are not a fundamental flaw of the RAPSEM algorithm itself, but rather a direct consequence of the low-power regime inherent in the original data. When supported by a sufficiently strong instrument, RAPSEM approaches the precision of standard SEM while maintaining its unique advantage of robustness against omitted variable bias.

Taken together, these results empirically demonstrate that RAPSEM requires larger samples and stronger interactions to achieve high precision, but offers estimates that are robust to the omission of observed confounders. In educational research settings where variables such as cognitive activation or personality traits frequently go unmeasured, this robustness represents a meaningful safeguard against bias that standard SEM does not provide.

\section{Discussion}

This article introduces a latent variable model for mediation analysis that is robust to unobserved confounding. The proposed approach builds on the $G$-estimation framework within a rank-preserving model of \textcite{Tenhave2007} and extends the general formulation of \textcite{Zheng2015a} by incorporating a two-stage method of moments for polynomial structural equation models \parencite{Wall2000}. This integration enables the identification of mediation effects of latent variables under the No  Unobserved Effect Modification assumption, rather than the stronger Sequential Ignorability assumption.

By relying on structural assumptions to identify the causal effects, the $G$-estimation approach is semi-parametric. While no specific distributional form is assumed for the error terms, the framework requires linear models for the outcome and mediator, subject to Assumption~\ref{assump:treat_cov_inter}. Consequently, it does not provide the general, nonparametric identification established by \textcite{Pearl2009, Pearl2012} via the Mediation Formula or by \textcite{Imai2010general}, which can be seen as a trade-off to avoid Sequential Ignorability. Indeed, if Sequential Ignorability is actually satisfied, adopting our framework is suboptimal, as the additional constraints result in a loss of statistical power and less efficient estimates compared to methods that fully exploit those stronger ignorability conditions.

We establish the consistency and asymptotic normality of the resulting estimator and evaluate its performance in simulation studies. Our results show that while standard methods exhibit substantial bias under unobserved confounding, RAPSEM remains unbiased. It achieves acceptable power for medium-to-large effects at sample sizes of $N > 500$, provided there are significant interaction effects between covariates and treatment on the mediator. These results align with previous findings \parencite[e.g.,][]{Zheng2015b, Brandt2020}. Here, we found that an increased reliability of the indicator variables will also improve power.

Given these results, our findings suggest that RAPSEM offers a promising framework for robustly identifying mediation effects in the presence of unobserved confounding, particularly in large-scale intervention studies. The need for relatively large sample sizes to achieve well-powered estimation is a limitation, but it also reflects the broader challenges of causal identification. Without strong assumptions such as Sequential Ignorability, mediation effects are difficult to identify, particularly in small-sample settings or when indicator variables have low reliability. In this sense, the reliance on larger datasets highlights an important practical consideration: findings from small studies may provide only limited evidence about mediation effects.

Beyond sample size, a critical point for the use of RAPSEM is the requirement for effect modification of the mediator to serve as a structurally emerging instrument, a condition that can be challenging to meet. However, this requirement should be seen in the light of the trade-off it provides. Sequential Ignorability requires the exhaustive identification and precise measurement of all confounders to avoid bias. In contrast, our approach only requires finding at least one significant mediator-covariate interaction. Importantly, the interaction effect does not necessarily need to be causal itself, and the instrument strength can be accumulated from different covariates.

Furthermore, a significant advantage of this framework is that the strength and relevance of these variables can be empirically tested using standard diagnostic statistics for instruments. This provides a more transparent and falsifiable alternative to the inherently untestable assumption of sequential ignorability, which can lead to significant hidden bias if violated. While an interaction was present in our empirical example, the effect was small, resulting in a weak instrument. This is unsurprising given that the study was not originally designed with this requirement in mind. Future studies can directly target this by proactively identifying and collecting data on variables hypothesized to interact with the treatment.

On the methodological side, the RAPSEM framework already provides the theoretical basis for implementations involving multiple treatments, treatment–mediator interactions, and non-linear measurement models, e.g., capturing dichotomous mediators or covariates. In contrast, adapting the framework for dichotomous outcomes requires more than a simple change to the measurement model, as the causal estimands and the identification strategy must be specifically adapted for non-linear transformations. Such a generalization \parencite[e.g., along the lines of][]{Dukes2018} represents a promising direction for future research. Further expansions may include higher-order polynomials and interaction terms among latent variables; for instance, introducing a nonlinear specification for baseline effects can increase power when linearity assumptions are violated \parencite{Brandt2020, Faleh2026}. These enhancements would broaden the method’s applicability and strengthen its ability to capture complex causal structures. 

Ultimately, as actual randomized trials in this context remain scarce, extending this approach to address confounding within observational data is a promising direction. In a similar vein, the extension of the RPM formulation to latent multilevel or longitudinal variable models can open new opportunities to broaden the flexibility of the approach. Such extensions, however, will require a very thorough evaluation and adaptation of the underlying (causal) assumptions.

\section*{Acknowledgement}

 \textbf{Financial Support.} This research was funded by the German Research Foundation (DFG) under grant BR 5175/2-1. The authors also acknowledge support from the state of Baden-Württemberg through bwHPC and from the DFG through grant INST 35/1597-1 FUGG.

  \textbf{Acknowledgement.} We thank Samuel Merk for his insights on our method and for connecting us with Heike Russ and Andreas Lachner. We extend our gratitude to Heike Russ and Andreas Lachner for sharing their high-quality open-access data and reviewing the manuscript. We also thank Stijn Vansteelandt and Yves Rosseel for their valuable feedback and suggestions.
  
  \textbf{AI Usage} Adhering to the ethical guidelines for generative AI in academic research by \textcite{PorsdamMann2024}, the authors acknowledge the use of Anthropic 4, Gemini 3, and Opus 5 in this work. These tools were employed to enhance the readability of the manuscript and to support code validation and testing, as well as plot creation. Every author has contributed substantially to the research, has verified the accuracy of the content, and accepts full responsibility for the integrity of the final work.

  \textbf{Data Availability Statement.} Data and code is available at \url{https://github.com/PsychometricsMZ/rapsem}.  

  \textbf{Competing Interests.} The authors declare no competing interests.

  \paragraph{Author Contributions}
Conceptualization: S.M., H.B. Methodology: S.M., R.F. Software, data curation, formal analysis, investigation, resources, visualization, writing original draft, project administration: S.M. Editing, validation: H.B., R.F. Funding acquisition, supervision: H.B. All authors approved the final submitted draft.

\printbibliography[title={References}]
\newpage
\appendix

\newpage
\renewcommand{\thesubsection}{\Alph{subsection}}
\section{Notation}
\label{sec:notation}

\begingroup
\renewcommand{\arraystretch}{1.15}
\begin{longtable}{@{}l p{0.60\textwidth} l@{}}
\caption{Notation used throughout the manuscript.}\label{tab:notation}\\
\toprule
Symbol & Meaning & Defined in \\
\midrule
\endfirsthead
\toprule
Symbol & Meaning & Defined in \\
\midrule
\endhead
\midrule \multicolumn{3}{r}{\emph{continued on next page}} \\
\endfoot
\bottomrule
\endlastfoot
 
\multicolumn{3}{@{}l}{\textbf{Observed data}}\\
$N$ & Sample size (indexed by $i$)& --- \\
$R_i$ & Randomized binary treatment & --- \\
$\mathbf{m}_i,\mathbf{x}_i,\mathbf{y}_i$ & Indicators of mediator, covariates, outcome & \eqref{eq:observed_vars_xi} \\
$\mathbf{z}_i$ & Vector of all indicators, $(\mathbf{m}_i,\mathbf{x}_i,\mathbf{y}_i)^{\top}$ & \eqref{eq:observed_vars_xi} \\
$P$ & Number of observed indicators, $\dim(\mathbf{z}_i)$ & \eqref{eq:observed_vars_xi} \\
$U_i$ & Unmeasured mediator-outcome confounder & --- \\
\addlinespace
 
\multicolumn{3}{@{}l}{\textbf{Latent variables}}\\
$\eta_{M_i},\eta_{Y_i}$, $\boldsymbol{\eta}_{\mathbf{X}_i}$ & Latent mediator, outcome and covariates & \eqref{eq:latent_vars_eta} \\
$\boldsymbol{\eta}_i$ & Vector of stacked latent variables$(\eta_{M_i}\ \ \boldsymbol{\eta}_{\mathbf{X}_i}\ \ \eta_{Y_i})^{\top}$ & \eqref{eq:latent_vars_eta} \\
$K$ & Number of latent variables, $\dim(\boldsymbol{\eta}_i)$ & \eqref{eq:latent_vars_eta} \\
$\boldsymbol{\eta}_{\mathrm{pred},i}$ & Latent outcome predictors, i.e.\ $(\eta_{M_i}\ \boldsymbol{\eta}_{\mathbf{X}_i})^{\top}$ & \eqref{eq:oberserved_moments} \\
$J$ & Highest order in which a component of $\boldsymbol{\eta}_{\mathrm{pred},i}$ enters $\boldsymbol{\Xi}_y$ & \eqref{eq:moment_construction} \\
$\tilde{\boldsymbol{\eta}}_i$ & Factor score at the \emph{true} measurement parameters & \eqref{eq:factor_score_estimator} \\
$\hat{\boldsymbol{\eta}}_i$ & Factor score at the \emph{estimated} measurement parameters & \eqref{eq:factor_score_estimator} \\
$\mathbf{H}$ & Weight matrix in the factor-score estimator & --- \\
$\mathbf{e}_i$ & Estimation error, $\tilde{\boldsymbol{\eta}}_i = \boldsymbol{\eta}_i + \mathbf{e}_i$ & \eqref{eq:fact_score_proxy} \\
$\boldsymbol{\Sigma}_{ee}$ & Variance of the estimation error with elements $\Sigma_{ee,\cdot \cdot}$ & \eqref{eq:sigma_ee} \\

\addlinespace

\multicolumn{3}{@{}l}{\textbf{Measurement model}}\\
$\boldsymbol{\epsilon}_i$ & Measurement error with covariance matrix $\operatorname{Var}(\boldsymbol{\epsilon}_i)$ & \eqref{eq:measurement_model} \\
$\boldsymbol{\tau},\boldsymbol{\tau}_{\mathrm{free}}$ & Indicator intercepts; free subvector, $(P-K)\times 1$ & \eqref{eq:cfa_parameters} \\
$\boldsymbol{\Lambda},\boldsymbol{\Lambda}_{\mathrm{free}}$ & Loading matrix; free submatrix, $(P-K)\times k$ & \eqref{eq:cfa_parameters} \\
$\boldsymbol{\Upsilon}_1$ & Free CFA parameters & \eqref{eq:upsilon1} \\
$\boldsymbol{\Upsilon}_2$ & Moments of $\mathbf{e}_i$ required by the 2SMM correction & --- \\
$\boldsymbol{\Upsilon}$ & $(\boldsymbol{\Upsilon}_1\ \ \boldsymbol{\Upsilon}_2)^{\top}$ & Thm.~\ref{thm:upsilon} \\
$\boldsymbol{\Delta}_i$ & Influence function of $\hat{\boldsymbol{\Upsilon}}$ & Thm.~\ref{thm:upsilon} \\
$\psi_{\boldsymbol{\Upsilon}_1}(\mathbf{z}_i)$ & Z-estimating function for $\boldsymbol{\Upsilon}$  & --- \\
$\mathbf{J}$ & Jacobian $\operatorname{E}[\partial\psi_{\boldsymbol{\Upsilon}}(\mathbf{z}_i)/\partial\boldsymbol{\Upsilon}^{\top}]$ & Ass.~\ref{assump:regularity_measurement_model} \\
$\mathbf{B}_i,\mathbf{q}_i,\mathbf{p}_i$ & Linearisation of $\hat{\boldsymbol{\eta}}_i-\tilde{\boldsymbol{\eta}}_i$ in the estimation error & Thm.~\ref{thm:eta} \\
\addlinespace

\multicolumn{3}{@{}l}{\textbf{Structural model}}\\
$r$ & Fixed level of the treatment & \eqref{eq:potential_outcome} \\
$\eta_m$ & Fixed level of the latent mediator & \eqref{eq:potential_outcome} \\
$\eta_{Y_i}^{r,\eta_m}$ & Potential outcome under $R_i=r$ and $\eta_{M_i}$ set to $\eta_m$ & \eqref{eq:potential_outcome} \\
$\theta_s,\ h_s(\cdot)$ & Structural parameters and basis functions & \eqref{eq:potential_outcome} \\
$S$ & Number of structural terms $h_s$ & \eqref{eq:potential_outcome} \\
$\theta_r$ & Controlled direct effect of treatment & \eqref{eq:cde} \\
$\theta_m$ & Controlled mediation effect (per unit of $\eta_m$) & \eqref{eq:cme} \\
$\boldsymbol{\beta}_x$ & Covariate coefficients in the outcome equation & \eqref{eq:observed_outcome} \\
$q_{yx},q_{mx},q_{mrx}$ & Admissible polynomial transformations of $\boldsymbol{\eta}_{\mathbf{X}_i}$ & \eqref{eq:concrete_model} \\
$g(\boldsymbol{\eta}_{\mathbf{X}_i})$ & Nuisance term modeled as $q_{yx}(\boldsymbol{\eta}_{\mathbf{X}_i})\,\boldsymbol{\beta}_x$ & \eqref{eq:potential_outcome} \\
$\boldsymbol{\xi}_{Y_i}$ & Outcome regressors $(R_i\ \ \eta_{M_i}\ \ q_{yx}(\boldsymbol{\eta}_{\mathbf{X}_i}))$ & \eqref{eq:observed_outcome} \\
$\boldsymbol{\Xi}_y$ & $N\times\dim(\boldsymbol{\alpha})$ design matrices with rows $\boldsymbol{\xi}_{Y_i}$ & \eqref{eq:observed_outcome} \\
$\boldsymbol{\alpha}$ & outcome structural parameters $(\theta_r\ \ \theta_m\ \ \boldsymbol{\beta}_x)^{\top}$ & \eqref{eq:observed_outcome} \\
$\zeta_{Y_i}$ & Structural errors of outcome equation & \eqref{eq:observed_outcome}\\
$\boldsymbol{\xi}_{M_i}$ & Mediator regressors  $(R_i\ \ q_{mx}(\boldsymbol{\eta}_{\mathbf{X}_i})\ \ R_i\,q_{mrx}(\boldsymbol{\eta}_{\mathbf{X}_i}))$ & \eqref{eq:mediator_model} \\
$\boldsymbol{\Xi}_m$ & $N\times\dim(\boldsymbol{\gamma})$ design matrices with rows $\boldsymbol{\xi}_{M_i}$ & \eqref{eq:mediator_model} \\
$\boldsymbol{\gamma}$ & $(\gamma_r\ \ \boldsymbol{\gamma}_x\ \ \boldsymbol{\gamma}_{xr})^{\top}$, mediator-equation parameters & \eqref{eq:mediator_model} \\
$\zeta_{M_i}$ & Structural errors of mediator equation & \eqref{eq:mediator_model} \\

\addlinespace

\multicolumn{3}{@{}l}{\textbf{$G$-estimation weights}}\\
$\mathbf{A}(R_i,\boldsymbol{\eta}_{\mathbf{X}_i})$ & Centred weight (instrument) matrix, $\operatorname{E}[a_j\mid\boldsymbol{\eta}_{\mathbf{X}_i}]=0$ & \eqref{eq:moment_condition} \\
$a_{r,i}$ & Treatment weight & \eqref{eq:med_weight_general} \\
$a_{m,i}$ & Mediator weight & \eqref{eq:med_weight_general} \\
$\mathbf{W}$ & Full weight matrix with rows $(a_{r,i}\ \ a_{m,i}\ \ q_{yx}(\boldsymbol{\eta}_{\mathbf{X}_i}))$ & \eqref{eq:g_estimation} \\
$\boldsymbol{\Omega}(\boldsymbol{\eta}_{\mathbf{X}})$ & Conditional error term covariance $\operatorname{Var}(\zeta^{r,\eta_m}_{Y_i}\mid\boldsymbol{\eta}_{\mathbf{X}_i})$ & \eqref{eq:optimal_weights} \\
$\bar{\boldsymbol{\alpha}}$ & Naive IV/$G$-estimator without factor score correction & \eqref{eq:observed_gparm} \\
\addlinespace

\multicolumn{3}{@{}l}{\textbf{Moment functions and estimators}}\\
$D_j(\cdot)$ & $j$th 2SMM correction term & \eqref{eq:correction_term_construction} \\
$\mathbf{M}(\cdot),\mathbf{m}(\cdot)$ & Per-unit corrected moment functions & \eqref{eq:m_conditional_expectation} \\
$\hat{\mathbf{M}},\hat{\mathbf{m}}$ & Sample averages of $\mathbf{M},\mathbf{m}$ at $(\hat{\boldsymbol{\eta}}_i,\hat{\boldsymbol{\Upsilon}}_2)$ & \eqref{eq:oberserved_moments} \\
$\hat{\mathbf{M}}_1,\hat{\mathbf{m}}_1$ & Uncorrected ($D_0$) part & mod.\ 2SMM \\
$\hat{\mathbf{M}}_2,\hat{\mathbf{m}}_2$ & Correction part, $\sum_{j\ge1}(-1)^jD_j$ & mod.\ 2SMM \\
$\mathbf{R}_1,\mathbf{R}_2$ & Blocks of the Fuller-type positive-definiteness modification & mod.\ 2SMM \\
$\hat{\lambda}$ & Largest eigenvalue of $\mathbf{R}_1^{-1}\mathbf{R}_2$ & mod.\ 2SMM \\
$\hat{\boldsymbol{\alpha}}$ & 2SMM estimator $\hat{\mathbf{M}}^{-1}\hat{\mathbf{m}}$ & \eqref{eq:structural_para_estimator} \\
$c$ & Fuller tuning constant $\in[0,\dim(\boldsymbol{\alpha})+5]$ & mod.\ 2SMM \\
$v$ & Ridge regularisation constant & \eqref{eq:final_param_est_theta} \\
$\check{\boldsymbol{\alpha}}$ & Final estimator $(\check{\mathbf{M}}+v\mathbf{I})^{-1}\check{\mathbf{m}}$ & \eqref{eq:final_param_est_theta} \\
$\mathbf{G},\mathbf{h}$ & $\operatorname{E}[\mathbf{w}_i\boldsymbol{\xi}_{y,i}^{\top}]$, $\operatorname{E}[\mathbf{w}_i\eta_{y_i}]$ & Thm.~\ref{thm:theta} \\
$\mathbf{l}(\cdot)$ & $\mathbf{m}(\cdot)-\mathbf{M}(\cdot)\boldsymbol{\alpha}$, the moment residual & Thm.~\ref{thm:theta} \\
$\overline{\mathbf{C}}$ & Expected derivative matrix linking $\mathbf{l}$ to $\hat{\boldsymbol{\Upsilon}}-\boldsymbol{\Upsilon}$ & Thm.~\ref{thm:theta} \\
$\mathbf{d}_i,\mathbf{S}$ & $\mathbf{l}(\tilde{\boldsymbol{\eta}}_i,\boldsymbol{\Upsilon}_2,\boldsymbol{\alpha})+\overline{\mathbf{C}}\boldsymbol{\Delta}_i$ and $\operatorname{Var}(\mathbf{d}_i)$ & Thm.~\ref{thm:theta} \\
\end{longtable}
\endgroup

\newpage
\section{Additional Assumptions}
\label{sec:appendix_assumption}

\begin{causalassumption}[Consistency] \label{assump:consistency}
    The realized mediator and outcome corresponding to a given treatment and mediator assignment equals the potential outcome under those values
    \begin{align*}
        \eta_{M_i} &= \eta_{M_i}^{r}  &&\text{whenever } R_i = r, \\
        \eta_{Y_i} &= \eta_{Y_i}^{r, \eta_m} &&\text{whenever } R_i = r \text{ and } \eta_{M_i} = \eta_m .
    \end{align*}
    This assumption ensures a unique mapping from treatment and mediator values to potential outcomes. In practice, consistency is typically justified under the Stable Unit Treatment Value Assumption (SUTVA), which comprises two conditions: first, that each treatment corresponds to a single well-defined version; and second, that there is no interference between units, meaning that one unit's treatment does not affect another unit's outcome.
\end{causalassumption}

\begin{causalassumption}[Positivity] \label{assump:positivity}
    For all $\boldsymbol{\eta}_x$ in the support of $\boldsymbol{\eta}_{\mathbf{X}}$, the probability of receiving each treatment level is strictly between 0 and 1:
    \[
        0 < P(R = r \mid \boldsymbol{\eta}_{\mathbf{X}} = \boldsymbol{\eta}_x) < 1
        \quad \forall r .
    \]
    In an RCT, this holds by design, since
    treatment assignment is independent of the covariates. For the mediator, positivity is required at the levels contrasted by the
    estimands (e.g.\ $\eta_m$ and $\eta_{m'}$ for the CME in \Cref{eq:cme}):
    \begin{itemize}
        \item If the mediator is discrete:
        \[
            P(\eta_M = m \mid R = r, \boldsymbol{\eta}_{\mathbf{X}} = \boldsymbol{\eta}_x) > 0
            \quad \forall r, \eta_m.
        \]
        \item If the mediator is continuous:
        \[
            f_{\eta_M \mid R, \boldsymbol{\eta}_{\mathbf{X}}}(m \mid r, \boldsymbol{\eta}_x) > 0
            \quad \forall r, \eta_m ,
        \]
        where $f_{\eta_M \mid R, \boldsymbol{\eta}_{\mathbf{X}}}$ denotes the conditional
        density of the mediator.
    \end{itemize}
\end{causalassumption}

\begin{causalassumption}[Causal Efficacy] \label{assump:structural_measurement}
    We assume that the latent factor model admits a
    structural interpretation in the sense of \textcite{VanderWeeleVansteelandt2022}.
    That is, the latent variables rather than their indicators are causally efficacious.
    Specifically, for each latent variable $\eta$
    with indicator set $\mathbf{z}_\eta$,
    (i) the indicators $\mathbf{z}_\eta$ have no causal effect on any subsequent variable, and
    (ii) they are affected by any other variable only
    through $\eta$.

    Property (ii) implies that the measurement parameters $(\boldsymbol{\tau}, \boldsymbol{\Lambda}, \boldsymbol{\Psi})$ are invariant across treatment arms.
    Since the factor score estimation error in \Cref{eq:fact_score_proxy} is a fixed linear
    transformation of $\boldsymbol{\epsilon}$, it inherits this independence, so that the
    instruments remain valid once the 2SMM correction accounts for the measurement error.
\end{causalassumption}

\begin{statassumption}[IID Sampling of Observed Data] \label{assump:iid_sampling}
    The observations $\{(r_i,  \mathbf{m}_i,  \mathbf{x}_i, \mathbf{y}_i)\}_{i=1}^n$ are independent and identically distributed (iid) draws from the joint distribution $P$.
    Note that this iid assumption pertains only to independence across units and places no restrictions on within-unit dependence. In particular, latent variables such as $\eta_{m_i}$ and $\eta_{y_i}$ may be arbitrarily dependent, allowing for unobserved confounding between the mediator and the outcome.
\end{statassumption}

\begin{statassumption}[Correct Model Specification] \label{assump:model_spec}
    The measurement model in \Cref{eq:measuremenq_model_general} and the structural dependence among latent variables in \Cref{eq:potential_outcome} are correctly specified as polynomial models that are linear in their parameters. 

    Correct specification of the function $g(\boldsymbol{\eta}_{\mathbf{X}_i})$ and of the mediator model in \Cref{eq:mediator_model} is not required for identification of the controlled effects. Misspecifications may affect the efficiency of estimation, but do not invalidate \Cref{eq:moment_condition}.
\end{statassumption}.

\begin{statassumption}[Measurement Model Regularity] \label{assump:regularity_measurement_model}    
    The observed indicators $\mathbf{z}$ satisfy
    \[
        \operatorname{E}\!\big[| \mathbf{z}_i |^2\big] < \infty.
    \]
    The true parameter $\boldsymbol{\Upsilon}$ lies in the interior of 
    a compact set $\mathcal{U}$, the estimating function $\boldsymbol{\Upsilon} \mapsto \psi_{\boldsymbol{\Upsilon}}(\mathbf{z}_i)$ is continuous in $\boldsymbol{\Upsilon}$ almost surely, and the Jacobian
    \[
        \mathbf{J} := \operatorname{E}\!\Big[ 
        \partial \psi_{\boldsymbol{\Upsilon}}(\mathbf{z}_i) / 
        \partial \boldsymbol{\Upsilon}^\top \Big],
    \]
    is nonsingular at $\boldsymbol{\Upsilon}$.
\end{statassumption}

\begin{statassumption}[Measurement Error] \label{assump:measurement_error}
    The measurement errors $\boldsymbol{\epsilon}_i$ are iid, independent of the latent variables $\boldsymbol{\eta}_i$, have zero mean, and satisfy
    \begin{equation*}
        \operatorname{E}[|\boldsymbol{\epsilon}_i|^{4J}] < \infty \text{.}\footnote{In our implementation, we additionally assume $\boldsymbol{\epsilon}_i$ to be normally distributed, allowing the 2SMM correction terms to be derived directly from $\boldsymbol{\Sigma}_{ee}$. An alternative approach, based on OLS-estimated higher-order error moments and not requiring distributional assumptions on $\boldsymbol{\epsilon}_i$, but yielding identical asymptotic properties, is described in \textcite{Wall2000}.}
    \end{equation*}
\end{statassumption}

\begin{statassumption}[Factor Score Finite Moments] \label{assump:finite_moments}
    The factor scores $\boldsymbol{\eta}_{\text{pred},i}$  satisfy finite moment conditions of sufficiently high order:
    \[
        \operatorname{E}[|\boldsymbol{\eta}_{\text{pred},i}|^{4J-2}] < \infty
    \]
\end{statassumption}

\begin{statassumption}[Structural Equation Error] \label{assump:structural_error}
    The structural equation errors $\boldsymbol{\zeta}_{i}$ comprising $\boldsymbol{\zeta}_{Y_i}$ and $\boldsymbol{\zeta}_{M_i}$ are iid, independent of $\boldsymbol{\epsilon}_i$, hence of $\mathbf{e}_i$, have zero mean, and finite variance:
    \[
        \operatorname{Var}[\boldsymbol{\zeta}_{i}] < \infty.
    \]
\end{statassumption}

\section{More details on \ref{assump:nem}}
\label{sec:appendix_rp_nem}

\ref{assump:nem} is implied by what econometricians call No Essential Heterogeneity \parencite{Heckman2006}. Extending the original single-treatment notion to a joint treatment-mediator setting, it states: individuals may have idiosyncratic responses depending on the treatment-mediator condition, but their selection into that condition cannot be mean-informative about these gains conditional on covariates.

A sufficient condition for both is full homogeneity of the treatment and mediator effects across both observed covariates $\boldsymbol{\eta}_{\mathbf{X}}$ and unmeasured confounders $U$:
\[
    \theta_r(\boldsymbol{\eta}_{\mathbf{X}_i}, U_i) = \theta_r, \quad
    \theta_m(\boldsymbol{\eta}_{\mathbf{X}_i}, U_i) = \theta_m.
\]
This homogeneity is, in turn, implied by the additive Rank Preserving assumption, the original approach of \cite{Tenhave2007} is built on. It links potential outcomes through a common shift:
\[
    \eta_{Y_i}^{r',\eta_{m'}} = \eta_{Y_i}^{r,\eta_m} + \theta_r\,(r' - r) + \theta_m\,(\eta_{m'} - \eta_m),
\]
such that if individual $i$ has a better outcome than $j$ under one treatment-mediator condition, they will also have a better outcome under any other condition:\footnote{The additive form is a special case of general rank preservation, $\eta_{Y_i}^{r',\eta_{m'}} = \varphi(\eta_{Y_i}^{r,\eta_m})$ for a strictly increasing function $\varphi$ with $\varphi' > 0$. The general form also preserves ranks but does not imply homogeneity, since the individual effect $\varphi(y) - y$ may vary with the outcome level and hence with $U_i$.}
\[
    \eta_{Y_i}^{r,\eta_m} > \eta_{Y_j}^{r,\eta_m} \iff \eta_{Y_i}^{r',\eta_{m'}} > \eta_{Y_j}^{r',\eta_{m'}}
    \quad \forall i, j \forall (r,\eta_m), (r',\eta_{m'}).
\]

The formulation \ref{assump:nem} is weakest, permitting heterogeneous effects as long as they are not systematically related to treatment-mediator assignment.

\section{Implementation}
\label{sec:appendix_implementation}

The R package \texttt{rapsem} allows estimation of structural models as specified in \Cref{eq:observed_outcome} and \Cref{eq:mediator_model}. While directly observed variables may be transformed arbitrarily, latent variables can only enter as polynomial terms. The current version supports latent variable terms up to dimension $J=1$ (i.e., computation of $D_0$ and $D_1$, but the framework can be extended to higher orders by incorporating additional correction terms $D_{j>1}$. Furthermore, the current implementation is based on the measurement model in \Cref{eq:measurement_model}, treating it as a confirmatory factor analysis (CFA) subject to the constraints in \Cref{eq:cfa_parameters}. However, it can be extended to accommodate alternative factor specifications and modeling techniques, such as Item Response Theory (IRT) for discrete indicators.

Estimation is carried out via the function \texttt{med\_est}, which takes as input the observed data and a \texttt{lavaan} model specifying the structural equation model, and returns results from both the standard regression approach and the $G$-estimation approach, each using factor score corrections. Factor intercepts, loadings, and residual variances are estimated by maximum likelihood with \texttt{lavaan}. By default, the measurement model of each latent variable is fitted separately (which requires at least three indicators per latent variable), since the factor scores do not depend on the factor covariances and misspecification in one measurement block cannot affect the others. Subsequently, the factor scores are computed according to \Cref{eq:factor_score_estimator}. Structural parameters are then estimated using the modified 2SMM regularized $G$-estimator in \Cref{eq:final_param_est_theta}, with default settings $c=5$ and $v=10^{-4}$ (sufficiently small for the effect size estimates to remain essentially unaffected). 

Since the analytic asymptotic variance in \Cref{thm:theta} does not account for the finite-sample modification and the ridge penalty, uncertainty is quantified by the nonparametric bootstrap (default: $100$ bootstrap samples): point estimates, standard errors, and confidence intervals are the mean, standard deviation, and percentile intervals of the bootstrap replicates.

\section{Factor Score Corrections for Simulation Study Model}
\label{sec:appendix_fact_corr}

In the concrete model defined in \Cref{eq:concrete_model}, the polynomial order is $J = 1$, so the moment expansion consists of $D_0$ and $D_1$. $D_0$ provides the uncorrected terms

\[
\begin{aligned}
   D_0(\hat{\boldsymbol{\eta}}_i, \hat{\boldsymbol{\Upsilon}}_2) &=
   \begin{pmatrix}
        a_{r,i} & \hat{a}_{m,i} & \hat{\eta}_{X_{1i}} & \hat{\eta}_{X_{2i}}
    \end{pmatrix}^\top
    \begin{pmatrix}
        R_i & \hat{\eta}_{M_i} & \hat{\eta}_{X_{1i}} & \hat{\eta}_{X_{2i}} & \hat{\eta}_{Y_i}
    \end{pmatrix} \\
    &=
    \begin{pmatrix}
        a_{r,i} R_i &
        a_{r,i} \hat{\eta}_{M_i} &
        a_{r,i} \hat{\eta}_{X_{1i}} &
        a_{r,i} \hat{\eta}_{X_{2i}} &
        a_{r,i} \hat{\eta}_{Y_i} \\

        \hat{a}_{m,i} R_i &
        \hat{a}_{m,i} \hat{\eta}_{M_i} &
        \hat{a}_{m,i} \hat{\eta}_{X_{1i}} &
        \hat{a}_{m,i} \hat{\eta}_{X_{2i}} &
        \hat{a}_{m,i} \hat{\eta}_{Y_i} \\

        \hat{\eta}_{X_{1i}} R_i &
        \hat{\eta}_{X_{1i}} \hat{\eta}_{M_i} &
        \hat{\eta}_{X_{1i}}^2 &
        \hat{\eta}_{X_{1i}} \hat{\eta}_{X_{2i}} &
        \hat{\eta}_{X_{1i}} \hat{\eta}_{Y_i} \\

        \hat{\eta}_{X_{2i}} R_i &
        \hat{\eta}_{X_{2i}} \hat{\eta}_{M_i} &
        \hat{\eta}_{X_{2i}} \hat{\eta}_{X_{1i}} &
        \hat{\eta}_{X_{2i}}^2 &
        \hat{\eta}_{X_{2i}} \hat{\eta}_{Y_i}
    \end{pmatrix}.
\end{aligned}
\]

The correction term $D_1(\hat{\boldsymbol{\eta}}_i, \hat{\boldsymbol{\Upsilon}}_2)$ involves expectations over the measurement error variance and covariances. Under the latent factor ordering assumed in \Cref{eq:latent_vars_eta}, in which the indices $1,\dots,4$ of $\hat{\boldsymbol{\Sigma}}_{ee}$ refer to $\eta_M$, $\eta_{X_1}$, $\eta_{X_2}$, and $\eta_Y$, and using the form of $\hat{a}_{m,i}$ from \Crefrange{eq:med_weight_general}{eq:med_weight_explicit}, it takes the form

\[
    D_1(\hat{\boldsymbol{\eta}}_i, \hat{\boldsymbol{\Upsilon}}_2)
    = \begin{pmatrix}
        0 & 0 & 0 & 0 & 0 \\
        0 &
        \begin{aligned}
            (\hat{\gamma}_{x1r}\hat{\boldsymbol{\Sigma}}_{ee,21} + \\[-1em]
            \hat{\gamma}_{x2r} \hat{\boldsymbol{\Sigma}}_{ee,31})a_{r,i}
        \end{aligned} &
        \begin{aligned}
            (\hat{\gamma}_{x1r}\hat{\boldsymbol{\Sigma}}_{ee,22} + \\[-1em]
            \hat{\gamma}_{x2r} \hat{\boldsymbol{\Sigma}}_{ee,32})a_{r,i}
        \end{aligned} &
        \begin{aligned}
            (\hat{\gamma}_{x1r}\hat{\boldsymbol{\Sigma}}_{ee,23} + \\[-1em]
            \hat{\gamma}_{x2r} \hat{\boldsymbol{\Sigma}}_{ee,33})a_{r,i}
        \end{aligned} &
        \begin{aligned}
            (\hat{\gamma}_{x1r}\hat{\boldsymbol{\Sigma}}_{ee,24} + \\[-1em]
        \hat{\gamma}_{x2r} \hat{\boldsymbol{\Sigma}}_{ee,34})a_{r,i}
        \end{aligned} \\
        0 & \hat{\boldsymbol{\Sigma}}_{ee,21} & \hat{\boldsymbol{\Sigma}}_{ee,22} & \hat{\boldsymbol{\Sigma}}_{ee,23} & \hat{\boldsymbol{\Sigma}}_{ee,24} \\
        0 & \hat{\boldsymbol{\Sigma}}_{ee,31} & \hat{\boldsymbol{\Sigma}}_{ee,32} & \hat{\boldsymbol{\Sigma}}_{ee,33} & \hat{\boldsymbol{\Sigma}}_{ee,34} \\
    \end{pmatrix}.
\]

This correction ensures that second-order bias due to measurement error is appropriately removed from the moment conditions when estimating the structural model parameters.
\section{Proof of Theorem \ref{thm:upsilon}}
\label{sec:append_proofs}

\subsection{Consistency}
\begin{proof}
By Assumptions~\ref{assump:model_spec}, $\boldsymbol{\Upsilon}$ is the unique solution of $\Psi(\boldsymbol{\Upsilon}):= \operatorname{E}[\psi_{\boldsymbol{\Upsilon}} (\mathbf{z}_i)] =\mathbf{0}$. Let $\hat{\boldsymbol{\Upsilon}}$ solve $\Psi_N(\boldsymbol{\Upsilon}) := \frac{1}{N}\sum_{i=1}^N \psi_{\boldsymbol{\Upsilon}}(\mathbf{z}_i) = \mathbf{0}$. 
Since $\psi_{\boldsymbol{\Upsilon}}(\mathbf{z}_i)$ is polynomial in $\mathbf{z}_i$ with coefficients that are continuous in $\boldsymbol{\Upsilon}$ (Assumption~\ref{assump:model_spec}), and $\mathcal{U}$ is compact (Assumption~\ref{assump:regularity_measurement_model}), there exist finite constants $C > 0$ and integer $q \geq 1$ such that
\[
    \sup_{\boldsymbol{\Upsilon} \in \mathcal{U}} 
    \|\psi_{\boldsymbol{\Upsilon}}(\mathbf{z}_i)\| 
    \leq C\bigl(1 + \|\mathbf{z}_i\|^q\bigr)
    \quad \text{a.s.}
\]
Since Assumptions~\ref{assump:regularity_measurement_model} and~\ref{assump:measurement_error} jointly ensure $\operatorname{E}[\|\mathbf{z}_i\|^q] < \infty$ for the relevant $q$ (since $\mathbf{z}_i = \boldsymbol{\tau} + \boldsymbol{\Lambda}\boldsymbol{\eta}_i + \boldsymbol{\epsilon}_i$, moments of $\|\mathbf{z}_i\|^q$ are controlled by those of $\boldsymbol{\eta}_i$ and $\boldsymbol{\epsilon}_i$, and the estimating equations of \textcite{Wall2000} satisfy $q \leq 4J-2$), the envelope condition
\[
    \operatorname{E}\!\Big[\sup_{\boldsymbol{\Upsilon}\in\mathcal{U}} 
    \|\psi_{\boldsymbol{\Upsilon}}(\mathbf{z}_i)\|\Big] < \infty
\]
holds, so $\{\psi_{\boldsymbol{\Upsilon}}(\cdot) : \boldsymbol{\Upsilon} \in 
\mathcal{U}\}$ is Glivenko--Cantelli. The uniform law of large numbers 
(Assumption~\ref{assump:iid_sampling}) thus ensures uniform convergence:
\[
    \sup_{\boldsymbol{\Upsilon}\in\mathcal{U}}
    \|\Psi_N(\boldsymbol{\Upsilon}) - \Psi(\boldsymbol{\Upsilon})\|
    \xrightarrow{p} 0.
\]
Almost sure continuity of $\psi_{\boldsymbol{\Upsilon}}(\mathbf{z}_i)$ in $\boldsymbol{\Upsilon}$ (Assumption~\ref{assump:regularity_measurement_model}), together with the envelope condition established above, allows an application of the dominated convergence theorem, which implies that $\Psi(\boldsymbol{\Upsilon}') = \operatorname{E}[\psi_{\boldsymbol{\Upsilon}'}(\mathbf{z}_i)]$ is continuous in $\boldsymbol{\Upsilon}'$ on $\mathcal{U}$. Since $\boldsymbol{\Upsilon}$ is the unique zero of $\Psi$ (Assumption~\ref{assump:model_spec}) and $\mathcal{U}$ is compact (Assumption~\ref{assump:regularity_measurement_model}), the set $\{\boldsymbol{\Upsilon}' : \|\boldsymbol{\Upsilon}' - \boldsymbol{\Upsilon}\| \geq \varepsilon\}$ is compact and contains no zero of $\Psi$ for any $\varepsilon > 0$. A continuous function on a compact set that is everywhere nonzero attains a strictly positive infimum, so
\[
    \inf_{\|\boldsymbol{\Upsilon}'-\boldsymbol{\Upsilon}\|>\varepsilon}
    \|\Psi(\boldsymbol{\Upsilon}')\| > 0 \quad \text{for every } \varepsilon > 0.
\]
Since $\Psi_N(\hat{\boldsymbol{\Upsilon}}) = \mathbf{0}$ by definition of $\hat{\boldsymbol{\Upsilon}}$, uniform convergence of $\Psi_N$ to $\Psi$ and the separation condition above together satisfy the conditions of \textcite[][Theorem~5.9]{VanderVaart1998}, giving
\[
    \hat{\boldsymbol{\Upsilon}} \xrightarrow{p} \boldsymbol{\Upsilon}.
\]
\end{proof}

\subsection{Asymptotic normality.}

\begin{proof}
Since $\hat{\boldsymbol{\Upsilon}}\xrightarrow{p}\boldsymbol{\Upsilon}$, 
we can linearize the sample estimating equation around $\boldsymbol{\Upsilon}$ using a first-order Taylor expansion:
    \[
        \frac{1}{N} \sum_{i=1}^N \psi_i(\hat{\boldsymbol{\Upsilon}}) = \frac{1}{N} \sum_{i=1}^N \psi_i(\boldsymbol{\Upsilon}) + \mathbf{J} (\hat{\boldsymbol{\Upsilon}} - \boldsymbol{\Upsilon}) + o_p\!\big( \lVert \hat{\boldsymbol{\Upsilon}} - \boldsymbol{\Upsilon} \rVert \big),
    \]
    where $\mathbf{J}$ is nonsingular by assumption. Since the left-hand side equals zero by definition of $\hat{\boldsymbol{\Upsilon}}$, rearranging yields
    \[
        \hat{\boldsymbol{\Upsilon}} - \boldsymbol{\Upsilon} = - \mathbf{J}^{-1} \frac{1}{N} \sum_{i=1}^N \psi_i(\boldsymbol{\Upsilon})  + o_p(N^{-1/2}).
    \]
    Defining
    \[
        \boldsymbol{\Delta}_i := - \mathbf{J}^{-1} \psi_i(\boldsymbol{\Upsilon}),
    \]
    we obtain the desired expansion.  By definition, $\operatorname{E}[\psi(\mathbf{z}_i,\boldsymbol{\Upsilon})]=0$, so $\operatorname{E}[\boldsymbol{\Delta}_i]=0$.  
    Since $\psi(\mathbf{z}_i,\boldsymbol{\Upsilon})$ is polynomial in $\boldsymbol{\eta}_{\text{pred},i}$ and linear in $(\boldsymbol{\epsilon}_i,\boldsymbol{\zeta}_i)$, Assumption~\ref{assump:regularity_measurement_model} ensures $\operatorname{Var}(\boldsymbol{\Delta}_i)<\infty$. Hence,
    \[
        \hat{\boldsymbol{\Upsilon}} - \boldsymbol{\Upsilon} = O_p(N^{-1/2}),
    \]
    establishing $\sqrt{N}$-consistency.
\end{proof}

\section{Proof of Theorem \ref{thm:eta}}
\begin{proof}
    From the definition of the latent factor score estimator in \Cref{eq:factor_score_estimator}, we can express
    \[
        \hat{\boldsymbol{\eta}}_i - \tilde{\boldsymbol{\eta}}_i = \hat{\mathbf{H}} f(\hat{\boldsymbol{\tau}}_{\mathrm{free}}, \hat{\boldsymbol{\Lambda}}_{\mathrm{free}}) - \mathbf{H} f(\boldsymbol{\tau}_{\mathrm{free}}, \boldsymbol{\Lambda}_{\mathrm{free}}),
    \]
    with 
    \[
        f(\boldsymbol{\tau}_{\mathrm{free}}, \boldsymbol{\Lambda}_{\mathrm{free}}) = -\mathbf{z}_i + \boldsymbol{\tau}_{\mathrm{free}} + \boldsymbol{\Lambda}_{\mathrm{free}} \mathbf{z}_i - \boldsymbol{\Lambda}_{\mathrm{free}} \mathbf{0}_{k \times 1}.
    \]
    Expanding this difference and regrouping by parameters yields
    \[
        \hat{\boldsymbol{\eta}}_i - \tilde{\boldsymbol{\eta}}_i = \mathbf{H}(\hat{\boldsymbol{\tau}}_{\mathrm{free}} - \boldsymbol{\tau}_{\mathrm{free}}) + \mathbf{H} (\mathbf{q}_i^\top \otimes \mathbf{I}_{p-k}) \operatorname{vec}(\hat{\boldsymbol{\Lambda}}_{\mathrm{free}} - \boldsymbol{\Lambda}_{\mathrm{free}}) - (\mathbf{p}_i^\top \otimes \mathbf{I}_k) \operatorname{vec}(\hat{\mathbf{H}} - \mathbf{H}),
    \]
    which can be written compactly as
    \[
        \hat{\boldsymbol{\eta}}_i - \tilde{\boldsymbol{\eta}}_i = \mathbf{B}_i \, (\hat{\boldsymbol{\Upsilon}}_1 - \boldsymbol{\Upsilon}_1),
    \]
    with $\mathbf{B}_i$, $\mathbf{q}_i$, and $\mathbf{p}_i$ defined in \Cref{thm:eta}.  By \Cref{thm:upsilon}, $\hat{\boldsymbol{\Upsilon}}_1 - \boldsymbol{\Upsilon}_1 = O_p(N^{-1/2})$, which implies
    \[
        \hat{\boldsymbol{\eta}}_i - \tilde{\boldsymbol{\eta}}_i = O_p(N^{-1/2}),
    \]
    establishing $\sqrt{N}$-consistency.
\end{proof}

\section{Proof of Theorem \ref{thm:theta}}

\subsection{Consistency}
\begin{proof}
    The identification argument provides us with
    \[
        \operatorname{E}[\mathbf{w}_i \zeta_{y_i}]
        = \mathbf{0},
    \]
    where
    \[
        \zeta_{y_i} = \eta_{y_i} - \boldsymbol{\xi}_{y_i}^{\top}\boldsymbol{\alpha}.
    \]
     By inserting this error term into the above equation, under Assumptions~\ref{assump:iid_sampling} and \ref{assump:finite_moments}, we obtain
    \[
        \operatorname{E}[\mathbf{w}_i \eta_{y_i}] = \operatorname{E}[\mathbf{w}_i \boldsymbol{\xi}_{y_i}^\top] \boldsymbol{\alpha}.
    \] 
    We define the population moments
    \[
        \mathbf{G} = \operatorname{E}[\mathbf{w}_i \boldsymbol{\xi}_{y_i}^\top],\quad 
        \mathbf{h} = \operatorname{E}[\mathbf{w}_i \eta_{y_i}].
    \]
    and thus write
    \[
        \mathbf{h} = \mathbf{G} \boldsymbol{\alpha}.
    \]
    By Assumption~\ref{assump:treat_cov_inter}, $\mathbf{G}$ is invertible. Hence, the equation 
    \[ \mathbf h-\mathbf G\boldsymbol\alpha= \mathbf{0}\] 
    has a unique solution, given by 
    \[ \boldsymbol\alpha=\mathbf G^{-1}-\mathbf h. \]
    To show that $\hat{\mathbf{M}} \overset{p}{\longrightarrow} \mathbf{G}$ and $\hat{\mathbf{m}} \overset{p}{\longrightarrow} \mathbf{h}$, we decompose each empirical moment into two terms:,
    \[
        \hat{\mathbf{M}} - \mathbf{G} = 
        \underbrace{\frac{1}{N}\sum_{i=1}^N \left[ M(\hat{\boldsymbol{\eta}}_{\mathrm{pred},i}, \hat{\boldsymbol{\Upsilon}}_2) - M(\tilde{\boldsymbol{\eta}}_{\mathrm{pred},i}, \boldsymbol{\Upsilon}_2) \right]}_{\text{(I)}}
        + \underbrace{\frac{1}{N}\sum_{i=1}^N M(\tilde{\boldsymbol{\eta}}_{\mathrm{pred},i}, \boldsymbol{\Upsilon}_2) - \mathbf{G}}_{\text{(II)}},
    \]
    and analogously for $\hat{\mathbf{m}} - \mathbf{h}$.
    Term~(II) involves i.i.d.\ summands, since $M(\tilde{\boldsymbol{\eta}}_{\mathrm{pred},i}, \boldsymbol{\Upsilon}_2)$ evaluated at the true parameters depends only on $\mathbf{z}_i$, which are i.i.d.\ by Assumption~\ref{assump:iid_sampling}. The standard LLN therefore gives Term~(II) $= o_p(1)$.
    For Term~(I), a first-order Taylor expansion of $M$ around $(\tilde{\boldsymbol{\eta}}_{\mathrm{pred},i}, \boldsymbol{\Upsilon}_2)$ yields
    \[
        M(\hat{\boldsymbol{\eta}}_{\mathrm{pred},i}, \hat{\boldsymbol{\Upsilon}}_2) - M(\tilde{\boldsymbol{\eta}}_{\mathrm{pred},i}, \boldsymbol{\Upsilon}_2) 
        = \frac{\partial M}{\partial \boldsymbol{\Upsilon}^\top}\bigg|_{\tilde{\boldsymbol{\eta}}_i,\, \boldsymbol{\Upsilon}_2} (\hat{\boldsymbol{\Upsilon}} - \boldsymbol{\Upsilon}) + o_p(N^{-1/2}).
    \]
    Averaging over $i$, applying the standard LLN to the i.i.d.\ gradient terms, and using $\hat{\boldsymbol{\Upsilon}} - \boldsymbol{\Upsilon} = o_p(1)$ from Theorem~\ref{thm:upsilon} gives Term~(I) $= o_p(1)$.
    Combining both terms,
    \[
        \begin{pmatrix}
            \hat{\mathbf{M}} \\
            \hat{\mathbf{m}}
        \end{pmatrix}
        \;\;\overset{p}{\longrightarrow}\;\;
        \begin{pmatrix}
            \mathbf{G} \\
            \mathbf{h}
        \end{pmatrix}.
    \]
    Finally, the estimator defined in \eqref{eq:structural_para_estimator} is consistent: as $N \rightarrow \infty$,
    \[
        \hat{\boldsymbol{\alpha}} = \hat{\mathbf{M}}^{-1} \hat{\mathbf{m}} \;\;\overset{p}{\longrightarrow}\;\; \mathbf{G}^{-1} \mathbf{h} = \boldsymbol{\alpha}.
    \]
\end{proof}

\subsection{Asymptotic normality}
\begin{proof}
    To show asymptotic normality, we rewrite
    \[
        \sqrt{N}(\hat{\boldsymbol{\alpha}} - \boldsymbol{\alpha}) = \sqrt{N}(\hat{\mathbf{M}}^{-1} \hat{\mathbf{m}} - \boldsymbol{\alpha}) = \hat{\mathbf{M}}^{-1}\sqrt{N}(\hat{\mathbf{m}} - \hat{\mathbf{M}} \boldsymbol{\alpha}),
    \]
    With the definition
    \[
        \mathbf{l}(\boldsymbol{\eta}_i, \boldsymbol{\Upsilon}_{2}, \boldsymbol{\alpha}) 
        = \mathbf{m}(\boldsymbol{\eta}_i, \boldsymbol{\Upsilon}_{2}) 
        - \mathbf{M}(\boldsymbol{\eta}_{\mathrm{pred},i}, \boldsymbol{\Upsilon}_{2})\, \boldsymbol{\alpha},
    \]
    the second term becomes
    \[
        \hat{\mathbf{m}} - \hat{\mathbf{M}} \boldsymbol{\alpha} = \frac{1}{N} \sum_{i=1}^N \mathbf{l}(\hat{\boldsymbol{\eta}}_i, \hat{\boldsymbol{\Upsilon}}_{2},\boldsymbol{\alpha}).
    \]
    A first-order Taylor expansion of the moment function $\mathbf{l}$ around $(\tilde{\boldsymbol{\eta}}_i, \boldsymbol{\Upsilon}_2, \boldsymbol{\alpha})$ yields
    \[
        \mathbf{l}(\hat{\boldsymbol{\eta}}_i, \hat{\boldsymbol{\Upsilon}}_2, \boldsymbol{\alpha}_0)
        = \mathbf{l}(\tilde{\boldsymbol{\eta}}_i, \boldsymbol{\Upsilon}_2, \boldsymbol{\alpha}_0)
        + \left. \frac{\partial \mathbf{l}}{\partial \boldsymbol{\eta}_i^\top} \right|_{\tilde{\boldsymbol{\eta}}_i, \boldsymbol{\Upsilon}_2, \boldsymbol{\alpha}_0} (\hat{\boldsymbol{\eta}}_i - \tilde{\boldsymbol{\eta}}_i)
        + \left. \frac{\partial \mathbf{l}}{\partial \boldsymbol{\Upsilon}_2^\top} \right|_{\tilde{\boldsymbol{\eta}}_i, \boldsymbol{\Upsilon}_2, \boldsymbol{\alpha}_0} (\hat{\boldsymbol{\Upsilon}}_2 - \boldsymbol{\Upsilon}_2)+ o_p(N^{-1/2}).
    \]
    Substituting the first-stage expansion $\hat{\boldsymbol{\eta}}_i - \tilde{\boldsymbol{\eta}}_i = \mathbf{B}_i (\hat{\boldsymbol{\Upsilon}} - \boldsymbol{\Upsilon})$ gives
    \[
        \frac{1}{N}\sum_{i=1}^{N}
        \mathbf{l}(\hat{\boldsymbol{\eta}}_i, \hat{\boldsymbol{\Upsilon}}_2, \boldsymbol{\alpha}_0)
        = \frac{1}{N}\sum_{i=1}^{N}\Big\{
        \mathbf{l}(\tilde{\boldsymbol{\eta}}_i, \boldsymbol{\Upsilon}_2, \boldsymbol{\alpha}_0)
        + \overline{\mathbf{C}} \, \boldsymbol{\Delta}_i \Big\} + o_p(N^{-1/2}),
    \]
    where
    \[
        \overline{\mathbf{C}} = \operatorname{E}\!
        \left[
            \begin{pmatrix}
                \frac{\partial \mathbf{l}}{\partial \boldsymbol{\eta}_i^\top} \mathbf{B}_i & 
                \frac{\partial \mathbf{l}}{\partial \boldsymbol{\Upsilon}_2^\top}
            \end{pmatrix}
        \right].
    \]
    By the central limit theorem for i.i.d.\ variables (Assumption~\ref{assump:iid_sampling}) and the finite-moment conditions (Assumptions~\ref{assump:finite_moments}, \ref{assump:structural_error}), we obtain
    \[
        \sqrt{N} \left( \frac{1}{N} \sum_{i=1}^{N} \mathbf{l}(\hat{\boldsymbol{\eta}}_i, \hat{\boldsymbol{\Upsilon}}_2, \boldsymbol{\alpha}_0) \right)
        \overset{d}{\longrightarrow} \mathcal{N}\big(\mathbf{0}, \mathbf{S} \big),
    \]
    with
    \[
        \mathbf{S} = \operatorname{Var}\big[ \mathbf{l}(\tilde{\boldsymbol{\eta}}_i, \boldsymbol{\Upsilon}_2, \boldsymbol{\alpha}_0) + \overline{\mathbf{C}} \, \boldsymbol{\Delta}_i \big].
    \]
    Finally, using Assumption~\ref{assump:treat_cov_inter} and applying the delta method gives
    \[
        \sqrt{N} (\hat{\boldsymbol{\alpha}} - \boldsymbol{\alpha}_0) \overset{d}{\longrightarrow} \mathcal{N}\big( \mathbf{0}, \mathbf{G}^{-1} \mathbf{S} \mathbf{G}^{-\top} \big).
    \]
\end{proof}

\section{Additional Plots for Simulation Study 1}
\label{sec:append_plots}

\begin{figure}
     \centering
      \begin{subfigure}[b]{\textwidth}
      \centering
     \includegraphics[width=0.65\textwidth]{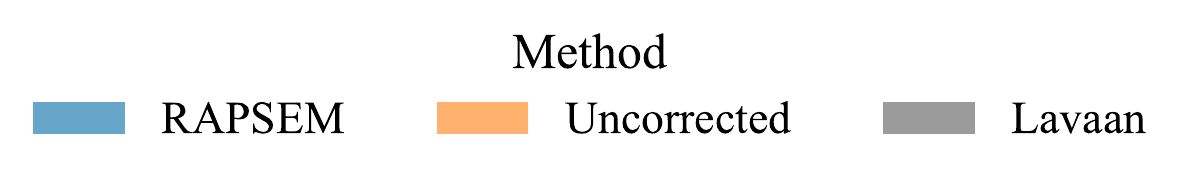}
     \end{subfigure}
     \begin{subfigure}[b]{0.31\textwidth}
         \centering
         \includegraphics[height=3.1cm]{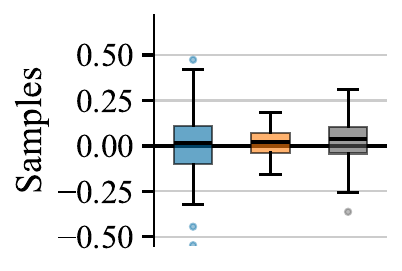}
         \caption{\centering $\delta_{u}=0.100,$\\$N=250$}
         \label{fig:conf_0.1_n250}
     \end{subfigure}
     \hspace{-1em}
     \begin{subfigure}[b]{0.23\textwidth}
         \centering
         \includegraphics[height=3.1cm]{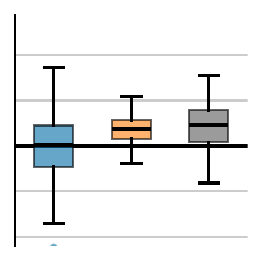}
         \caption{\centering$\delta_{u}=0.245,$\\$N=250$}
         \label{fig:conf_0.245_n250}
     \end{subfigure}
     \hspace{-1em}
     \begin{subfigure}[b]{0.23\textwidth}
         \centering
         \includegraphics[height=3.1cm]{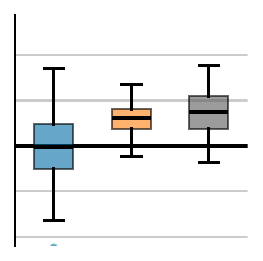}
         \caption{\centering $\delta_{u}=0.316,$\\$N=250$}
         \label{fig:conf_0.316_n250}
     \end{subfigure}
     \hspace{-1em}
     \begin{subfigure}[b]{0.23\textwidth}
         \centering
         \includegraphics[height=3.1cm]{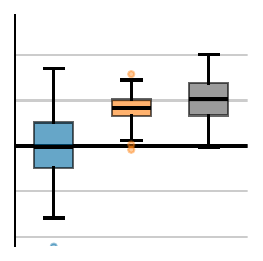}
         \caption{\centering $\delta_{u}=0.374,$\\$N=250$}
         \label{fig:conf_0.374_n250}
     \end{subfigure}
     \par
     \begin{subfigure}[b]{0.31\textwidth}
         \centering
         \includegraphics[height=3.1cm]{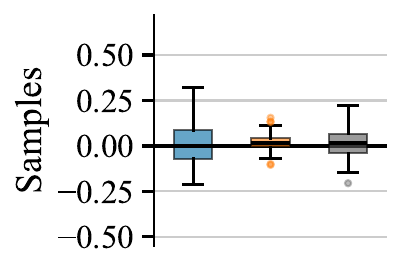}
         \caption{\centering $\delta_{u}=0.100,$\\$N=500$}
         \label{fig:conf_0.1_n500}
     \end{subfigure}
     \hspace{-1em}
     \begin{subfigure}[b]{0.23\textwidth}
         \centering
         \includegraphics[height=3.1cm]{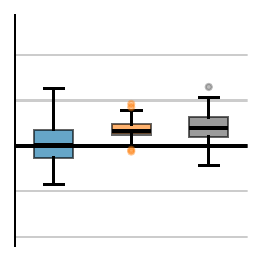}
         \caption{\centering $\delta_{u}=0.245,$\\$N=500$}
         \label{fig:conf_0.245_n500}
     \end{subfigure}
     \hspace{-1em}
     \begin{subfigure}[b]{0.23\textwidth}
         \centering
         \includegraphics[height=3.1cm]{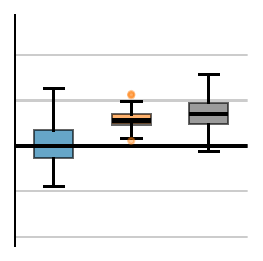}
         \caption{\centering $\delta_{u}=0.31,$\\$N=500$}
         \label{fig:conf_0.316_n500}
     \end{subfigure}
     \hspace{-1em}
     \begin{subfigure}[b]{0.23\textwidth}
         \centering
         \includegraphics[height=3.1cm]{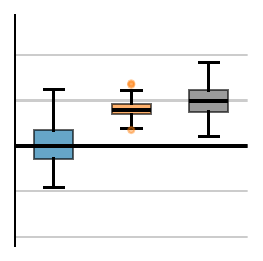}
         \caption{\centering $\delta_{u}=0.374,$\\$N=500$}
         \label{fig:conf_0.374_n500}
     \end{subfigure}
     \par
     \begin{subfigure}[b]{0.31\textwidth}
         \centering
         \includegraphics[height=3.1cm]{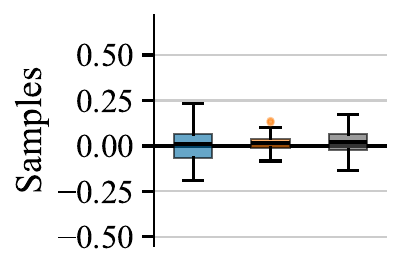}
         \caption{\centering $\delta_{u}=0.100,$\\$N=750$}
         \label{fig:conf_0.1_n750}
     \end{subfigure}
     \hspace{-1em}
     \begin{subfigure}[b]{0.23\textwidth}
         \centering
         \includegraphics[height=3.1cm]{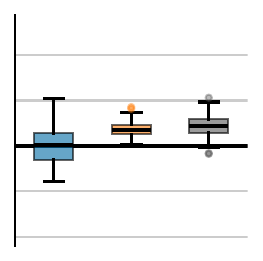}
         \caption{\centering $\delta_{u}=0.245,$\\$N=750$}
         \label{fig:conf_0.245_n750}
     \end{subfigure}
     \hspace{-1em}
     \begin{subfigure}[b]{0.23\textwidth}
         \centering
         \includegraphics[height=3.1cm]{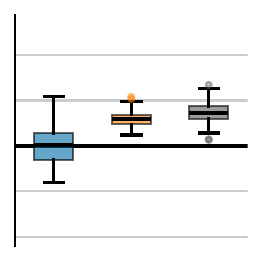}
         \caption{\centering $\delta_{u}=0.316,$\\$N=750$}
         \label{fig:conf_0.316_n750}
     \end{subfigure}
     \hspace{-1em}
     \begin{subfigure}[b]{0.23\textwidth}
         \centering
         \includegraphics[height=3.1cm]{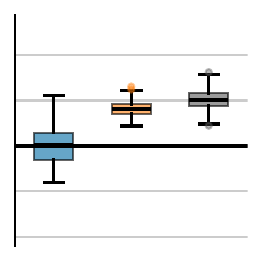}
         \caption{\centering $\delta_{u}=0.374,$\\$N=750$}
         \label{fig:conf_0.374_n750}
     \end{subfigure}
     \par
     \begin{subfigure}[b]{0.31\textwidth}
         \centering
         \includegraphics[height=3.1cm]{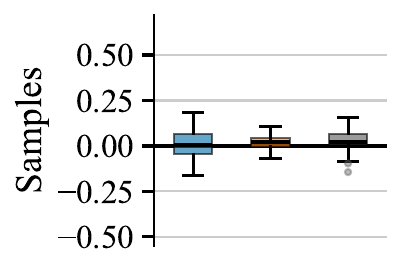}
         \caption{\centering $\delta_{u}=0.100,$\\$N=1000$}
         \label{fig:conf_0.1_n1000}
     \end{subfigure}
     \hspace{-1em}
     \begin{subfigure}[b]{0.23\textwidth}
         \centering
         \includegraphics[height=3.1cm]{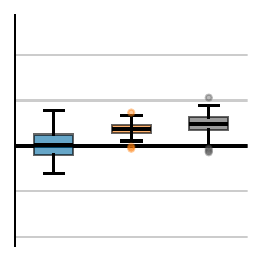}
         \caption{\centering $\delta_{u}=0.245,$\\$N=1000$}
         \label{fig:conf_0.245_n1000}
     \end{subfigure}
     \hspace{-1em}
     \begin{subfigure}[b]{0.23\textwidth}
         \centering
         \includegraphics[height=3.1cm]{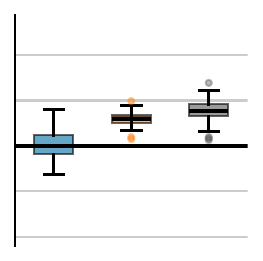}
         \caption{\centering $\delta_{u}=0.316,$\\$N=1000$}
         \label{fig:conf_0.316_n1000}
     \end{subfigure}
     \hspace{-1em}
     \begin{subfigure}[b]{0.23\textwidth}
         \centering
         \includegraphics[height=3.1cm]{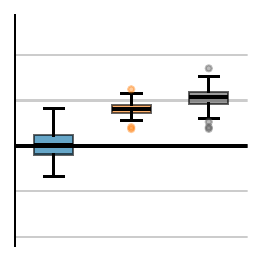}
         \caption{\centering $\delta_{u}=0.374,$\\$N=1000$}
         \label{fig:conf_0.374_n1000}
     \end{subfigure}
     \caption{Sampling distributions of $\theta_m$ under varying confounding levels and sample sizes.}
     \label{fig:bias_histograms}
\end{figure}

\Cref{fig:bias_histograms} displays the sampling distributions of $\hat{\theta}_m$ across the setting of Simulation Study 1. This illustrates the bias and variance patterns discussed in the main text: RAPSEM distributions are centered at zero throughout, while standard SEM distributions shift progressively rightward with increasing unobserved confounding effect size $\delta_u$. The wider spread of RAPSEM's distributions at small sample sizes is also clearly visible, narrowing relative to standard SEM as the sample size $N$ increases.

\begin{figure}
     \centering
     \begin{subfigure}[b]{0.31\textwidth}
         \centering
         \includegraphics[height=3.5cm]{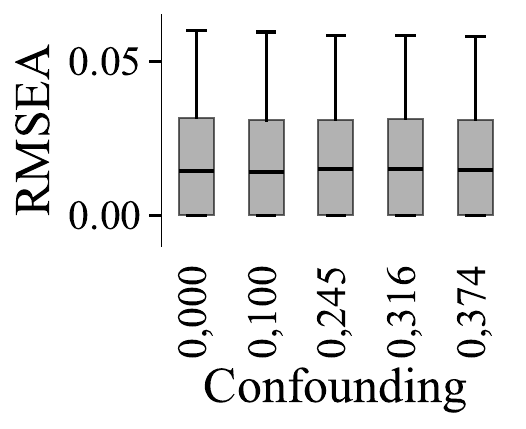}
         \caption{\centering RMSEA, $N=250$}
         \label{fig:RMSEA_n250}
     \end{subfigure}
     \hspace{-1em}
     \begin{subfigure}[b]{0.23\textwidth}
         \centering
         \includegraphics[height=3.5cm]{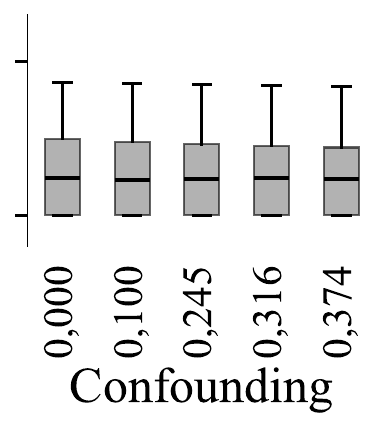}
         \caption{\centering RMSEA, $N=500$}
         \label{fig:RMSEA_n500}
     \end{subfigure}
     \hspace{-1em}
     \begin{subfigure}[b]{0.23\textwidth}
         \centering
         \includegraphics[height=3.5cm]{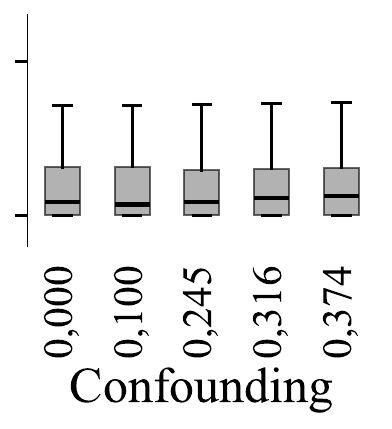}
         \caption{\centering RMSEA, $N=750$}
         \label{fig:RMSEA_n750}
     \end{subfigure}
     \hspace{-1em}
     \begin{subfigure}[b]{0.23\textwidth}
         \centering
         \includegraphics[height=3.5cm]{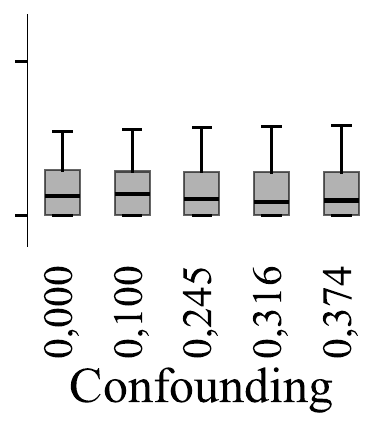}
         \caption{\centering RMSEA, $N=1000$}
         \label{fig:RMSEA_n1000}
     \end{subfigure}
     \par
     \begin{subfigure}[b]{0.31\textwidth}
         \centering
         \includegraphics[height=3.5cm]{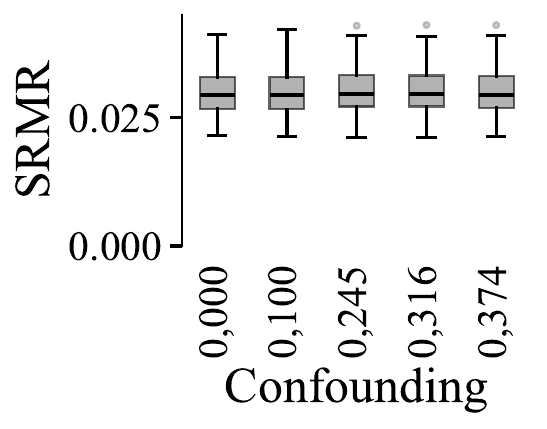}
         \caption{\centering SRMR, $N=250$}
         \label{fig:SRMR_n250}
     \end{subfigure}
     \hspace{-1em}
     \begin{subfigure}[b]{0.23\textwidth}
         \centering
         \includegraphics[height=3.5cm]{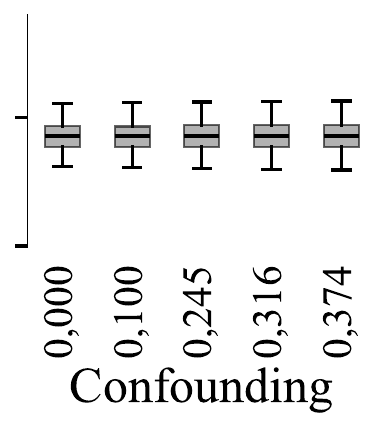}
         \caption{\centering SRMR, $N=500$}
         \label{fig:SRMR_n500}
     \end{subfigure}
     \hspace{-1em}
     \begin{subfigure}[b]{0.23\textwidth}
         \centering
         \includegraphics[height=3.5cm]{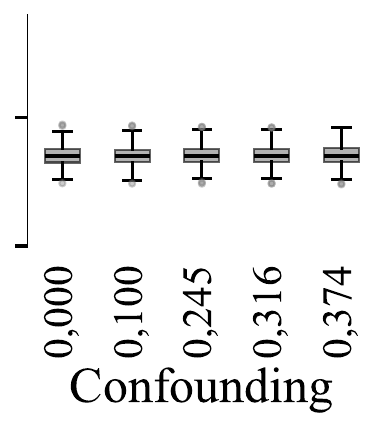}
         \caption{\centering SRMR, $N=750$}
         \label{fig:SRMR_n750}
     \end{subfigure}
     \hspace{-1em}
     \begin{subfigure}[b]{0.23\textwidth}
         \centering
         \includegraphics[height=3.5cm]{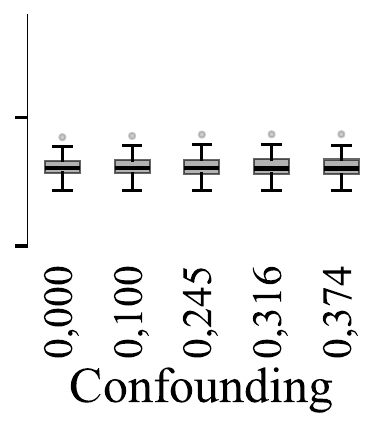}
         \caption{\centering SRMR, $N=1000$}
         \label{fig:SRMR_n1000}
     \end{subfigure}
     \par
     \begin{subfigure}[b]{0.31\textwidth}
         \centering
         \includegraphics[height=3.5cm]{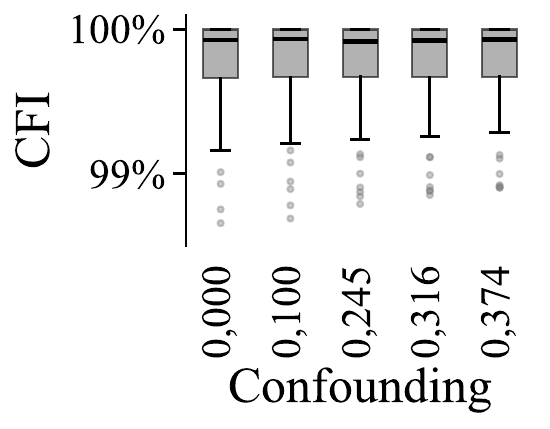}
         \caption{\centering CFI, $N=250$}
         \label{fig:CFI_n250}
     \end{subfigure}
     \hspace{-1em}
     \begin{subfigure}[b]{0.23\textwidth}
         \centering
         \includegraphics[height=3.5cm]{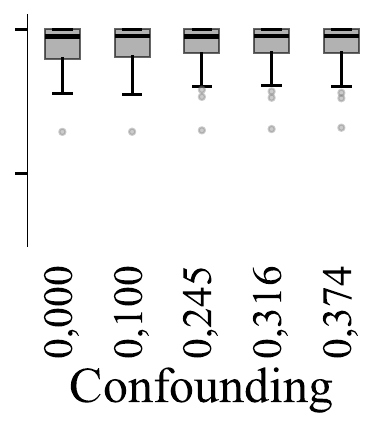}
         \caption{\centering CFI, $N=500$}
         \label{fig:CFI_n500}
     \end{subfigure}
     \hspace{-1em}
     \begin{subfigure}[b]{0.23\textwidth}
         \centering
         \includegraphics[height=3.5cm]{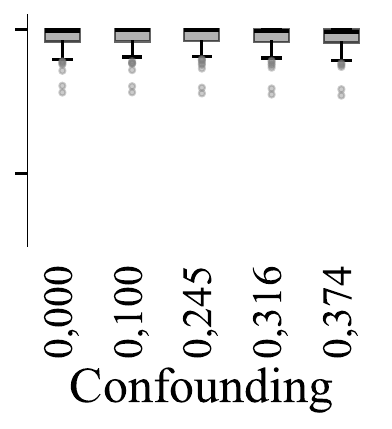}
         \caption{\centering CFI, $N=750$}
         \label{fig:CFI_n750}
     \end{subfigure}
     \hspace{-1em}
     \begin{subfigure}[b]{0.23\textwidth}
         \centering
         \includegraphics[height=3.5cm]{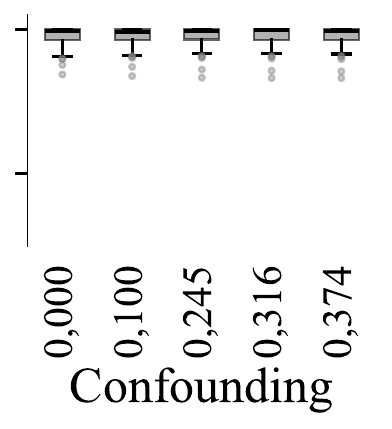}
         \caption{\centering CFI, $N=1000$}
         \label{fig:CFI_n1000}
     \end{subfigure}
     \par
     \begin{subfigure}[b]{0.31\textwidth}
         \centering
         \includegraphics[height=3.5cm]{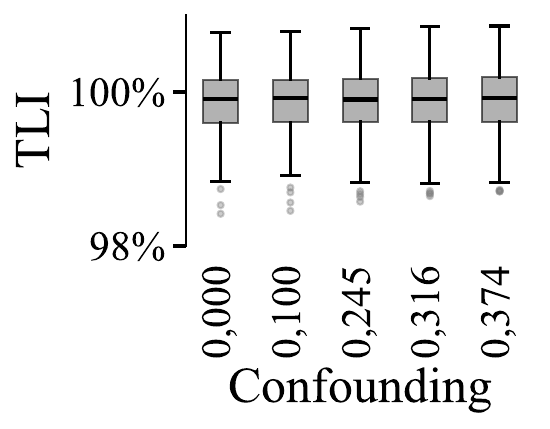}
         \caption{\centering TLI, $N=250$}
         \label{fig:TLI_n250}
     \end{subfigure}
     \hspace{-1em}
     \begin{subfigure}[b]{0.23\textwidth}
         \centering
         \includegraphics[height=3.5cm]{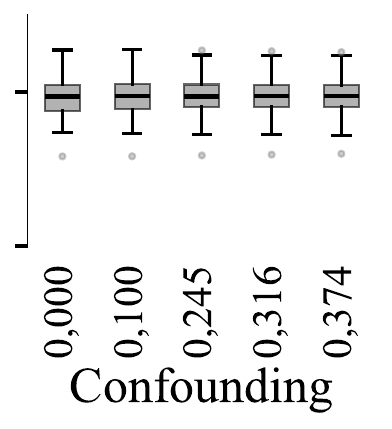}
         \caption{\centering TLI, $N=500$}
         \label{fig:TLI_n500}
     \end{subfigure}
     \hspace{-1em}
     \begin{subfigure}[b]{0.23\textwidth}
         \centering
         \includegraphics[height=3.5cm]{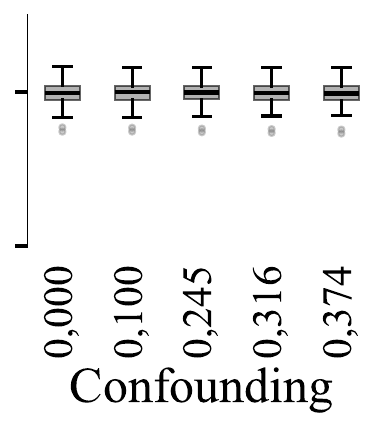}
         \caption{\centering TLI, $N=750$}
         \label{fig:TLI_n750}
     \end{subfigure}
     \hspace{-1em}
     \begin{subfigure}[b]{0.23\textwidth}
         \centering
         \includegraphics[height=3.5cm]{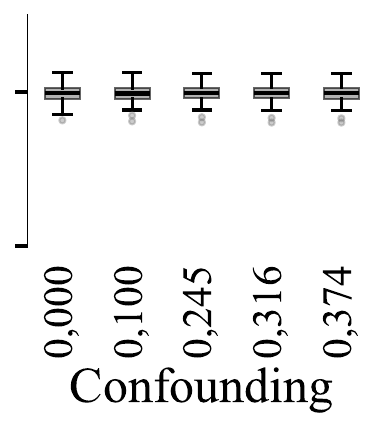}
         \caption{\centering TLI, $N=1000$}
         \label{fig:TLI_n1000}
     \end{subfigure}
     \caption{Lavaan model fit indices across simulation runs under varying confounding levels and sample size.}
     \label{fig:model_fit}
\end{figure}

\Cref{fig:model_fit} shows the fit indices for the lavaan model in Simulation Study 1. For all model runs, they remain within the common thresholds for good model fit (below $0.06$ for RMSEA, below $0.08$ for SRMR, and above $95\%$ for CFA and TLI). Notably, model fit hardly changes for different levels of confounding.

\newpage
\section{Reliabilities in the Empirical Example}
\label{sec:append_rel}

To assess the internal consistency of the latent factors, Table~\ref{tab:reliability} reports the pooled McDonald's $\omega$ reliability coefficients along with their standard errors.

\begin{table}[ht!]
  \centering
  \caption{Pooled McDonald's $\omega$ Reliability Coefficients for the full Model M1.}
  \label{tab:reliability}
  \begin{tabular}{lcc}
    \toprule
    \textbf{Factor} & $\hat{\omega}$ & \textbf{SE} \\
    \midrule
         conceptual knowledge (outcome) &  0.65 & 0.04 \\
         task interest (mediator) & 0.86 & 0.01 \\
         interest in physics & 0.84 & 0.01 \\
         conscientiousness in physics & 0.75 & 0.02 \\
         perceived teacher support & 0.63 & 0.04 \\
         perceived cognitive activation & 0.79 & 0.03 \\
    \bottomrule
  \end{tabular}
\end{table}

\end{document}